\documentclass[twocolumn,twocolappendix]{aastex701}
\usepackage{apjfonts}
\usepackage{rotating}
\usepackage[toc,page]{appendix}
\usepackage{amsmath,color,graphicx,tablefootnote}
\usepackage{graphics,subfigure,hyperref,float}
\usepackage[rightcaption]{sidecap}
\usepackage{bm,comment}

\newcommand{\CM}{\checkmark}
\newcommand{\W}{$\lambda$}
\newcommand{\D}{$\Delta$}

\newcommand{\r}{\mathrm}
\newcommand{\ns}{\normalsize}

\newcommand{\ssn}{\scriptscriptstyle}
\newcommand{\dr}{\displaystyle\mathrm}
\newcommand{\sr}{\scriptstyle\mathrm}
\newcommand{\sxr}{\scriptscriptstyle\mathrm}

\newcommand{\s}{\hskip0.16667em\relax}
\newcommand{\g}{\nobreak\hskip0.16667em\relax}

\def\beq{\begin{equation}}
\def\eeq{\end{equation}}
\def\alwaysmath#1{{\ifmmode{#1}\else{$#1$}\fi}}
\def\iso#1#2{\mbox{${}^{#2}{\rm #1}$}}
\def\he#1{\iso{He}{#1}}
\def\li#1{\hbox{\alwaysmath{{}^{#1}}{\rm Li}}}

\def\sun{${\,_\odot}$}
\def\10830{He\,{\small I}\s$\lambda$10830}
\def\3889{He\,{\small I}\s$\lambda$3889}
\def\y+{\ensuremath{\mathrm{y}^{+}}}
\def\ydp{\ensuremath{\mathrm{y}^{++}}}
\def\X2{\ensuremath{\chi^{2}}}
\newcommand{\Yp}{\ensuremath{\mathrm{Y_{p}}}}
\newcommand{\yp}{\ensuremath{\mathrm{y_{p}}}}
\newcommand{\yavg}{\ensuremath{\mathrm{\langle y \rangle}}}
\newcommand{\Yavg}{\ensuremath{\mathrm{\langle Y \rangle}}}
\newcommand{\Te}{\ensuremath{\mathrm{T_{e}}}}
\newcommand{\Dy}{\ensuremath{\mathrm{\Delta y/\Delta(O/H)}}}
\newcommand{\DY}{\ensuremath{\mathrm{\Delta Y/\Delta(O/H)}}}
\newcommand{\OHc}{\ensuremath{\mathrm{O/H < 4 \times 10^{-5}}}}
\newcommand{\OHe}{\ensuremath{\mathrm{O/H < 8 \times 10^{-5}}}}
\newcommand{\OHge}{\ensuremath{\mathrm{O/H > 8 \times 10^{-5}}}}
\def\T4{\ensuremath{\mathrm{T_{4}}}}
\newcommand{\n}{\ensuremath{\mathrm{n_{e}}}}
\newcommand{\Proj}{the LBT \Yp\ Project}
\newcommand{\cm}{\ensuremath{\mathrm{cm^{-3}}}}

\newcommand\hii{H\,{\small II} }
\newcommand\hei{He\,{\small I}\s}
\newcommand\heins{He\,{\small I}}
\newcommand\heil{He\,{\small I}\s$\lambda$}
\newcommand\heiil{He\,{\small II}\s$\lambda$}
\newcommand\hi{H\,{\small I} }
\newcommand\oiii{O\,{\small III}}
\newcommand\siii{S\,{\small III}}
\newcommand\sii{S\,{\small II}}
\newcommand\nai{Na\,{\small I}}
\newcommand\fev{Fe\,{\small V}}
\newcommand\nev{Ne\,{\small V}}

\defcitealias{benj2002}{BSS02}
\defcitealias{kuri2025}{KI25}
\defcitealias{Planck2018}{Planck Collaboration}
\defcitealias{Atacama2025}{Atacama Cosmology Telescope}
\defcitealias{SPT3G2025}{SPT-3G Collaboration}

\setcitestyle{notesep={}}

\shorttitle{The LBT \Yp\ Project IV}
\shortauthors{Aver et al.}

\begin{document}

\title{The LBT \Yp\ Project IV: A New Value of the Primordial Helium Abundance}

\author[0009-0006-2077-2552]{Erik Aver}
\affiliation{Department of Physics, Gonzaga University, 502 E Boone Ave., Spokane, WA, 99258, USA}
\email{aver@gonzaga.edu}
\author[0000-0003-0605-8732]{Evan D.\ Skillman}
\affiliation{Minnesota Institute for Astrophysics, University of Minnesota, 116 Church St. SE, Minneapolis, MN 55455, USA}
\email{skill001@umn.edu}
\author[0000-0003-1435-3053]{Richard W.\ Pogge}
\affiliation{Department of Astronomy, The Ohio State University, 140 W 18th Ave., Columbus, OH, 43210, USA}
\affiliation{Center for Cosmology \& AstroParticle Physics, The Ohio State University, 191 West Woodruff Avenue, Columbus, OH 43210, USA}
\email{pogge.1@osu.edu}
\author[0000-0002-0361-8223]{Noah S.\ J.\ Rogers}
\affiliation{Center for Interdisciplinary Exploration and Research in Astrophysics (CIERA), Northwestern University, 1800 Sherman Avenue, Evanston, IL 60201, USA}
\email{noah.rogers@northwestern.edu}
\author[0000-0003-4912-5157]{Miqaela K.\ Weller}
\affiliation{Department of Astronomy, The Ohio State University, 140 W 18th Ave., Columbus, OH, 43210, USA}
\email{weller.133@buckeyemail.osu.edu}
\author[0000-0001-7201-5998]{Keith A.\ Olive}
\affiliation{William I.\ Fine Theoretical Physics Institute, School of Physics and Astronomy, University of Minnesota, Minneapolis, MN 55455, USA}
\email{olive@umn.edu}
\author[0000-0002-4153-053X]{Danielle A.\ Berg}
\affiliation{Department of Astronomy, The University of Texas at Austin, 2515 Speedway, Stop C1400, Austin, TX 78712, USA}
\email{daberg@austin.utexas.edu}
\author[0000-0001-8483-603X]{John J.\ Salzer}
\affiliation{Department of Astronomy, Indiana University, 727 East Third Street, Bloomington, IN 47405, USA}
\email{josalzer@iu.edu}
\author[0000-0002-2901-5260]{John H.\ Miller, Jr.}
\affiliation{Minnesota Institute for Astrophysics, University of Minnesota, 116 Church St. SE, Minneapolis, MN 55455, USA}
\email{mill9614@umn.edu}
\author[0000-0002-6972-6411]{Jos\'e Eduardo M\'endez-Delgado}
\affiliation{Instituto de Astronom\'ia, Universidad Nacional Aut\'onoma de M\'exico, Ap. 70-264, 04510 CDMX, M\'exico}
\email{jmendez@astro.unam.mx}
%


\begin{abstract}
We present a new determination of the primordial helium abundance based on new, high-quality Large Binocular Telescope (LBT) observations of 54 metal-poor \hii regions.  These regions have been observed and analyzed uniformly.  We also describe a number of updates to our methodology, including updated helium emissivities.  Enabled by the large, high-quality dataset, we examine our sample targets for potential systematic errors, which could bias their results.  We perform a standard 95\% confidence level \X2 cut and find that a significantly larger fraction (47/54 = 87\%) of our sample qualifies than for previous datasets.  We also screen for quality and reliability, flagging targets which may introduce significant systematic errors, producing a dataset of 41 targets.  In a significant breakthrough for the field, that dataset includes 15 high SNR targets with low metallicity (\OHc).  Due to this low-metallicity dataset, for the first time, a weighted average for determining the primordial helium abundance (\Yp) is well-justified and produces a robust result.  By weighted average of our 15 low-metallicity targets, we determine $\Yp = 0.2458 \pm 0.0013$.  This result achieves an unprecedented precision of 0.5\%, and it is in good agreement with the BBN result, $\Yp = 0.2467 \pm 0.0002$, based on the Planck determination of the baryon density.  
\end{abstract}



\section{Introduction} \label{Intro}


Observations of the cosmic microwave background (CMB) and its anisotropy spectrum have provided unparalleled information about the early universe up to a redshift $z\simeq 1100$ (\citetalias{Planck2018} \citeyear{Planck2018}). Examples of high precision determinations include the Hubble parameter, cold dark matter density, baryon density, and total density. With higher resolution experiments \citep{ACT:2025fju,SPT3G2025}, the primordial helium abundance, \Yp , and the effective number of neutrino degrees of freedom, $N_{\rm eff}$,  may soon join this list.  Big bang nucleosynthesis (BBN), however, enables one to probe the early Universe back to redshifts of order $z \sim 10^{10}$ when the temperature of the radiation background was of order 1 MeV. At these temperatures, protons and neutrons are processed to form the light element isotopes, D, \he3, \he4, and \li7 \citep{olive2000, Iocco:2008va,cyburt2016, Pitrou:2018cgg, fields2020}.   

Standard Model physics up to the MeV scale is well understood and if provided with precision determinations of the light element abundances, the physics of the early Universe can be tested.  To do so requires accurate {\em primordial abundance determinations}, that is, abundances in astrophysical environments which have not been contaminated by either stellar production, astration, or depletion. For example, \he3 is both produced and destroyed in stars \citep{Olive:1994fq,Dearborn:1995ex,Olive:1996tt} making it difficult to extract a primordial abundance from available observations \citep{Vangioni-Flam:2002cvh}.  \li7 abundance determinations in low mass halo dwarf stars were long thought to provide a good measure of the primordial abundance \citep{Spite:1982dd,Yang:1983gn}.  The degree to which \li7 has been destroyed in its host star is unclear \citep{Fields:2022mpw}.  In contrast, there are several high precision determinations of D/H in high redshift ($z\sim 3$) quasar absorption systems \citep{cook2014,cook2016,cook2018, riem2015,riem2017, bala2016,zava2018,guarneri2024MNRAS.529..839G,kislitsyn2024MNRAS.528.4068K}, which have enabled precision tests of the compatibility between BBN and CMB observations. Our goal here is to provide a high precision determination of the primordial \he4 abundance. 

\Yp\ needs to be measured to high accuracy to provide any diagnostic power.  In the currently favored method of measuring relative abundances in ionized plasmas associated with low-metallicity star forming regions, this means accounting for a number of effects that can alter the observed ratios of the H and He emission lines.  In addition to measuring the appropriate values of temperature and density, which determine the emissivities required to convert the recombination emission line ratios into relative abundances, there are a number of other factors that need to be accounted for.  These have been identified as reddening by the intervening ISM, the stellar absorption profiles underlying the emission lines, the collisional excitation of H lines (which requires measuring the neutral hydrogen fraction), and the radiative transfer gains and losses in the He triplet lines due to the presence of the $2^3S$ metastable level.  All of these effects have various histories of corrections. Those histories will not be enumerated here, but we strive to be as clear as possible concerning how we deal with each of these corrections, as discussed in Section \ref{Model}.

The absence of direct measurements of the physical quantities which enter into the model may lead to significant systematic uncertainties in the derived helium abundance.  Historically, reasonable values for the physical parameters were adopted ignoring their inherent uncertainties. Our method for accounting for and estimating the size of the systematic errors is described in Section \ref{Model}. The physical parameters are treated on an equal footing with the helium abundance and, together, they are all determined simultaneously by minimizing the $\chi^2$ likelihood function, which compares the observed emission line fluxes with model predictions.
A Markov Chain Monte Carlo (MCMC) exploration of the parameter space then allows us to determine the uncertainties in each parameter including the helium abundance. This method typically leads to larger calculated uncertainties in the helium abundance, thereby motivating higher precision to reduce these (more realistic) uncertainties. 

Based on the success in significantly lowering the uncertainty on a helium abundance measurement in the metal-poor galaxy Leo\,P \citep{aver2021}, we proposed to obtain new, high quality optical and NIR spectra of as many known low-metallicity galaxies as possible.  For that project, we prioritize on low values of measured oxygen abundance and high values of hydrogen emission line flux and equivalent width.  Many known low-metallicity galaxies are too faint to provide sufficiently small uncertainties on their helium abundances with a reasonable amount of telescope time, so not all known low-metallicity galaxies were pursued.  Nonetheless, we were able to compile a sample of high quality spectra enabling a significant step forward in the determination of \Yp.

This is the fourth paper in the series of the LBT \Yp\ project.  Papers~I-III \citep{Skillman2026, Rogers2026, Weller2026} presented an overview of the project, our new LBT MODS optical spectra, and our new LBT LUCI NIR spectra, respectively.  This paper is followed by Paper~V \citep{Yeh2026} demonstrating the impact of our new \Yp\ determination, and Paper~VI \citep[][, in prep]{Rogers2026b} will present a complete survey of all of the metal abundances derived from our new LBT MODS spectra.

This paper is organized as follows.  First, in Section \ref{Model}, we discuss the basic theoretical model for the analysis of \he4 data, including recent updates that have been made.  Model equations and data for those updates are provided in the \hyperref[Appendix]{Appendix}. In Section \ref{Sample}, we provide the requirements for inclusion in the sample used for determining \Yp.  The results of the \X2 cut for our sample are discussed in Section \ref{chi2cut}.  We then examine that qualifying dataset for systematic errors in Section \ref{Testing}, and we review the best-fit physical parameters and the effects of key model updates in Section \ref{Results}.  In Section \ref{Yp}, we then derive the resulting primordial helium abundance.  Two avenues for future investigation and improvements to further reduce potential systematic errors are discussed in Section \ref{Future}.  Finally, Section \ref{Conclusion} offers a discussion of the results and prospects for improvement.


\section{Model} \label{Model}


\hii regions are emission nebulae photoionized by one or more massive stars (O \& B stars).  Correspondingly, their spectra include nebular emission lines overlaid on a stellar continuum.  In seeking the helium abundance, $y^{+}=\frac{n(\r{He^+})}{n(\r{H^+})}$, we model the primary emission region using a radiative transfer model based on the ``Case B'' approximation (optically thick to Lyman lines).  For hydrogen and helium, recombination emission dominates, but collisional excitation also contributes, with hydrogen collisional excitation from the small fraction of neutral hydrogen in the ionized region, $\xi$.  Self-absorption and re-emission of photons within the nebula also occurs.  The radiative transfer model assumes a uniform \hii region with an average electron density, $n_e$, and average electron temperature, $T_e$, as well as optical depth, $\tau$.  The stellar continuum juxtaposes absorption features under the nebular helium and hydrogen emission lines, including separate parameters for the Balmer and Paschen lines, $a_{He}$, $a_H$, and $a_P$, respectively \citep{aver2021}.  Dust along the line of sight also scatters the emitted photons (interstellar reddening, $C(\r{H\beta})$).  

We determine the helium abundance in an individual \hii region based on a MCMC analysis. The MCMC method is an algorithmic procedure for sampling from a statistical distribution \citep{mark1906,metr1953}.  
We define a \X2 distribution from the difference between
flux ratios, 
\beq
\chi^2 = \sum_{\lambda} \frac{\left(\frac{F(\lambda)}{F(H\beta|P\gamma)} - {\frac{F(\lambda)}{F(H\beta|P\gamma)}}_{meas}\right)^2} { \sigma(\lambda)^2},
\label{eq:X2}
\eeq
where the emission line fluxes, $F(\lambda)$, are measured and calculated for a set of H and He lines, and $\sigma(\lambda)$ is the measured uncertainty in the flux ratio at each wavelength. 
The optical/near-IR emission line fluxes from the LBT/MODS spectra are calculated relative to H$\beta$, while the infrared flux of \10830 is calculated relative to the IR Paschen line, P$\gamma$.  Thus, by $F(H\beta|P\gamma)$, we mean $F(H\beta)$ for all lines other than \10830, and $F(P\gamma)$ for the latter. 

Up to eight helium line ratios are employed:  \W\W4026, 4388, 4471, 4922, 5876, 6678, and 7065, relative to H$\beta$, and \W10830, relative to P$\gamma$.  \heil5016 was previously employed \citep{aver2021}, and \heil7281 is available in our model, but, as discussed below (\S \ref{Eduardo}), there is evidence for significant systematic bias in these two emission lines \citep{izot2007,mend2025}.  As a result, they have been excluded from our analysis here.  Up to fifteen hydrogen line ratios are employed:  H$\alpha$, H$\gamma$, H$\delta$, H9, H10, H11, H12, P8, P9, P10, P11, P12, P13, P14, and P15, relative to H$\beta$.  Finally, the blended line \3889 + H8, relative to H$\beta$, is also utilized. Depending on the observed target's S/N, not all of the preceding lines are detected and used (see Section \ref{RemovedLines}).  

The observed line ratios are used to fit up to nine model parameters, as introduced above \citep{aver2021}: $T_e$, $n_e$, $\tau$, $C(H\beta)$, $a_{He}$, $a_H$, $a_P$, $\xi$, and $y^+$.  Thus, if all of the above emission lines are detected and employed, we are left with 15 degrees of freedom, corresponding to a total of 24 observed line ratios and 9 parameters to fit. 

The MCMC scans of our 9-dimensional parameter space map out the \X2 distribution given above.  We conduct a frequentist analysis, and the \X2 is minimized to determine the best-fit solution for the nine physical parameters, including \y+, as well as determining the ``goodness-of-fit''.  Uncertainties in each quantity are estimated by calculating a 1D marginalized likelihood and finding the 68\% confidence interval from the increase in the \X2 from the minimum.  

Our entire model and the data employed, including the model equations for the helium and hydrogen flux ratios and relevant parameterizations, are given in full and detailed in the extensive appendices in \citet{aver2021}.  Please see that work for details and a comprehensive accounting.  Since that work, the following updates have been made to our model:  
\begin{enumerate}
    \item The helium and hydrogen emissivities have been updated to employ the most recent calculations based on the latest atomic data and modeling improvements \citep{delz2022,stor2015}.  These helium and hydrogen emissivities have been computed on higher-resolution temperature and density grids. 
    \item The radiative transfer model has been updated to employ the most recent calculations based on the latest atomic data and modeling improvements \citep{kuri2025}.  Previously, limited-range fitting equations were employed \citep{benj2002}.  In addition to the updates to the calculations, the new radiative transfer equations were fit over an expanded parameter range, largely eliminating potential extrapolation. 
    \item Three hydrogen Paschen series emission lines, P13, P14, and P15, were added to the model, increasing the maximum number of Paschen lines employed from five to eight, spanning from P8 to P15 (all in the NIR from the MODS Optical + NIR Spectrum).  
    \item The radiative transfer treatment of the blended emission line H8 + \3889 was revised.  In particular, the radiative transfer correction to H8 due to absorption by \3889 was removed.  The two lines are separated by 0.4~\AA, and the number density of He atoms compared to H atoms is less than one tenth.  As a result, the absorption of H8-emitted photons by He is expected to be relatively rare with negligible impacts on the H8 or He fluxes (\hei \W\W3889, 7065, 10830).  
    \item The neutral hydrogen fraction, $\xi$, is not included as a fit parameter in our model for temperatures below 14,000\,K.  Below 14,000\,K, the potential emission from the collisional excitation is negligible, due to its exponential dependence on temperature.  
\end{enumerate}

Further discussion of the new helium and hydrogen emissivities, the new radiative transfer calculations, the excluded \heil5016 \& \W7281 emission lines, and the treatment of neutral hydrogen collisional excitation is given below (\S \ref{Emiss}, \S \ref{RadTrans}, \S \ref{Eduardo}, \& \ref{NHCC}, respectively).  The corresponding updates to the equations and data provided in the Appendices in \citet{aver2021} are provided here in Appendices \ref{Appendix:Blended}-\ref{Appendix:P13-15}.

\subsection{He \& H Emissivities} \label{Emiss}

\begin{figure}[t]
\resizebox{\columnwidth}{!}{\includegraphics{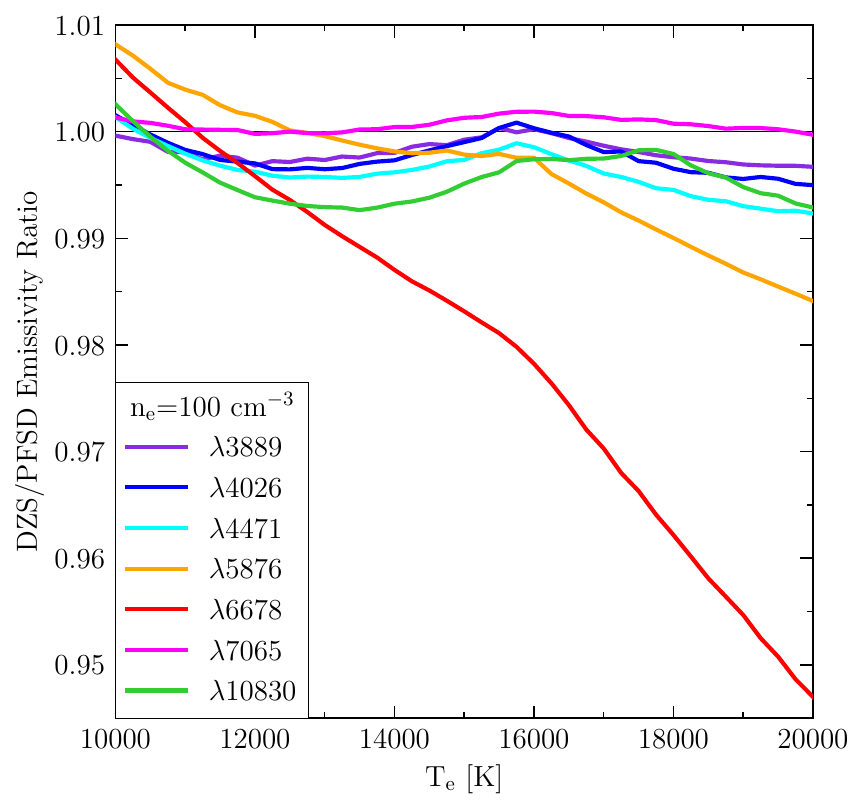}}
\caption{
The \citet{delz2022} emissivities plotted relative to those of \citet{port2012, port2013} as a function of temperature at $\n = 100~\r{cm}^{-3}$ for the core He emission lines, \hei \W\W3889, 4026, 4471, 5876, 6678, 7065, 10830.  At low density, all of the He emission lines employed in this work show changes of $\lesssim 1\%$, and typically a slight decrease, except for \heil6678 which shows a more significant decrease.  
}
\label{figure:DSZ-PFSD-T}
\end{figure}

\begin{figure}[t]
\resizebox{\columnwidth}{!}{\includegraphics{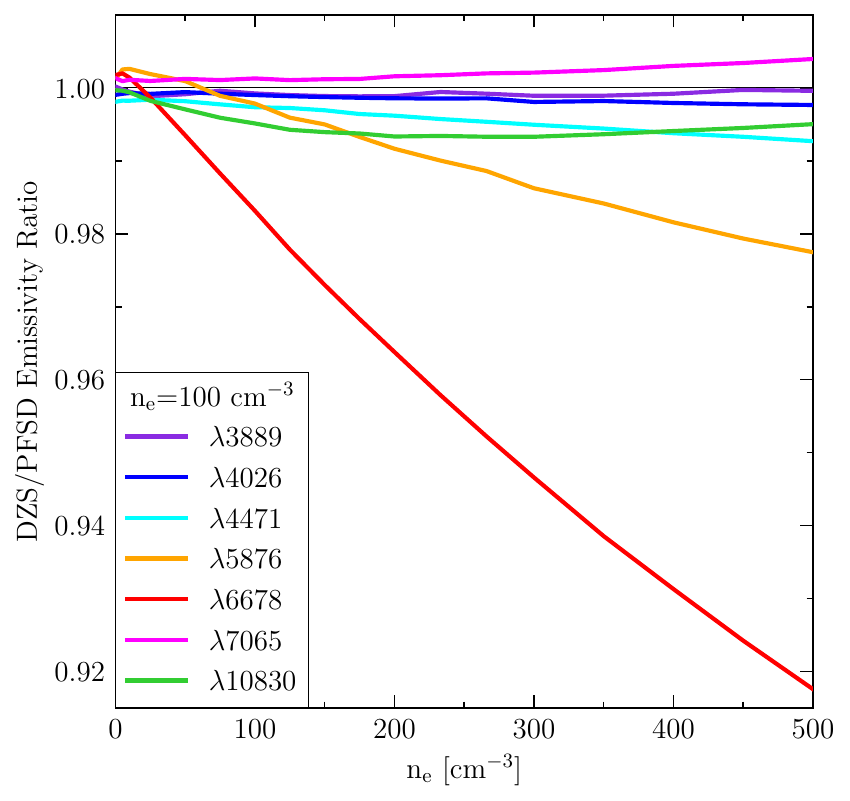}}
\caption{
The \citet{delz2022} emissivities plotted relative to those of \citet{port2012, port2013} as a function of density at $\Te = 15,000\,\r{K}$ for the core He emission lines, \hei \W\W3889, 4026, 4471, 5876, 6678, 7065, 10830.  Across the range of densities spanning our qualifying sample, all of the He emission lines employed in this work show changes of $\lesssim 1\%$, and typically a slight decrease, except for \heil6678 which shows a more significant decrease.  
}
\label{figure:DSZ-PFSD-D}
\end{figure}

\citet{delz2022} have published updated He emissivities, based on improved collision rates and which correct some errors in previous He emissivity calculations \citep[please see][ for details]{delz2022}.  Extending that work, G.\:Del~Zanna graciously agreed to calculate the updated He emissivities on a higher-resolution temperature and density grid to help limit interpolation errors (Private Communication).  Those He emissivities calculated on the finer grid are included with and employed in this work.  As discussed in \citet{delz2022} and shown below, the largest deviations from the previously employed \citet{port2012, port2013} emissivities occur for \heil6678, and the deviations increase with density.  Figures \ref{figure:DSZ-PFSD-T} \& \ref{figure:DSZ-PFSD-D} compare the \citet{delz2022} and \citet{port2012, port2013} emissivities versus temperature (at $\n = 100~\r{cm}^{-3}$) and density (at $\Te = 15,000\,\r{K}$), respectively.  With the exception of \heil6678 and for the low densities most relevant for our sample, all of the He emission lines employed in this work show changes of $\lesssim 1\%$.  The effect of the updated He emissivities on the helium abundance is discussed in Section \ref{EmissComp}.

\citet{stor2015} have published updated H emissivities.  Extending that work, P.\:Storey graciously agreed to calculate the updated H emissivities on a higher-resolution temperature and density grid (Private Communication).  Those H emissivities calculated on the finer grid are included with and employed in this work.  At the low densities applicable to the \hii regions in this work, the updated \citet{stor2015} H emissivities and the previously employed \citet{humm1987} emissivities are essentially identical, with the largest deviation across our range of densities and temperatures $<0.05\%$.  

Given the higher-resolution grid for the H emissivities, the normalizing emissivity values for Balmer series H$\beta$ and Paschen series P$\gamma$, E(H$\beta$) \& E(P$\gamma$), were also updated to employ the latest data and for consistency.  The previously employed H$\beta$ emissivity fit equation did not include density dependence and was based on CLOUDY modeling calculations by R.L.~Porter \citep{aver2010, aver2021}.  The difference varies with temperature and density, but the \citet{stor2015} H$\beta$ emissivity data shows a systematic increase ranging from $\sim0.1\%$ to $\sim0.25\%$ compared to the previous fit equation across our dataset.  The previous P$\gamma$ fit equation \citep{aver2021} was based on \citet{humm1987}, with the new \citet{stor2015} data nearly identical for the low densities relevant for our work.  However, the higher-resolution temperature and density data allowed for a more accurate fit equation.  The updated fit equations are provided in Appendix \ref{Appendix:HEmiss}.

\subsection{Radiative Transfer} \label{RadTrans}

Starting with \citet{oliv2004}, we have used the results of \citet[][, BSS02]{benj2002} for the radiative transfer correction, $f_\tau$ \citep[see][ for model implementation]{aver2021}.  \citetalias{benj2002} included limited-range fitting equations, parameterized in terms of optical depth for \3889 at line center (i.e., $\tau = \tau_{\ssn{3889}}$).  These fit equations have been employed for modeling the radiative transfer effects on the He triplet emission lines.  The equations were fit over the parameter ranges $12,000\,\r{K} < \Te < 20,000\,\r{K}$, $1\,\cm < \n < 300~\cm$, and $0 < \tau < 2$.  Given that limited range and the presence of significantly higher optical depth targets in our sample, we stopped using the limited-range fitting equations for this project.  To avoid extrapolation and following the approach in \citet{berg2026}, we switched to interpolation using a grid based on the trilinear interpolation program developed by \citetalias{benj2002}.   

However, very recently, updated radiative transfer calculations have been provided by \citet[][, KI25]{kuri2025}, and we have adopted them.  The new \citetalias{kuri2025} radiative transfer calculations are based on the latest atomic data, as well as improvements in modeling, including an expanded number of energy levels and an improved treatment of collisional transitions.  

\citetalias{kuri2025} provide fit equations for the radiative transfer correction, $f_\tau$, again parameterized in terms of optical depth for \3889 ($\tau = \tau_{\ssn{3889}}$).  Those equations were fit over the parameter ranges $8,000\,\r{K} < \Te < 22,000\,\r{K}$, $1\,\cm < \n < 10,000~\cm$, and $0 < \tau < 10$ with an accuracy of $\lesssim0.1\%$.  Those expanded parameter ranges nearly eliminate the possibility of extrapolation beyond the fit range for our targets, except perhaps for rare targets with higher optical depths, $\tau>10$.  However, systems with such high optical depths are suspect and may be less reliable, given the large radiative transfer effects and the potential for more complex dynamics.  As such, targets with $\tau>10$ are less desirable for helium abundance determinations, and we caution against relying on them (see \S \ref{tau} below).  The \citetalias{kuri2025} fit equations employed in this work are reproduced in Appendix \ref{Appendix:RadTrans}.  

Figure \ref{figure:ftau_all} shows the radiative transfer correction term, $f_\tau$, across all He triplet lines utilized in our model, plotted versus optical depth ($\tau = \tau_{\ssn{3889}}$), with those fit equations adopted from \citetalias{kuri2025}.  Absorption within the \hii region and subsequent re-emission---but frequently at longer wavelengths---decreases \3889 emission, while increasing net \heil7065 emission \citep[see description in][]{oste2006}.  As Figure \ref{figure:ftau_all} shows, these two emission lines are by far the most strongly affected by radiative transfer.

\begin{figure}[t!]
\resizebox{\columnwidth}{!}{\includegraphics{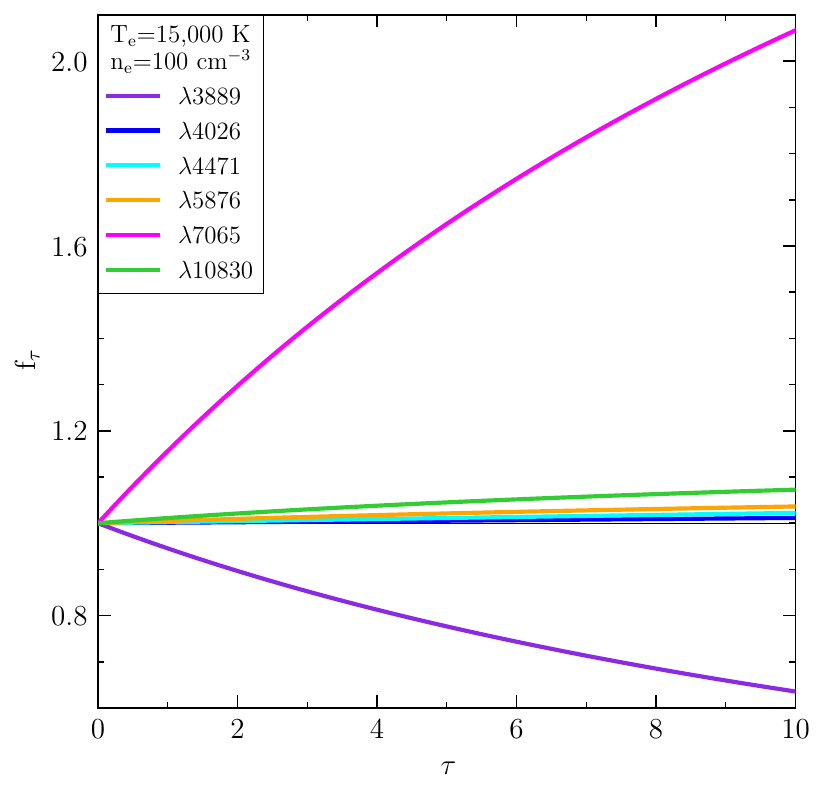}}
\caption{
The \citet{kuri2025} radiative transfer corrections, $f_\tau$, plotted as a function of optical depth ($\tau = \tau_{\ssn{3889}}$) at $\n = 100~\r{cm}^{-3}$ and $\Te = 15,000\,\r{K}$ for the He triplet emission lines, \hei \W\W3889, 4026, 4471, 5876, 7065, 10830.  As shown, radiative transfer results in a net decrease in \3889 emission and a net increase in \heil7065 emission, with those two lines by far the most strongly affected.  
}
\label{figure:ftau_all}
\end{figure}

\begin{figure}[t!]
\resizebox{\columnwidth}{!}{\includegraphics{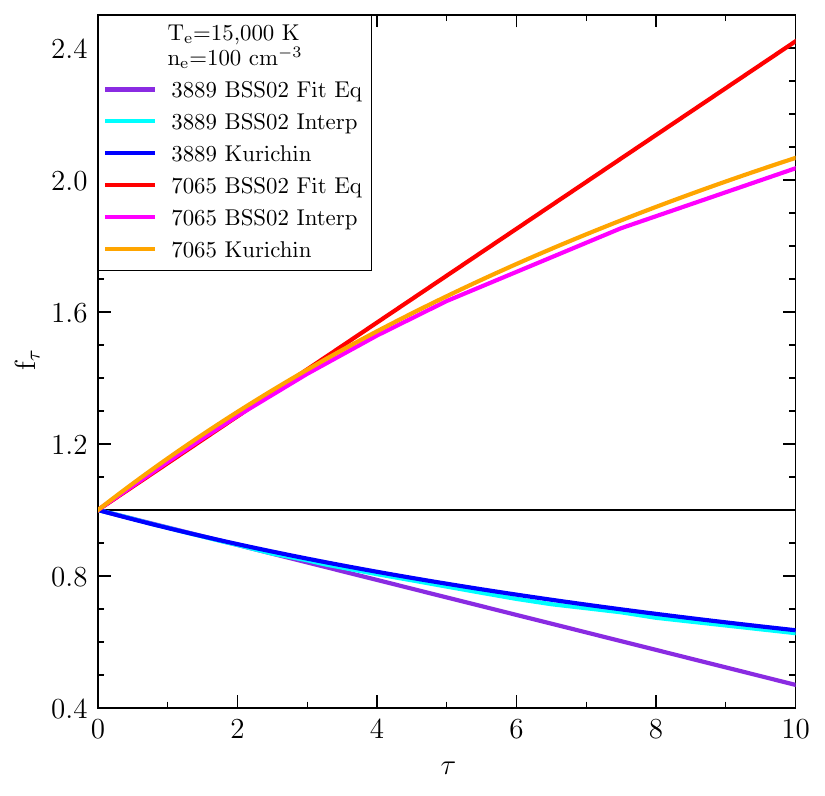}}
\caption{
The \citet{kuri2025} and \citet{benj2002} radiative transfer corrections, $f_\tau$, plotted as a function of optical depth ($\tau = \tau_{\ssn{3889}}$) at $\n = 100~\r{cm}^{-3}$ and $\Te = 15,000\,\r{K}$ for \3889 and \heil7065, the two He emission lines most strongly affected by radiative transfer.  The plot compares the limited-range fitting function and the trilinear interpolation program from \citetalias{benj2002} and the updated results of \citetalias{kuri2025}.  As is the case for all He triplet lines affected by radiative transfer, the magnitude of the radiative transfer effect is overestimated by extrapolation based on the limited-range fitting function, which was fit for $0 < \tau < 2$.  The \citetalias{benj2002} interpolation and \citetalias{kuri2025} fit equations are more similar, though with differences that increase for higher optical depths (and also for higher densities, especially for \heil7065).  
}
\label{figure:ftau_3889_7065}
\end{figure}

Figure \ref{figure:ftau_3889_7065} compares the \citetalias{benj2002} limited-range fit equation, the \citetalias{benj2002} interpolation grid, and the \citetalias{kuri2025} fit equations as optical depth increases.  As shown in Figure \ref{figure:ftau_3889_7065}, the dominant effect of moving from the limited-range fit equations from \citetalias{benj2002} to the \citetalias{benj2002} interpolation grid or the \citetalias{kuri2025} fit equations is to decrease the radiative transfer effect at higher optical depths (compared to extrapolating beyond $\tau > 2$ using the limited-range fit equations).  This applies across all triplet He lines.  The \citetalias{benj2002} limited-range fit equations are linear with $\tau$, so it is not surprising that the radiative transfer effects depart from that linear extrapolation.  These differences grow steadily for $\tau > 2$, exceeding 10\% for \3889 for $\tau > 6$, and reaching 35\% and 15\% at $\tau = 10$ for \3889 and \heil7065, respectively.

As a result, the best-fit optical depth values for targets returning $\tau > 2$, but especially for $\tau > 4$, are larger using the \citetalias{benj2002} interpolation grid or \citetalias{kuri2025} fit equations, compared to the best-fit optical depth resulting from the \citetalias{benj2002} limited-range fit equations.  This comparative increase is a consequence of comparing to the results based on an extrapolation of the limited-range fit equations beyond their fit domain.  The \citetalias{kuri2025} differences with the underlying \citetalias{benj2002} results are much smaller, though still significant, as is discussed next and illustrated in Section \ref{RadTransComp}.  

Compared to interpolation based on the \citetalias{benj2002} calculations, the new calculations of \citetalias{kuri2025} are similar for most emission lines, with differences of less than 1\% across the fit range.  The differences are larger for \3889 and \heil7065, and tend to increase with optical depth, as is shown in Figure \ref{figure:ftau_3889_7065} with $\n = 100~\r{cm}^{-3}$ and $\Te = 15,000\,\r{K}$.  For \3889 and \heil7065, these differences between the \citetalias{benj2002} interpolation and the \citetalias{kuri2025} calculations exceed 1\% for $\tau > 5$, reaching 1.5\% at $\tau = 8$.  At higher densities, larger differences open up for the more collisionally-sensitive \heil7065 emission line, in particular.  For $\n = 300~\cm$, the difference grows to 10\% at $\tau = 10$.  

As discussed above, \heil7065 is the most strongly affected by radiative transfer, and it shows the largest relative changes from the updated calculations by \citetalias{kuri2025}.  As shown in Figure \ref{figure:ftau_3889_7065}, the \citetalias{kuri2025} calculations slightly increase the magnitude of the boost from radiative transfer for \heil7065, compared to the \citetalias{benj2002} calculations (i.e., the \citetalias{benj2002} interpolation).  Correspondingly, for targets with higher optical depth values, the adoption of the updated \citetalias{kuri2025} radiative transfer model results in a moderate decrease in the best-fit optical depth, compared to the \citetalias{benj2002} radiative transfer model.  This finding will be further examined and discussed in Section \ref{RadTransComp}.

\subsection{He\,{\small \textit{I}} \W5016 \& \W7281} \label{Eduardo}

\citet{izot2007} and later \citet{mend2025} found significant, systematic discrepancies in observed \hei \W5016 \& \W7281 fluxes compared to standard ``Case B'' modeling and assumed temperature homogeneity.  Correspondingly, \citet{mend2025} investigates two hypotheses for these discrepancies:  (1) deviations from ``Case B'' due to photon loss and/or (2) significant temperature inhomogeneities.  Both scenarios were found to be possible sources of the observed discrepancies.  Regardless, the observed discrepancies, with both lines systematically biased low compared to the model, makes them unsuitable for calculating their corresponding helium abundance using the available atomic data and model.  The inclusion of those lines would systematically bias the result.  The finding from \citet{mend2025} was reinforced by our own analysis, which showed that the measured fluxes for both emission lines were systematically low compared to the model prediction based on the best-fit parameter solution for the full set of He \& H lines (see \S \ref{Model} above for a summary of our model and methodology and \citet{aver2021} for a full description).  As such, we exclude both lines from our analysis.  

\subsection{Neutral Hydrogen Collisional Correction} \label{NHCC}

\begin{figure}[t!]
\resizebox{\columnwidth}{!}{\includegraphics{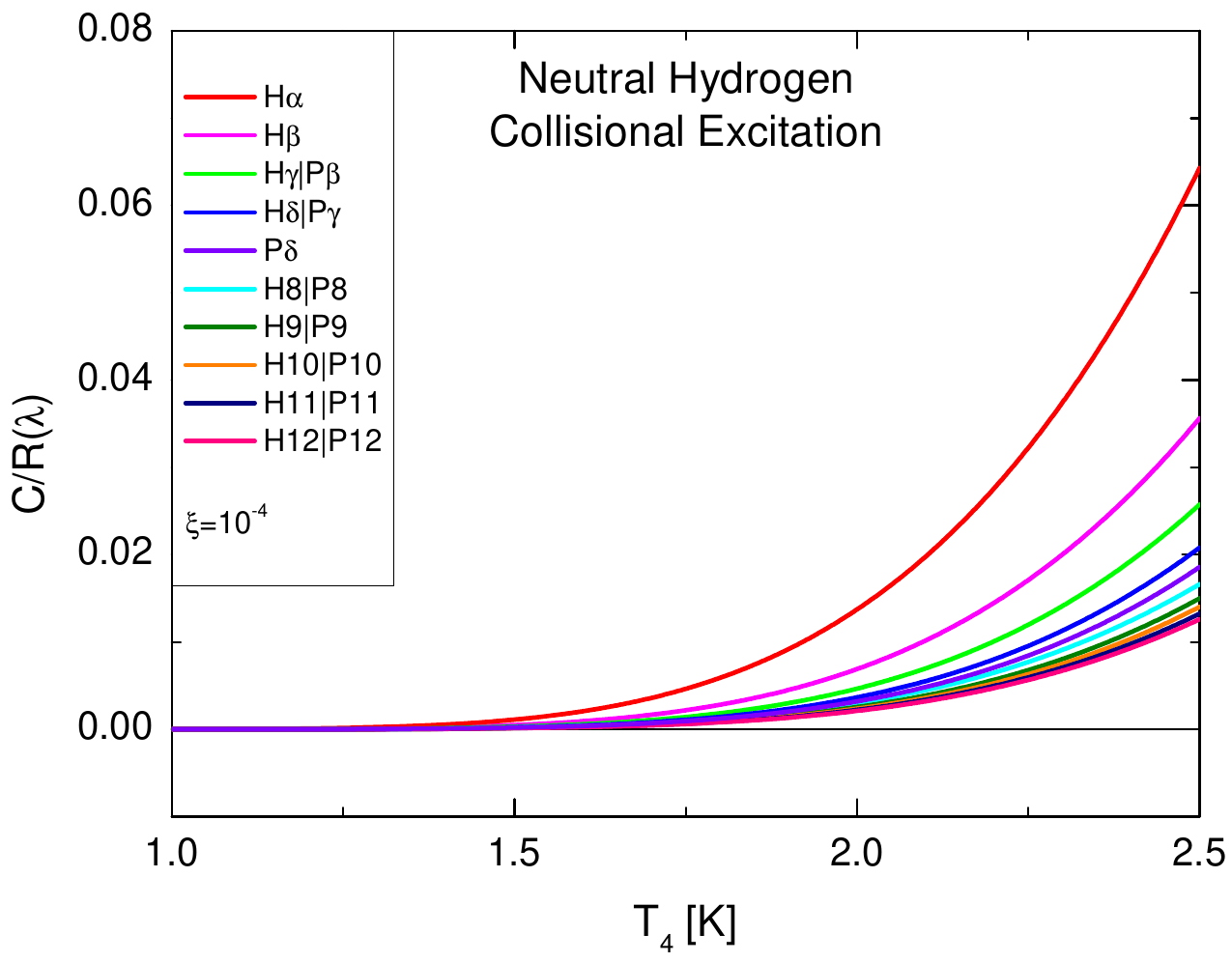}}
\caption{
The collisional excitation relative to recombination rate for the hydrogen emission lines due to neutral hydrogen.  The behavior is dominantly exponential with temperature, with, as expected, larger corrections for the lower level (lower energy) emission lines.  Reproduced from \citet{aver2021}.
}
\label{H_CR_T}
\end{figure}

As shown in Figure \ref{H_CR_T}, emission from the collisional excitation of neutral hydrogen has a strong, exponential dependence on the electron temperature \citep{aver2010,aver2021}.  As a result, at low temperatures, the potential contribution from neutral hydrogen is negligible.  However, as discussed in \citet{aver2012}, this extremely low collisional emission rate can drive the model's best-fit parameter solution toward increasingly large and completely non-physical fractions of neutral hydrogen in the \hii region.  As we have improved and expanded our model with additional H emission lines, this degenerate result from our model has been diminished and the best-fit for neutral hydrogen fraction, $\xi$, has become better constrained \citep{aver2021}.  

However, given that, \textit{physically}, the collisional excitation rate is completely negligible compared to the recombination rate at low temperatures as discussed in \citet{aver2010,aver2021} and shown in Figure \ref{H_CR_T}, trying to solve for the fraction of neutral hydrogen becomes physically meaningless. Correspondingly, the correction for neutral hydrogen collisional excitation is not included in our model below electron temperatures of 14,000\,K ($\r{T}_4<1.4$).  As a result, for temperatures below 14,000\,K, the neutral hydrogen fraction, $\xi$, is no longer included as one of our model parameters, and our number of model parameters decreases from 9 to 8.

\section{Sample Selection} \label{Sample}


Papers I, II, and III discuss our target selection, optical and near-IR MODS observations, and near-IR LUCI observations, respectively.  The He \& H emission line flux ratios utilized in our model (see Section \ref{Model}) are measured from the MODS spectra and reported in Paper~II, with the exception of \10830, where the values are measured from the LUCI NIR spectra and reported in Paper~III or are taken from the literature.  Paper~II also includes the determination of T(OIII) and O/H for those observed targets.  For determining the primordial helium abundance, \Yp, the following criteria were required for our dataset:  

\begin{enumerate}
    \item $\bm{\r{O/H} < 15 \times 10^{-5}}$.  Ideally, one would assemble a large dataset of only the lowest metallicity XMP galaxies \citep[$\r{O/H} < 2.24 \times 10^{-5}$,][]{guse2015, mcQu2020}, but, as discussed in Paper~I, such targets are rare and tend to be faint.  Nevertheless, targeting a reduced metallicity baseline limits potential systemic error introduced by enrichment of He and/or of an assumed linear relationship between He and O (between Y \& O/H).  Correspondingly, we prioritized the lowest metallicity targets from the literature, which still met our criteria for producing high SNR spectra (namely $\r{F(H\beta) \gtrsim 3 \times 10^{-15}~erg~s^{-1} cm^{-2}}$ and $\r{W(H\beta) \gtrsim 100}$, see Paper~I).  We particularly emphasized targets with \OHc\ for their impact on determining \Yp, and prioritized them over observing more evenly across our metallicity range.  As we exhausted identified, high SNR targets, we worked our way up in metallicity with a set cut-off of $\r{O/H} < 15 \times 10^{-5}$.  This is still a relatively constrained metallicity baseline compared to many other \Yp\ analyses \citep[e.g.,][]{izot2007, fern2019, kuri2021}.  Furthermore, our sample coverage at low metallicity is much higher than other \Yp\ analyses \citep[e.g.,][]{izot2007, fern2019, hsyu2020, kuri2021, yana2025}.  Two targets, UM\,420 and Mrk\,5, were proposed based on literature O/H values that were less than our cutoff, but returned O/H values exceeding it.  As such, they were excluded from our He analysis dataset.  
    \item \textbf{He\,{\small I}}\,$\bm{\lambda}$\textbf{4026 detected} ($>3\sigma$).  The weaker He emission lines are more sensitive to the effects of underlying stellar absorption, and, relative to its strength, the weak emission line \heil4026 is the most sensitive (excluding \3889, which is blended with H8 and is a stronger line).  Therefore, \heil4026 is a strong diagnostic for determining the underlying absorption.  Furthermore, as shown in \citet{aver2012}, the absence of \heil4026 can bias the result, due to overestimating the underlying absorption for the helium lines, and hence overestimating the helium abundance.  Two observed targets, HSC\s J2314+0154 and KUG\s0743+513, were too faint for \heil4026 detections.  As a result, they were excluded from our He analysis dataset.  Note that, when detected, we employ the fainter He emission lines, \hei \W4388 \& \W4922, which are also quite sensitive to underlying absorption, but when \heil4026 is not detected, not surprisingly, \hei \W4388 \& \W4922 are also not detected.  
    \item \textbf{He\,{\small I}}\,$\bm{\lambda}$\textbf{10830 measured}.  As discussed in Paper II, \10830 is much more sensitive to collisional excitations, and thus the electron density, than any of our other observed He emission lines \citep{izot2014,aver2015}.  Correspondingly, \citet{aver2015} demonstrated the crucial role \10830 plays in helping break the temperature-density degeneracy \citep{oliv2004, aver2010, aver2011}.  Therefore, \10830 is a required emission line for our analysis.  For most of our targets, \10830 was measured from our LUCI spectra, but for 11 targets, literature \10830 values were available and utilized.  Paper~I provides a table with that breakdown by target.  The MODS spectra for some targets showed clear signs of high excitation and other dynamics, which do not conform to the general \hii region assumptions our model rests upon.  They are thus less suitable for our analysis, and LUCI observations were not always pursued.  There were two such targets: WISEA\s J120503.54+455151.0 and SHOC\s133.  Similarly, a LUCI observation was not pursued for HSC\s J2314+0154, which was too faint and lacked a \heil4026 detection (see above).  A LUCI spectrum was also not pursued for I\s Zw\s18\s NW, as opposed to I\s Zw\s18\s SE, where I\s Zw\s18\s NW has been shown to be unsuitable for helium abundance determinations due to strong underlying absorption, Wolf-Rayet stars, and a strong temperature gradient, as well as significant Galactic \nai\ absorption for \heil5876 \citep{izot1999}.  Finally, the morphology of one target, HS\s2236+1344, shows a double knot structure without a reliable \10830 measurement.  Therefore, lacking \10830, five targets were excluded from our He analysis dataset (one of which also lacked \heil4026).  
\end{enumerate}

These criteria yielded a starting dataset of 54 targets for He abundance analysis and, after testing for statistical reliability and flagging for potential systematic biases, determining \Yp.


\section{\X2 "Goodness of Fit" Testing} \label{chi2cut}


In evaluating the results from our sample, we first perform a 95\% confidence level \X2 cut for statistical reliability.  For 15 degrees of freedom (for example, for 24 line ratios and 9 fit parameters), we require $\chi^2 < 25.0$. Assuming reasonable flux uncertainties (not systematically underestimated), targets with best-fit (i.e., minimized) \X2 values exceeding the 95\% confidence level for their number of degrees of freedom are very likely not statistically reliable, and it is probable that systematic errors are undermining their returned physical conditions.  This could be because of systematic effects in the measurements (e.g., \nai\ absorption), systematic errors in the model, or a physical system with conditions not captured by the model (e.g., a supernova remnant and very high excitation states).  

From our starting dataset of 54 observations, 7 have best-fit \X2 values that exceed the 95\% confidence level ``goodness-of-fit'' test, based on their number of degrees of freedom (which varies by target). These 7 targets are listed in Table~\ref{table:Excluded}, along with the model-determined \X2, the 95\% confidence level (CL) value for \X2, and the number of degrees of freedom.  This leaves 47 targets, given in Table~\ref{table:Params}, which pass our standard 95\% confidence level \X2 cut for statistical reliability.  In the absence of systematic errors and based on random error alone, one would expect 3 targets to exceed the 95\% threshold.  However, some of our targets clearly exhibit systematic effects not captured by our model.


\begin{deluxetable}{lcccc}[t!]
\label{table:Excluded}
\centering
\tabletypesize{\footnotesize}
\tablewidth{\textwidth}
\tablehead{Object   & \X2       & 95\% CL     & d.o.f.  & [\fev]}
\startdata
\hline
HS~0811+4913	&	30.4	&	23.7	&	14     &           \\
HS~0837+4717	&	63.1	&	25.0	&	15     &   \CM     \\
J0519+0007	&	24.6	&	23.7	&	14     &   \CM     \\
LEDA~101527	&	22.3	&	21.0	&	12     &   \CM     \\
Mrk~36	&	23.5	&	19.7	&	11     &           \\
VCC~1744	&	42.6	&	26.3	&	16     &           \\
WJ0851+5841	&	79.5	&	21.0	&	12     &   \CM     \\
\enddata
\caption{Targets excluded on the basis of their best-fit \X2 (column 2) exceeding the 95\% CL for the number of degrees of freedom available for each target.  Targets with [\fev]\s\W4227 detections are marked.  [\fev] emission indicates unexpectedly high ionization states not attributable to photoionization by massive stars.  As such, these targets are not expected to be well fit by our model.}
\end{deluxetable}


\begin{deluxetable*}{lcrcclcccrcccc}[pth!]
\label{table:Params}
\centering
\tablecaption{Physical Parameters and He$^+$/H$^+$ Abundance Best-Fit Solutions for the Qualifying Dataset}
\tabletypesize{\footnotesize}
\tablehead{
Object 			& $\y+=\r{He^+/H^+}$     		   &  \multicolumn{1}{r}{\n\,[\cm]}   		      & a$_{\r{He}}$\,[\AA]     		   & $\tau$     			& \multicolumn{1}{c}{\Te\,[K]}     		   & C(H$\beta$)     	  	& a$_\r{H}$\,[\AA]		     & a$_\r{P}$\,[\AA]		     & $\xi$ $\times$ 10$^4$     & T(OIII)\,[K]      & $\chi^2$       & 95\% CL     & d.o.f. 
}
\startdata
\multicolumn{13}{c}{Final Dataset Not Flagged} \\
\hline
DDO~68           & 0.07706 $^{+0.00280}_{-0.00208}$ &      58 $^{+     18}_{-     17}$ &  0.20 $^{+ 0.14}_{- 0.11}$ &  0.00 $^{+ 0.10}_{- 0.00}$ & 18500 $^{+1400}_{-1400}$ &  0.07 $^{+ 0.02}_{- 0.02}$ &  0.39 $^{+ 0.14}_{- 0.16}$ &  0.00 $^{+ 0.13}_{- 0.00}$ &        0 $^{+       6}_{-       0}$ & 18900 & 16.4 & 19.7 &  11  \\
HS~0029+1748     & 0.07869 $^{+0.00131}_{-0.00133}$ &     148 $^{+     35}_{-     31}$ &  0.11 $^{+ 0.08}_{- 0.08}$ &  0.52 $^{+ 0.29}_{- 0.27}$ & 12600 $^{+1000}_{- 900}$ &  0.27 $^{+ 0.02}_{- 0.02}$ &  1.41 $^{+ 0.16}_{- 0.16}$ &  0.93 $^{+ 0.21}_{- 0.21}$ & 		\nodata			 & 12800 & 17.4 & 23.7 &  14  \\
HS~0122+0743     & 0.08214 $^{+0.00145}_{-0.00137}$ &     172 $^{+     28}_{-     23}$ &  0.15 $^{+ 0.12}_{- 0.12}$ &  1.57 $^{+ 0.35}_{- 0.30}$ & 17900 $^{+1200}_{-1200}$ &  0.29 $^{+ 0.01}_{- 0.02}$ &  0.65 $^{+ 0.19}_{- 0.17}$ &  1.07 $^{+ 0.31}_{- 0.32}$ &        0 $^{+       3}_{-       0}$ & 17800 & 13.7 & 22.4 &  13  \\
HS~0134+3415     & 0.07777 $^{+0.00100}_{-0.00109}$ &     271 $^{+     52}_{-     43}$ &  0.12 $^{+ 0.09}_{- 0.09}$ &  1.72 $^{+ 0.31}_{- 0.31}$ & 12900 $^{+ 900}_{- 900}$ &  0.16 $^{+ 0.01}_{- 0.01}$ &  1.06 $^{+ 0.13}_{- 0.15}$ &  0.00 $^{+ 0.24}_{- 0.00}$ &		\nodata			 & 16300 & 14.3 & 26.3 &  16  \\
HS~1353+4706     & 0.08036 $^{+0.00111}_{-0.00113}$ &       1 $^{+      1}_{-      1}$ &  0.18 $^{+ 0.13}_{- 0.13}$ &  0.37 $^{+ 0.24}_{- 0.12}$ & 17700 $^{+ 900}_{-1200}$ &  0.17 $^{+ 0.01}_{- 0.02}$ &  0.57 $^{+ 0.20}_{- 0.23}$ &  0.00 $^{+ 0.30}_{- 0.00}$ &        0 $^{+       3}_{-       0}$ & 19600 & 17.7 & 23.7 &  14  \\
HS~1442+4250     & 0.08327 $^{+0.00295}_{-0.00323}$ &      30 $^{+     12}_{-      9}$ &  0.28 $^{+ 0.12}_{- 0.11}$ &  0.00 $^{+ 0.11}_{- 0.00}$ & 17400 $^{+ 600}_{- 800}$ &  0.09 $^{+ 0.02}_{- 0.02}$ &  0.92 $^{+ 0.19}_{- 0.18}$ &  0.15 $^{+ 0.30}_{- 0.15}$ &       54 $^{+      32}_{-      21}$ & 16600 & 14.1 & 22.4 &  13  \\
I~Zw~18~SE         & 0.08243 $^{+0.00260}_{-0.00278}$ &       7 $^{+     11}_{-      7}$ &  0.31 $^{+ 0.09}_{- 0.09}$ &  0.18 $^{+ 0.34}_{- 0.18}$ & 17500 $^{+1400}_{-1500}$ &  0.13 $^{+ 0.03}_{- 0.02}$ &  1.12 $^{+ 0.13}_{- 0.13}$ &  0.14 $^{+ 0.35}_{- 0.14}$ &       14 $^{+      25}_{-      11}$ & 18400 &  9.1 & 21.0 &  12  \\
J0133+1342      & 0.08042 $^{+0.00206}_{-0.00194}$ &      59 $^{+     20}_{-     15}$ &  0.33 $^{+ 0.09}_{- 0.09}$ &  0.54 $^{+ 0.33}_{- 0.28}$ & 15800 $^{+1200}_{-1300}$ &  0.18 $^{+ 0.02}_{- 0.02}$ &  0.75 $^{+ 0.14}_{- 0.14}$ &  0.56 $^{+ 0.22}_{- 0.24}$ &       16 $^{+      38}_{-      16}$ & 17500 & 17.1 & 25.0 &  15  \\
J0807+3414      & 0.08725 $^{+0.00196}_{-0.00241}$ &     250 $^{+     39}_{-     34}$ &  0.23 $^{+ 0.13}_{- 0.15}$ &  6.20 $^{+ 0.58}_{- 0.64}$ & 16800 $^{+1200}_{- 900}$ &  0.17 $^{+ 0.02}_{- 0.01}$ &  0.16 $^{+ 0.18}_{- 0.16}$ &  0.00 $^{+ 0.53}_{- 0.00}$ &       21 $^{+      20}_{-      14}$ & 17400 & 10.0 & 25.0 &  15  \\
J1044+0353      & 0.07844 $^{+0.00190}_{-0.00206}$ &     163 $^{+     29}_{-     22}$ &  0.13 $^{+ 0.10}_{- 0.11}$ &  1.55 $^{+ 0.36}_{- 0.33}$ & 18100 $^{+1100}_{-1300}$ &  0.15 $^{+ 0.02}_{- 0.02}$ &  0.84 $^{+ 0.16}_{- 0.14}$ &  0.78 $^{+ 0.39}_{- 0.41}$ &       11 $^{+      13}_{-       7}$ & 19600 & 19.2 & 23.7 &  14  \\
J1148+2546      & 0.08399 $^{+0.00214}_{-0.00168}$ &     145 $^{+     29}_{-     25}$ &  0.40 $^{+ 0.07}_{- 0.07}$ &  2.40 $^{+ 0.40}_{- 0.41}$ & 14900 $^{+1100}_{- 900}$ &  0.20 $^{+ 0.02}_{- 0.02}$ &  0.90 $^{+ 0.14}_{- 0.13}$ &  0.83 $^{+ 0.21}_{- 0.23}$ &       17 $^{+      55}_{-      17}$ & 14300 & 10.4 & 22.4 &  13  \\
J1323-01325     & 0.08480 $^{+0.00210}_{-0.00215}$ &      67 $^{+     16}_{-     14}$ &  0.26 $^{+ 0.08}_{- 0.10}$ &  2.51 $^{+ 0.35}_{- 0.38}$ & 16600 $^{+1100}_{-1000}$ &  0.10 $^{+ 0.02}_{- 0.01}$ &  0.76 $^{+ 0.13}_{- 0.13}$ &  0.02 $^{+ 0.39}_{- 0.02}$ &       32 $^{+      29}_{-      18}$ & 18400 &  8.9 & 23.7 &  14  \\
J1331+4151      & 0.07690 $^{+0.00136}_{-0.00128}$ &      62 $^{+     21}_{-     17}$ &  0.10 $^{+ 0.08}_{- 0.07}$ &  1.01 $^{+ 0.37}_{- 0.32}$ & 14800 $^{+1200}_{-1200}$ &  0.15 $^{+ 0.02}_{- 0.02}$ &  0.79 $^{+ 0.10}_{- 0.11}$ &  0.74 $^{+ 0.17}_{- 0.17}$ &        0 $^{+      18}_{-       0}$ & 16500 & 12.5 & 21.0 &  12  \\
J2104-0035      & 0.08267 $^{+0.00580}_{-0.00581}$ &      11 $^{+     18}_{-      9}$ &  0.16 $^{+ 0.23}_{- 0.16}$ &  0.45 $^{+ 0.59}_{- 0.41}$ & 18800 $^{+1900}_{-2200}$ &  0.13 $^{+ 0.04}_{- 0.04}$ &  1.27 $^{+ 0.31}_{- 0.31}$ &  0.49 $^{+ 0.77}_{- 0.49}$ &       28 $^{+      77}_{-      28}$ & 18900 &  9.6 & 21.0 &  12  \\
J2213+1722      & 0.07787 $^{+0.00144}_{-0.00170}$ &       1 $^{+      6}_{-      1}$ &  0.23 $^{+ 0.08}_{- 0.08}$ &  0.40 $^{+ 0.49}_{- 0.23}$ & 14200 $^{+1300}_{-1400}$ &  0.10 $^{+ 0.02}_{- 0.02}$ &  1.09 $^{+ 0.13}_{- 0.13}$ &  0.86 $^{+ 0.34}_{- 0.34}$ & 		\nodata 		& 16700 & 18.5 & 21.0 &  12  \\
KUG~1138+327     & 0.08836 $^{+0.00132}_{-0.00186}$ &      14 $^{+      9}_{-      8}$ &  0.44 $^{+ 0.04}_{- 0.05}$ &  0.00 $^{+ 0.12}_{- 0.00}$ & 15700 $^{+ 400}_{- 700}$ &  0.04 $^{+ 0.01}_{- 0.01}$ &  1.28 $^{+ 0.10}_{- 0.10}$ &  0.00 $^{+ 0.11}_{- 0.00}$ &       49 $^{+      37}_{-      19}$ & 14800 & 16.0 & 19.7 &  11  \\
LEDA~2790884     & 0.08615 $^{+0.00259}_{-0.00304}$ &      61 $^{+     39}_{-     33}$ &  0.25 $^{+ 0.14}_{- 0.15}$ &  0.05 $^{+ 0.29}_{- 0.05}$ & 16800 $^{+1100}_{-1300}$ &  0.11 $^{+ 0.02}_{- 0.02}$ &  0.88 $^{+ 0.22}_{- 0.21}$ &  0.00 $^{+ 0.40}_{- 0.00}$ &       70 $^{+      91}_{-      33}$ & 18200 & 15.1 & 22.4 &  13  \\
Leo~P            & 0.08386 $^{+0.00222}_{-0.00241}$ &      33 $^{+     18}_{-     10}$ &  0.41 $^{+ 0.19}_{- 0.18}$ &  0.00 $^{+ 0.27}_{- 0.00}$ & 17900 $^{+1000}_{-1600}$ &  0.10 $^{+ 0.02}_{- 0.02}$ &  0.12 $^{+ 0.19}_{- 0.12}$ &  0.00 $^{+ 0.34}_{- 0.00}$ &        0 $^{+       5}_{-       0}$ & 17200 & 10.7 & 22.4 &  13  \\
SBS~0335-052E    & 0.08470 $^{+0.00250}_{-0.00194}$ &      69 $^{+     13}_{-     12}$ &  0.30 $^{+ 0.05}_{- 0.05}$ &  4.44 $^{+ 0.50}_{- 0.43}$ & 21100 $^{+1200}_{-1300}$ &  0.21 $^{+ 0.02}_{- 0.02}$ &  0.62 $^{+ 0.07}_{- 0.07}$ &  0.74 $^{+ 0.22}_{- 0.23}$ &        1 $^{+       2}_{-       1}$ & 20900 & 20.9 & 21.0 &  12  \\
SBS~0926+606     & 0.08043 $^{+0.00182}_{-0.00188}$ &     121 $^{+     44}_{-     35}$ &  0.27 $^{+ 0.11}_{- 0.11}$ &  0.43 $^{+ 0.39}_{- 0.41}$ & 11400 $^{+1100}_{-1000}$ &  0.14 $^{+ 0.02}_{- 0.02}$ &  1.45 $^{+ 0.14}_{- 0.14}$ &  0.46 $^{+ 0.34}_{- 0.34}$ & 		\nodata			 & 11400 &  5.8 & 16.9 &   9  \\
SBS~0940+544     & 0.08146 $^{+0.00255}_{-0.00321}$ &      89 $^{+     30}_{-     22}$ &  0.32 $^{+ 0.26}_{- 0.28}$ &  0.48 $^{+ 0.39}_{- 0.40}$ & 16800 $^{+1700}_{-1800}$ &  0.14 $^{+ 0.03}_{- 0.02}$ &  0.82 $^{+ 0.25}_{- 0.27}$ &  0.00 $^{+ 0.80}_{- 0.00}$ &       19 $^{+      38}_{-      19}$ & 19200 & 10.1 & 23.7 &  14  \\
SBS~0946+558     & 0.08392 $^{+0.00131}_{-0.00128}$ &      43 $^{+     15}_{-     13}$ &  0.20 $^{+ 0.07}_{- 0.07}$ &  0.15 $^{+ 0.25}_{- 0.15}$ & 15100 $^{+1000}_{-1000}$ &  0.21 $^{+ 0.01}_{- 0.01}$ &  1.19 $^{+ 0.10}_{- 0.11}$ &  1.23 $^{+ 0.09}_{- 0.11}$ &        0 $^{+      13}_{-       0}$ & 13400 & 11.0 & 22.4 &  13  \\
SBS~0948+532     & 0.08219 $^{+0.00122}_{-0.00127}$ &     167 $^{+     37}_{-     33}$ &  0.38 $^{+ 0.07}_{- 0.07}$ &  0.93 $^{+ 0.30}_{- 0.30}$ & 11800 $^{+ 800}_{- 800}$ &  0.15 $^{+ 0.02}_{- 0.02}$ &  0.93 $^{+ 0.17}_{- 0.17}$ &  0.61 $^{+ 0.25}_{- 0.24}$ & 		\nodata			 & 12500 & 15.5 & 23.7 &  14  \\
SBS~1030+583     & 0.07953 $^{+0.00337}_{-0.00138}$ &      40 $^{+     10}_{-     11}$ &  0.36 $^{+ 0.05}_{- 0.04}$ &  0.00 $^{+ 0.10}_{- 0.00}$ & 16100 $^{+ 700}_{- 800}$ &  0.08 $^{+ 0.01}_{- 0.02}$ &  1.32 $^{+ 0.11}_{- 0.09}$ &  0.73 $^{+ 0.16}_{- 0.21}$ &        2 $^{+      18}_{-       2}$ & 15500 & 18.1 & 18.3 &  10  \\
SBS~1135+581     & 0.08256 $^{+0.00074}_{-0.00072}$ &       1 $^{+      3}_{-      1}$ &  0.15 $^{+ 0.06}_{- 0.06}$ &  0.76 $^{+ 0.33}_{- 0.17}$ & 12400 $^{+ 700}_{- 700}$ &  0.22 $^{+ 0.01}_{- 0.01}$ &  1.25 $^{+ 0.10}_{- 0.10}$ &  1.17 $^{+ 0.08}_{- 0.09}$ & 		\nodata			 & 12300 & 15.4 & 22.4 &  13  \\
SBS~1152+579     & 0.08183 $^{+0.00111}_{-0.00098}$ &     409 $^{+     64}_{-     49}$ &  0.22 $^{+ 0.05}_{- 0.06}$ &  3.87 $^{+ 0.49}_{- 0.37}$ & 15000 $^{+ 700}_{-1100}$ &  0.25 $^{+ 0.01}_{- 0.02}$ &  0.46 $^{+ 0.09}_{- 0.09}$ &  0.70 $^{+ 0.11}_{- 0.13}$ &        0 $^{+      17}_{-       0}$ & 15500 & 15.2 & 23.7 &  14  \\
SBS~1159+545     & 0.07813 $^{+0.00135}_{-0.00122}$ &     171 $^{+     29}_{-     24}$ &  0.15 $^{+ 0.08}_{- 0.07}$ &  1.36 $^{+ 0.36}_{- 0.28}$ & 16500 $^{+1100}_{-1200}$ &  0.13 $^{+ 0.01}_{- 0.02}$ &  0.09 $^{+ 0.12}_{- 0.09}$ &  0.15 $^{+ 0.26}_{- 0.15}$ &        0 $^{+       8}_{-       0}$ & 17600 & 11.4 & 23.7 &  14  \\
SBS~1211+540     & 0.07761 $^{+0.00237}_{-0.00157}$ &      62 $^{+     15}_{-     13}$ &  0.27 $^{+ 0.10}_{- 0.09}$ &  0.00 $^{+ 0.12}_{- 0.00}$ & 15900 $^{+ 700}_{- 800}$ &  0.07 $^{+ 0.02}_{- 0.02}$ &  1.69 $^{+ 0.15}_{- 0.16}$ &  0.84 $^{+ 0.27}_{- 0.31}$ &        5 $^{+      22}_{-       5}$ & 17400 & 10.8 & 23.7 &  14  \\
SBS~1249+493     & 0.07686 $^{+0.00194}_{-0.00159}$ &      53 $^{+     20}_{-     15}$ &  0.22 $^{+ 0.08}_{- 0.07}$ &  0.98 $^{+ 0.43}_{- 0.32}$ & 15700 $^{+1300}_{-1400}$ &  0.16 $^{+ 0.02}_{- 0.03}$ &  1.23 $^{+ 0.12}_{- 0.11}$ &  1.21 $^{+ 0.36}_{- 0.41}$ &        0 $^{+      29}_{-       0}$ & 15900 &  8.9 & 21.0 &  12  \\
SBS~1331+493     & 0.08275 $^{+0.00077}_{-0.00096}$ &      65 $^{+      9}_{-      8}$ &  0.17 $^{+ 0.08}_{- 0.08}$ &  0.00 $^{+ 0.06}_{- 0.00}$ & 17000 $^{+ 400}_{- 600}$ &  0.13 $^{+ 0.01}_{- 0.01}$ &  0.67 $^{+ 0.15}_{- 0.17}$ &  0.00 $^{+ 0.04}_{- 0.00}$ &        0 $^{+       1}_{-       0}$ & 15700 & 21.8 & 22.4 &  13  \\
SBS~1415+437     & 0.07870 $^{+0.00155}_{-0.00170}$ &     101 $^{+     16}_{-     12}$ &  0.31 $^{+ 0.06}_{- 0.07}$ &  0.00 $^{+ 0.06}_{- 0.00}$ & 17300 $^{+ 500}_{- 700}$ &  0.11 $^{+ 0.01}_{- 0.01}$ &  0.75 $^{+ 0.13}_{- 0.12}$ &  0.00 $^{+ 0.10}_{- 0.00}$ &       52 $^{+      22}_{-      15}$ & 15900 &  9.8 & 21.0 &  12  \\
SBS~1420+544     & 0.08421 $^{+0.00156}_{-0.00176}$ &     324 $^{+     50}_{-     42}$ &  0.16 $^{+ 0.08}_{- 0.08}$ &  4.95 $^{+ 0.46}_{- 0.46}$ & 17300 $^{+1200}_{-1100}$ &  0.17 $^{+ 0.01}_{- 0.01}$ &  0.08 $^{+ 0.15}_{- 0.08}$ &  0.00 $^{+ 0.33}_{- 0.00}$ &       15 $^{+      12}_{-       8}$ & 17900 & 11.6 & 23.7 &  14  \\
SBS~1437+370     & 0.08237 $^{+0.00155}_{-0.00151}$ &      93 $^{+     19}_{-     16}$ &  0.29 $^{+ 0.06}_{- 0.07}$ &  0.00 $^{+ 0.10}_{- 0.00}$ & 13500 $^{+ 600}_{- 600}$ &  0.12 $^{+ 0.01}_{- 0.01}$ &  1.15 $^{+ 0.10}_{- 0.09}$ &  0.89 $^{+ 0.18}_{- 0.19}$ & 		\nodata			 & 13900 &  6.6 & 25.0 &  15  \\
UGC~4483         & 0.07847 $^{+0.00281}_{-0.00188}$ &      29 $^{+     12}_{-     11}$ &  0.36 $^{+ 0.10}_{- 0.10}$ &  0.00 $^{+ 0.19}_{- 0.00}$ & 17300 $^{+1000}_{-1200}$ &  0.13 $^{+ 0.01}_{- 0.02}$ &  1.07 $^{+ 0.15}_{- 0.16}$ &  1.30 $^{+ 0.36}_{- 0.40}$ &        1 $^{+      11}_{-       1}$ & 17300 &  9.4 & 19.7 &  11  \\
UGC~5541         & 0.08004 $^{+0.00264}_{-0.00258}$ &      11 $^{+     11}_{-     10}$ &  0.43 $^{+ 0.09}_{- 0.09}$ &  0.00 $^{+ 0.11}_{- 0.00}$ & 15900 $^{+ 600}_{- 800}$ &  0.09 $^{+ 0.02}_{- 0.02}$ &  0.86 $^{+ 0.11}_{- 0.12}$ &  0.41 $^{+ 0.27}_{- 0.29}$ &       15 $^{+      24}_{-      15}$ & 15400 & 20.7 & 22.4 &  13  \\
UM~133           & 0.08434 $^{+0.00250}_{-0.00221}$ &      38 $^{+     13}_{-     11}$ &  0.40 $^{+ 0.09}_{- 0.09}$ &  0.00 $^{+ 0.14}_{- 0.00}$ & 16400 $^{+ 600}_{- 900}$ &  0.15 $^{+ 0.01}_{- 0.02}$ &  0.77 $^{+ 0.12}_{- 0.11}$ &  0.83 $^{+ 0.29}_{- 0.28}$ &       23 $^{+      21}_{-      13}$ & 15700 & 10.4 & 23.7 &  14  \\
UM~161           & 0.07987 $^{+0.00165}_{-0.00130}$ &     110 $^{+     28}_{-     22}$ &  0.30 $^{+ 0.12}_{- 0.11}$ &  1.39 $^{+ 0.33}_{- 0.32}$ & 14300 $^{+1100}_{-1000}$ &  0.23 $^{+ 0.02}_{- 0.02}$ &  0.68 $^{+ 0.15}_{- 0.14}$ &  0.00 $^{+ 0.49}_{- 0.00}$ &        0 $^{+      50}_{-       0}$ & 15600 &  7.1 & 25.0 &  15  \\
UM~461           & 0.07870 $^{+0.00140}_{-0.00091}$ &      94 $^{+     22}_{-     20}$ &  0.10 $^{+ 0.08}_{- 0.07}$ &  3.32 $^{+ 0.39}_{- 0.39}$ & 14100 $^{+ 900}_{- 900}$ &  0.22 $^{+ 0.01}_{- 0.01}$ &  0.66 $^{+ 0.13}_{- 0.16}$ &  0.39 $^{+ 0.20}_{- 0.27}$ &        0 $^{+      47}_{-       0}$ & 16500 & 15.6 & 23.7 &  14  \\
WJ1136+4709     & 0.08489 $^{+0.00190}_{-0.00168}$ &      42 $^{+     12}_{-     10}$ &  0.23 $^{+ 0.08}_{- 0.08}$ &  0.18 $^{+ 0.25}_{- 0.18}$ & 17000 $^{+1000}_{-1000}$ &  0.14 $^{+ 0.02}_{- 0.02}$ &  0.53 $^{+ 0.09}_{- 0.09}$ &  0.85 $^{+ 0.18}_{- 0.22}$ &        3 $^{+      12}_{-       3}$ & 15400 & 18.8 & 22.4 &  13  \\
WJ1327+4022     & 0.07532 $^{+0.00182}_{-0.00145}$ &      37 $^{+     19}_{-     14}$ &  0.11 $^{+ 0.13}_{- 0.11}$ &  0.34 $^{+ 0.34}_{- 0.30}$ & 14700 $^{+1300}_{-1400}$ &  0.11 $^{+ 0.02}_{- 0.02}$ &  0.94 $^{+ 0.17}_{- 0.16}$ &  0.27 $^{+ 0.31}_{- 0.27}$ &        0 $^{+      33}_{-       0}$ & 17200 &  5.9 & 18.3 &  10  \\
WJ2310-0211     & 0.08640 $^{+0.00170}_{-0.00198}$ &     198 $^{+     25}_{-     28}$ &  0.35 $^{+ 0.06}_{- 0.06}$ &  4.98 $^{+ 0.42}_{- 0.52}$ & 17300 $^{+1000}_{- 900}$ &  0.21 $^{+ 0.02}_{- 0.01}$ &  0.78 $^{+ 0.11}_{- 0.10}$ &  0.00 $^{+ 0.25}_{- 0.00}$ &       14 $^{+      10}_{-       9}$ & 16400 & 10.9 & 19.7 &  11  \\
\hline
\multicolumn{13}{c}{Final Dataset with Flags} \\
\hline
AGC~198691       & 0.08076 $^{+0.00572}_{-0.00356}$ &     237 $^{+    267}_{-    237}$ &  0.37 $^{+ 0.20}_{- 0.18}$ &  0.30 $^{+ 0.97}_{- 0.30}$ & 14000 $^{+2800}_{-2700}$ &  0.11 $^{+ 0.02}_{- 0.06}$ &  1.65 $^{+ 0.28}_{- 0.27}$ &  0.92 $^{+ 0.68}_{- 0.89}$ &        0 $^{+     585}_{-       0}$ & 20000 & 10.8 & 12.6 &   6  \\
HS~1028+3843     & 0.08786 $^{+0.00118}_{-0.00079}$ &     341 $^{+     37}_{-     34}$ &  0.02 $^{+ 0.11}_{- 0.02}$ & 13.25 $^{+ 0.65}_{- 0.71}$ & 17300 $^{+ 800}_{- 800}$ &  0.24 $^{+ 0.01}_{- 0.01}$ &  0.00 $^{+ 0.04}_{- 0.00}$ &  0.00 $^{+ 0.09}_{- 0.00}$ &        0 $^{+       1}_{-       0}$ & 16500 & 17.4 & 23.7 &  14  \\
HS~1222+3741     & 0.09171 $^{+0.00332}_{-0.00396}$ &      95 $^{+     23}_{-     17}$ &  0.61 $^{+ 0.07}_{- 0.08}$ &  0.00 $^{+ 0.30}_{- 0.00}$ & 17400 $^{+ 800}_{-1400}$ &  0.02 $^{+ 0.02}_{- 0.02}$ &  1.62 $^{+ 0.11}_{- 0.11}$ &  0.00 $^{+ 0.20}_{- 0.00}$ &       37 $^{+      47}_{-      14}$ & 15000 &  5.8 & 19.7 &  11  \\
J0118+3512      & 0.07520 $^{+0.00204}_{-0.00223}$ &      54 $^{+     33}_{-     21}$ &  0.16 $^{+ 0.14}_{- 0.15}$ &  0.00 $^{+ 0.13}_{- 0.00}$ & 13000 $^{+1200}_{-1300}$ &  0.25 $^{+ 0.02}_{- 0.02}$ &  0.93 $^{+ 0.18}_{- 0.17}$ &  1.41 $^{+ 0.53}_{- 0.56}$ & 		\nodata				 & 17500 & 10.3 & 19.7 &  11  \\
Mrk~71           & 0.08369 $^{+0.00108}_{-0.00092}$ &     213 $^{+     28}_{-     25}$ &  0.24 $^{+ 0.07}_{- 0.07}$ &  2.30 $^{+ 0.37}_{- 0.30}$ & 18500 $^{+ 900}_{-1000}$ &  0.25 $^{+ 0.01}_{- 0.01}$ &  0.64 $^{+ 0.12}_{- 0.13}$ &  0.35 $^{+ 0.21}_{- 0.22}$ &        0 $^{+       1}_{-       0}$ & 15800 & 21.8 & 23.7 &  14  \\
UGC~6456         & 0.08375 $^{+0.00252}_{-0.00281}$ &     100 $^{+     27}_{-     17}$ &  0.78 $^{+ 0.18}_{- 0.20}$ &  0.00 $^{+ 0.35}_{- 0.00}$ & 16600 $^{+ 700}_{-1400}$ &  0.15 $^{+ 0.01}_{- 0.02}$ &  1.09 $^{+ 0.19}_{- 0.19}$ &  0.74 $^{+ 0.36}_{- 0.37}$ &       13 $^{+      30}_{-      12}$ & 15500 & 20.7 & 22.4 &  13  \\
\enddata
\end{deluxetable*}


In particular, multiple targets include emission lines characteristic of unexpectedly high ionization states, such as [\fev]\s\W4227 and, in some cases, [\nev]\s\W3425, which are not typically expected in low-metallicity \hii regions. These types of objects have been identified as extreme emission line galaxies (EELGs) as discussed by \citet{berg2021}, and present the challenge of identifying a source of the high ionizing energy photons (which cannot be produced through standard photoionization).  Since these high ionization states may indicate the presence of a supernova remnant or another source of ionization not attributable to photoionization by massive stars, they are not expected to be well fit by our model.  

Multiple targets also exhibit very broad multi-component wings on H$\alpha$ and [\oiii]\s\W5007, and, extending with decreasing prominence, to H$\beta$, H$\gamma$, and [\oiii]\s\W4959.  These wings are very challenging to fit, raise questions about which components should be included in the measured flux for our analysis, and the wings indicate complicated kinematic structures, which are not accounted for in our model.  Multiple targets also exhibit clear Wolf-Rayet (WR) features, where the presence of WR stars may contribute to a poor model fit.  Additionally, the line profiles for several targets are also significantly skewed and asymmetric.  It is not clear what the cause of that asymmetry is, but it undermines the reliability of the measured fluxes.  

Of the 7 targets exceeding their 95\% threshold, two have features which clearly signal that we would not expect our model to apply to the physical conditions underlying their emission.  HS\s0837+4717 exhibits extremely strong wings on its strong emission lines (the strongest in our sample), as well as weak [\fev] emission.  While for Mrk\,36, the emission line profiles  are very strongly skewed.  In addition to those two, whose spectra are clearly discrepant and ill-suited for analysis using standard \hii region modeling, three other targets show weaker cases of the above pathologies (pathologies for our purposes).  J0519+0007, LEDA\s101527, and WISEA\s J085115.60+584055.7 include [\fev] emission (weaker for LEDA\s101527).  If we count all five of those targets as likely to result in poor model fits due to their anomalies, we are left with two targets---HS\s0811+4913 and VCC\s1744---excluded by the \X2 test.  This is in line with what would be expected based on random errors.

\subsection{Excluded Emission Lines} \label{RemovedLines}

In determining the best-fit parameter solution for each target, we examine the distribution of \X2 values for our set of emission lines.  Large \X2 contributions from a given line may indicate a systematic effect undermining the accuracy and reliability of its measured flux.  Further evaluation is required to determine whether removing that emission line from our analysis is justified.  

In particular, the Paschen sequence of lines, P8-P15, is observed in the NIR, where telluric contamination is common.  Water vapor absorption lines may be juxtaposed with our emission lines, as may the very strong night sky emission lines, which must be fit and subtracted.  As a result, Paschen lines may be decreased due to absorption features, and the accuracy of a given Paschen line measurement may be reduced by the night sky emission line subtraction.  Our error function for estimating the uncertainty on our measured fluxes varies with wavelength and is constructed to capture the increased uncertainty in the NIR.  Nevertheless, a Paschen line may be systematically decreased due to atmospheric absorption, and subtraction of the night sky emission lines may affect the local continuum and the line.   

Given those effects, when a given Paschen line is strongly discrepant compared to the rest of the Paschen sequence, we remove it from our analysis.  After correcting for reddening and underlying absorption, each of which depend on a single model parameter, the ratios of the theoretical fluxes for the Paschen lines only vary weakly with temperature.  The density dependence is almost negligible.  At the higher orbital excitation levels ($n \ge 8$) for the utilized Paschen lines, the contributions from neutral hydrogen collisional excitation are relatively small, and those contributions do not decrease substantially as the level increases (i.e., they are roughly constant across these higher Paschen lines).  Therefore, a Paschen line that is clearly discrepant from the others is likely contaminated and removing it is justified.  For example, for J1323-01325, the \X2 contribution from P9 \W9230 was 16.9 (the next highest \X2 contribution was 1.2), and removing P9 \W9230 reduced the total \X2 from 27.3 to 8.9.  

The tight wavelength spacing and strong underlying absorption of the higher Balmer lines, H8-H12, also makes accurately measuring their fluxes challenging.  Paralleling the explanation above for the Paschen series, the theoretical flux ratios are also well-anchored.  Correspondingly, similar to the Paschen series, though less common in frequency, we also remove higher Balmer lines that are clearly discrepant compared to the rest of Balmer sequence, especially when the ability to accurately fit the local continuum is compromised.  

However, we do not remove any of our ``core'' lines from our analysis based on their \X2 values.  The core lines (line ratios) are \hei \W\W4026, 4471, 5876, 6678, 7065, 10830; the Balmer lines, H$\alpha$, H$\gamma$, H$\delta$; and the blended line \3889 + H8.  These core lines are the heart of our model, providing the strongest diagnostic power for determining the physical conditions (model parameters) of the \hii region \citep{oliv2001,oliv2004,aver2010,aver2011,aver2012,aver2013,aver2015}.  With the exception of \10830, they are also isolated lines located in a region of the spectrum with the highest spectral sensitivity and with many fewer night sky emission features and telluric absorption bands (compared to the NIR).  

The only cases where a core line is removed from our analysis is the (rare) situation where we can see clear evidence of a systematic effect significantly undermining the reliability of the measured flux for that line.  These come in three cases:  (1) Atmospheric molecular oxygen A or B band absorption impacting \heil6678, (2) Galactic \nai\ absorption and/or night sky emission coincident with \heil5876, and (3) Very strong, broad emission wings on H$\alpha$ that are stronger and broader than would be expected based on [\oiii]\s\W5007 or H$\beta$.  

The atmospheric A \& B O$_2$ absorption bands are easily identified in our spectra, with a clear drop in the continuum level.  For galaxies with redshift ranges of $z=0.120 - 0.134$ \& $z=0.025 - 0.039$, \heil6678 will fall within the A or B bands, respectively.  In those cases, the measured \heil6678 flux will be systematically decreased and is not accurate.  There are two targets where \heil6678 is excluded for this reason:  HS\s1028+3843 ($z=0.0295$) and HS\s1353+4706 ($z=0.0281$).  As a validation of our methodology of determining the best-fit parameters by minimizing \X2 (see \S \ref{Model}), it should be noted that the returned \X2 contributions from \heil6678 for these targets (prior to removing \heil6678) were very large---32.4 for HS\s1028+3843 and 9.6 for HS\s1353+4706---signaling the very strong disagreement of the measured \heil6678 flux with the predicted flux.  That predicted \heil6678 flux is, in effect, determined by the other He emission lines, since their combined \X2 minimization is what determines the model parameters.  As such, for these two targets, the measured \heil6678 flux is in strong disagreement with all of the other He emission lines for any set of model parameters.  As expected, the measured \heil6678 flux for these two targets was also significantly lower than the model's predicted value.  
Upon removing \heil6678 the total \X2 for HS\s1028+3843 dropped from 64.7 to 17.4, and from 31.3 to 17.7 for HS\s1353+4706.  These dramatic decreases highlight just how extreme the discrepancy between the measured \heil6678 flux was with all of the other He emission lines, as well as highlighting the diagnostic utility of the \X2 distribution.  

Galactic \nai\s\W5889.95 and \nai\s\W5895.92 absorption will decrease the observed \heil5876 flux for redshifts of 0.00243 \& 0.00344 with a previously identified conservative range of $\pm$ 0.00033 \citep{oliv2004, aver2022}.  The magnitude of the decrease will vary and may be negligible.  At the same time, the same atmospheric \nai\ night sky emission lines will also overlay \heil5876, creating the potential for a systematic error to be introduced from the night sky subtraction.  For two of our targets, I\s Zw\s18\s SE with $z=0.002505$ and LEDA\s2790884 with $z=0.002425$, not only do their redshifts fall within the \nai\ absorption range, but the spectrum showed evidence for the absorption in the local continuum for \heil5876.  This finding was bolstered by the measured \heil5876 flux being significantly lower than the model's predicted value, as would be expected for \nai\ absorption.  Furthermore, this result is consistent with the relatively large foreground Galactic extinction for each target: $A_V = 0.091$ for I\s Zw\s18\s SE and $A_V = 0.126$ for LEDA\s2790884 \citep{schl2011}.  Therefore, \heil5876 was removed from our analysis for I\s Zw\s18\s SE and LEDA\s2790884.  

Four targets, HS\s1442+4250, SDSS\s J210455.31-003522.2, LEDA\s2790884, and SBS\s1415+437 show very strong, broad emission wings for H$\alpha$, and that, unusually, are stronger and broader than would be expected based on [\oiii]\s\W5007 or H$\beta$.  This behavior is not expected, lacking an obvious explanation, and it may indicate complicated kinematics within the \hii region, or potential Raman scattering \citep{henn2021}.  Regardless, it is not clear what portion of the H$\alpha$ emission should be included for comparison to the predicted model flux.  Furthermore, very strong wings on H$\alpha$ pose significant challenges for accurately fitting and measuring its emission line flux with its multiple components.  Therefore, H$\alpha$ is not reliable for these four galaxies and was excluded for our analysis.  Supporting the finding that H$\alpha$ is anomalous and unreliable for analysis with our model, all four of these galaxies showed unusually large \X2 contributions from H$\alpha$, indicating, in effect, significant disagreement between H$\alpha$ and all of the other H emission lines for any set of model parameters.  Upon removing H$\alpha$, all of the other Balmer lines that had been in tension snapped into good agreement with their measured values for the revised set of best-fit model parameters.  It is also worth noting that, though H$\alpha$ was in strong disagreement with the other lines and contributed a very large \X2, the same was not true of \heil6678, \heil7065, or the Paschen series lines.  This rules out a potential MODS red vs.\ blue channel calibration issue as the source of the discrepant H$\alpha$.

\subsection{\X2 Qualifying Fraction Comparison} 

Compared to previous analyses, the fraction of our dataset passing the \X2 cut is a dramatic improvement compared to previous studies.  47 out of our 54 targets qualify, equating to 87\%.  Though, ideally, 95\% would have passed, as discussed above (\S \ref{chi2cut}), systematic effects in our spectra and/or that are fundamentally unaccounted for in our \hii region emission modeling for a subset of our targets means that we should not expect 95\% to qualify.  Furthermore, an 87\% qualifying fraction is a vast improvement upon previous analyses. In \citet{aver2012}, the \X2 qualifying fraction using the HeBCD dataset from \citet{izot2007} was only 36\%, a very unsettling result.  Restricting to the subset of generally brighter HeBCD targets which included \10830 observations from \citet{izot2014} increased that qualifying fraction to 61\%, but that was still much lower than would be expected.  

Using different criteria (see Paper I), not \X2 testing, the qualification rate in \citet{hsyu2020} varies, but for their preferred sample criteria applied to the HeBCD dataset with \10830 observations, the qualification rate is 29\%.  That includes also excluding targets based on their model parameter values (where we flag such targets), but the majority of the excluded targets are based on the \citet{hsyu2020} criteria designed to play a similar role to a \X2 cut.  For their combination of the optical-only HeBCD, PHLEK, and SDSS observations, the qualifying rate decreases to 7\%.  Relaxing their criteria allows more targets to qualify, but, for their combined dataset, the highest qualifying rate \citet{hsyu2020} find is still only 50\%.  

High, best-fit \X2 values indicate one of three things, or a combination thereof:  (1) systematic errors with the measured fluxes, (2) systematic errors in the model, and/or (3) the reported uncertainties for the measured fluxes are underestimated.  Disentangling the source is challenging and impeded by the lack of information about various steps in the observational and/or measurement process, as well as the potential systematic errors in the atomic data underpinning the model equations.  

Therefore, one of the major goals of \Proj\ was to collect our own high-quality, high SNR observations to produce a uniformly observed dataset (see Paper~I), carefully reduce those observations ourselves (see Papers~II \& III), measure the emission line fluxes ourselves (see Papers~II \& III), and apply our own error model for calculating the flux uncertainties (see Papers~I \& II).  That dataset, \Proj\ dataset, resulted in a \X2 qualifying fraction of 87\%.  The dramatic improvement underscores the quality of our dataset and increases our confidence in the applied model, and thus the results.


\section{Sample Testing} \label{Testing}


The large, high-quality dataset provided by our LBT observations allows us to better investigate potential systematic biases and test for reliability.  We screen for potential systematic bias in the helium abundance that is correlated with other physical parameters.  In evaluating corrections for systematic effects in our model, we also examine parameter values which correspond to large systematic model corrections.  Since the underlying model equations have unknown uncertainties and, by necessity, make simplifying assumptions, large model corrections for systematic effects carry the risk of introducing significant systematic error and reducing the reliability of the returned physical conditions.  

In these cases, where we see evidence for potential systematic bias in the helium abundance and/or large model corrections which may introduce systematic error compromising the reliability of the returned physical conditions, we flag those targets.  We still report the best-fit parameter values for flagged targets, but they are not included in our preferred final dataset for determining \Yp.

\subsection{Optical Depth ($\tau$)} \label{tau}

As discussed above in Section \ref{RadTrans}, our radiative transfer model comes from \citetalias{kuri2025}.  It assumes a spherically symmetric \hii region.  It does not account for any non-spherical geometry, expansion, velocity structure, or potential temperature or density inhomogeneities.  Therefore, we note the potential impact of any model inaccuracy for high values of optical depth, where the effects of radiative transfer are larger.  As those corrections for radiative transfer increase, the magnitude of any potential error introduced by a potential systematic error in the model increases.  Furthermore, the \citetalias{kuri2025} fit equations for $f_\tau$ are fit over the range $0 < \tau < 10$ (see Section \ref{RadTrans} \& Appendix \ref{Appendix:RadTrans}).  For values of optical depth with $\tau > 10$, the radiative transfer correction is an extrapolation, which may introduce increasing degrees of systematic error.  

\begin{figure}[t!]
\resizebox{\columnwidth}{!}{\includegraphics{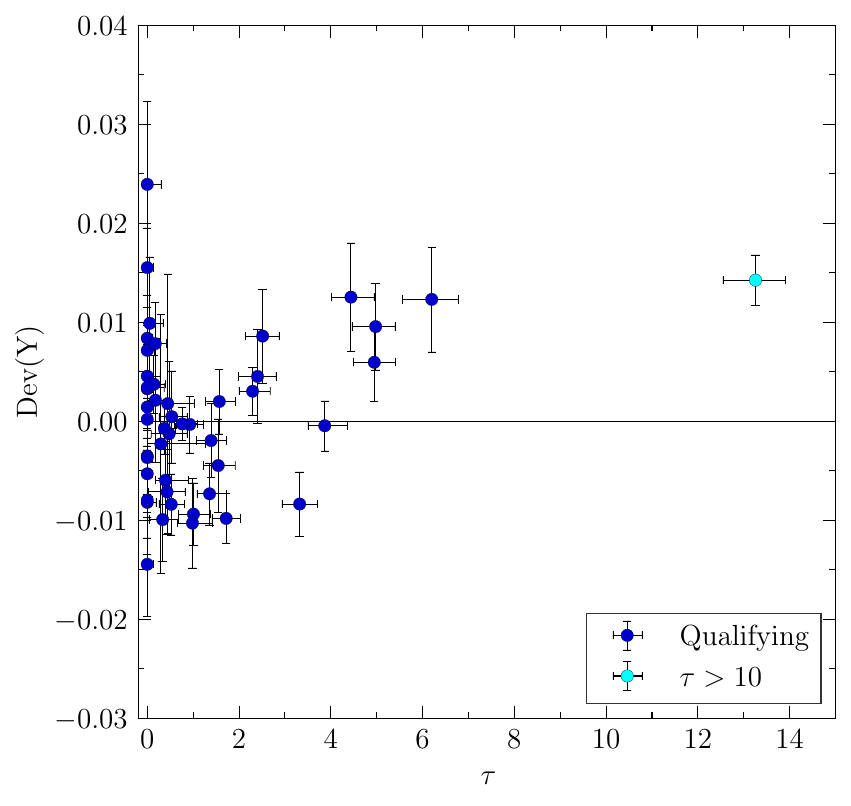}}
\caption{
The deviation in the helium abundance for our qualifying sample (47 targets) from their linear trend (i.e., their residuals) plotted versus their best-fit optical depth value ($\tau $).  The value of $\tau$ corresponds to the optical depth for \3889, with the radiative transfer correction for each helium triplet emission line parameterized in terms of the optical depth for \3889.  One point is a clear outlier, and the fit domain for the radiative transfer equations is $0 < \tau < 10$.  Therefore, the target with $\tau > 10$ is flagged.  The flagged target is plotted in cyan.  
}
\label{figure:dY-tau}
\end{figure}

In particular, because radiative transfer at larger values of optical depth significantly decreases \3889 emission and increases \heil7065 and \10830 emission (see Figure \ref{figure:ftau_all} above), it is possible radiative transfer corrections could systematically bias the helium abundance.  Figure \ref{figure:dY-tau} shows the deviation in the helium abundance mass fraction, Y, for our targets from their linear Y vs.\ O/H trend (i.e., their residuals, see Section \ref{Regression}) versus their best-fit optical depth value.  

There is no physical reason to expect a relationship between optical depth and helium abundance.  Nonetheless, for the highest values of optical depth, there is some potential evidence of an upward bias in the helium abundance, but the limited number of points with $\tau > 2$ limits the conclusions we can draw.  Additionally, the target with the largest optical depth, HS\s1028+3843 with $\tau = 13.25$, appears to be a clear outlier in the sample.  Correspondingly, it carries the risk of introducing significant systematic error as an outlier with the largest radiative transfer correction, as well as due to its best-fit optical depth extending beyond the fit domain for the radiative transfer equations ($0 < \tau < 10$).

Therefore, we flag HS\s1028+3843, the one target with an optical depth greater than the fit equation's domain, $\tau>10$.  One could reasonably argue for flagging the five targets with $\tau > 4$ or the ten targets with $\tau > 2$.  However, in each case, those points fall well within the range of deviations established by the much larger low optical depth subsample.  As such, and given the limited number of targets with $\tau > 4$ or even $\tau > 2$, those subsets do not exhibit a definitive upward bias in the helium abundance.  Furthermore, we do not have any reason to suspect the \citetalias{kuri2025} equations may be less accurate as optical depth increases within their fit domain.  Therefore, to be conservative and retain more of our sample in our preferred dataset for our subsequent analysis, we only flag the single target with $\tau > 10$.  

\citet{aver2012} flagged targets with $\tau > 4$, as justified by the reasoning above regarding any systematic errors introduced by the radiative transfer equations becoming more significant combined with an examination of the distribution of optical depth values in that dataset \citep[the HeBCD dataset from][]{izot2007}.  There are two key differences for this work.  First, \citet{aver2012} utilized the \citet{benj2002} limited-range fit equations with a fit domain of $0 < \tau < 2$, rather than the new \citetalias{kuri2025} calculations with a fit domain of $0 < \tau < 10$.  Second, our \X2 qualifying dataset is larger, and points with $\tau > 4$ are not outliers in our dataset.

\subsection{Underlying Helium Stellar Absorption (a$_{He}$)} \label{UAHe}

The underlying stellar absorption model employed in this work is detailed in \citet{aver2021}.  It is based on the spectral energy distributions from BPASS \citep{eldr2009, eldr2017, stan2018} for ages of 1, 2, 3, 4, \& 5 Myr and for both single and binary star systems.  For each case, the absorption equivalent width was measured, and the set of values for each line was averaged.  The underlying absorption ratios vary with those starburst ages and between the single and binary star systems.  As a result, it is possible the model may introduce systematic error, and any such error would increase in magnitude for larger best-fit underlying absorption values (i.e., $a_{He}$ for the helium emission lines).  

\begin{figure}[t!]
\resizebox{\columnwidth}{!}{\includegraphics{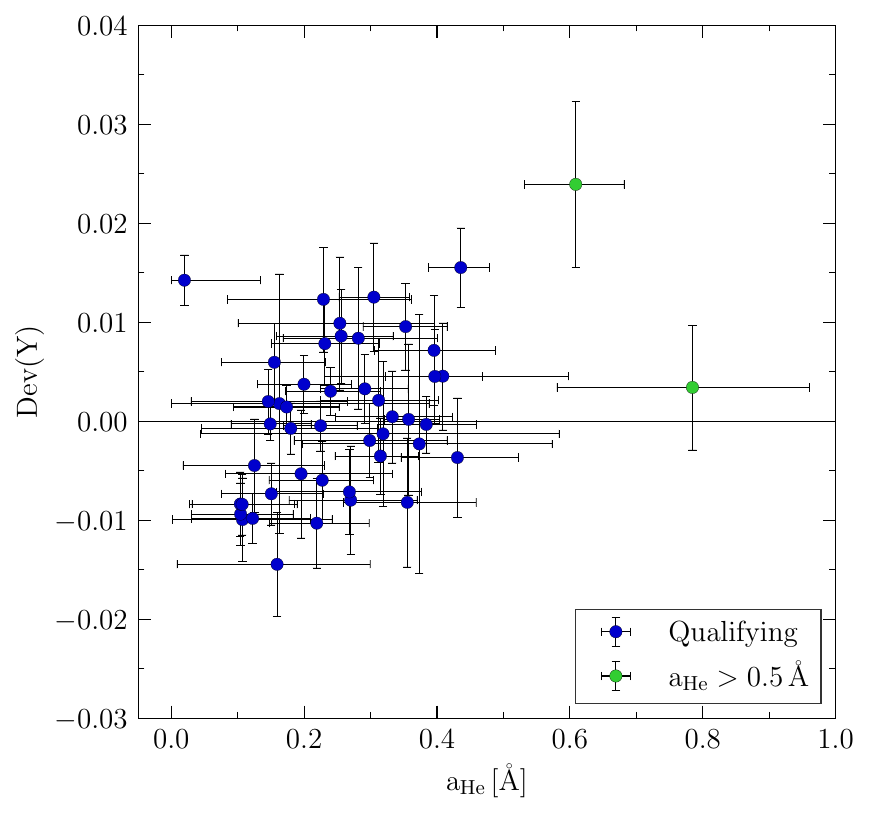}}
\caption{
The deviation in the helium abundance for our qualifying sample (47 targets) from their linear trend (i.e., their residuals) plotted versus their best-fit underlying stellar absorption parameter for the helium lines ($a_{He}$).  The value of $a_{He}$ corresponds to the underlying absorption for \heil4471.  The correction for every other helium line is scaled off that value according to the scaling coefficients given in \citet{aver2021}.  For one of the two targets with the highest underlying helium absorption, there is evidence of a potential upward bias in the derived helium abundance, which could bias our results.  Furthermore, the two targets with the highest underlying helium absorption are outliers from the rest of the sample.  Therefore, the two targets with a$_{He} > 0.5$\,\AA\ are flagged.  Those flagged targets are plotted in green.  
}
\label{figure:dY-aHe}
\end{figure}

The deviation in Y is plotted versus the best-fit underlying stellar absorption for the helium lines, $a_{He}$, in Figure \ref{figure:dY-aHe}. 
The value of $a_{He}$ corresponds to the underlying absorption for \heil4471.  The correction for every other helium line is scaled off that value, according to the scaling coefficients given in \citet{aver2021}.  Figure \ref{figure:dY-aHe} shows two clear outliers with $a_{He}>0.5$\,\AA.  Those two targets are HS\s1222+3741 and UGC\s6456.  Due to their greater potential for systematic error, we flag both.

In \citet{aver2012}, the flagging threshold for underlying helium absorption was twice as high: $a_{He}>1.0$\,\AA.  However, that work, utilizing the HeBCD dataset from \citet{izot2007}, showed significantly higher values for $a_{He}$ overall.  For \Proj\ dataset presented in this work, the average value for $a_{He}$ is 0.266\,\AA, while in \citet{aver2012} it is 0.522\,\AA.  Reassuringly, these lower best-fit underlying helium absorption values from \Proj's dataset are more consistent with our measurements from BPASS \citep{aver2015}, where the value (for \heil4471) ranged between 0.1 and 0.2\,\AA.  

The cause for this significant decrease in the typical underlying absorption value is not known.  The scaling coefficients for underlying helium absorption used in \citet{aver2012} were updated in \citet{aver2021}, with this work using the latter, but the changes in those coefficients were not large, and the overall pattern across the primary He lines was the same.  Therefore, that model change does not seem like the likely source.  Our revised treatment of the blended H8 + \3889 line could also have affected the best-fit $a_{He}$ \citep{aver2021}, but \3889 is less sensitive to underlying absorption than weaker He lines (e.g., \heil4026), and the treatment of underlying absorption for \3889 was not changed.  Differences in the extraction apertures for the HeBCD observations \citep{izot2004, izot2007} and our observations may be the source, as well as potential methodological differences in reducing the spectra.  

Regardless, the $a_{He}>0.5$\,\AA\ flagging criteria adopted here is conservative and results in the same number of flagged targets as were flagged in \citet{aver2012} based on its criteria (i.e., 2 flagged targets for $a_{He}$ in both).  The distribution of best-fit underlying Balmer and Paschen stellar absorption parameters, $a_H$ \& $a_P$, did not exhibit any unusually large values or clear outliers.

\subsection{Temperature Comparison -- T(He), T(\oiii), T(\siii)} \label{Tcomp}

In our model, the electron temperature is a fit parameter, with that temperature dominantly determined by the helium emission lines, due to their much stronger temperature sensitivity compared the hydrogen lines \citep{peim2000, oliv2004}.  In general, one would expect that this helium temperature, T(He), should closely track the temperature determined by the [\oiii] emission, T(\oiii), due to the O$^{++}$ ionization zone encompassing almost the entirety of the He$^+$ zone ($\sim$95\%).  For the idealized, spherically symmetric \hii region, the He$^+$ zone could have a slightly lower average electron temperature than the O$^{++}$ zone due to its larger volume, but this effect would be small due to the extent of their overlap.  

However, in the presence of temperature fluctuations throughout the emission region, the higher temperature regions will contribute an elevated share of the total [\oiii] emission, thus weighting the [\oiii] flux more strongly to the higher temperature regions than their share of the volume.  This effect could bias T(\oiii) systematically upward compared to T(He), as has been observed in some studies \citep{peim2000, peim2002, oliv2004, aver2011, mend2025}, though not others \citep{guse2006, guse2007}.  \citet{peim2002} found that T(He) was was always less than T(\oiii) by 6-10\%, using their best-fit values from the He emission lines, and 3-11\% in their photoionization modeling.  In our model, we use T(\oiii) as a weak prior on determining the best-fit value for T(He) \citep{aver2011, aver2021}.   

\begin{figure}[t!]
\resizebox{\columnwidth}{!}{\includegraphics{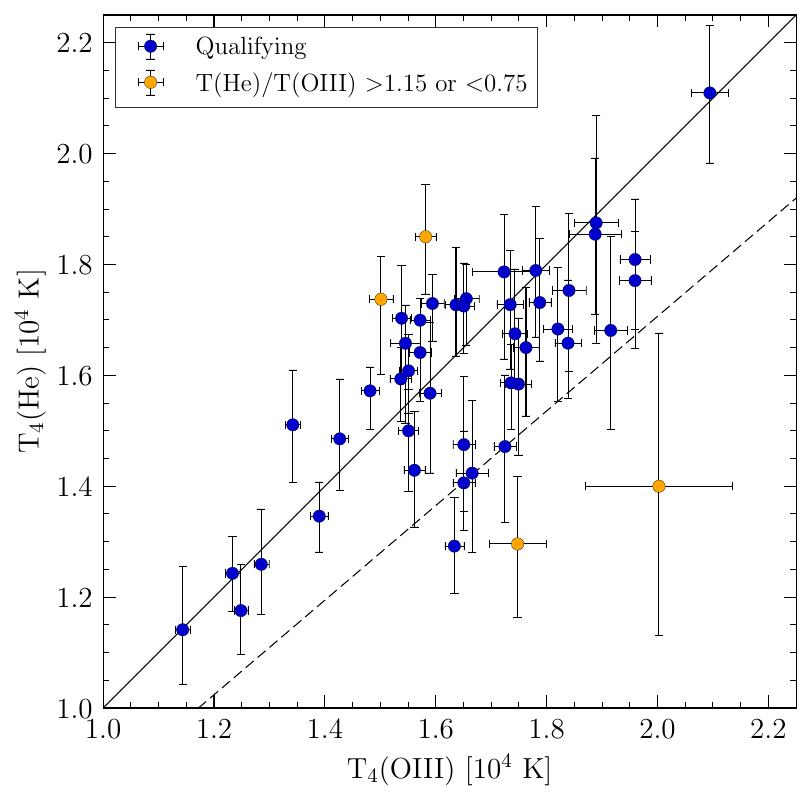}}
\caption{
The best-fit electron temperature from our model as determined (primarily) by the He emission lines, T(He), plotted versus the electron temperature determined by the [\oiii] emission lines, T(\oiii).  Our qualifying sample (47 targets) is plotted.  The solid line plotted is for equality, and the dashed line represents the expected relationship between T(He) and T(\oiii) for temperature fluctuations as parameterized by $t^2=0.03$ \citep[e.g.,][]{peim2000,garc2007}\footnote{The  ``$t^2$'' parameter is the RMS deviation from the average nebular temperature. It is a measure of the internal temperature variations of the gas as quantified in the formalism proposed by \citet{peim1967}.}.  A definitive upward bias in T(\oiii) is not observed, though there is some evidence for that upward bias at higher temperatures ($\gtrsim15,000\,\r{K}$).  Two targets where T(He) exceeds T(\oiii) and two targets where T(He) is lower than T(\oiii) appear to be outliers from the rest of the sample.  Those flagged targets are plotted in orange.  
}
\label{figure:THe_TOIII}
\end{figure}

We would expect T(He) to track T(\oiii) closely, though possibly with a value that is slightly lower, though not necessarily, given differing estimates of the typical magnitude of temperature inhomogeneities in \hii regions \citep{peim2000, peim2002, guse2006, guse2007}.  Figure \ref{figure:THe_TOIII} compares T(He) and T(\oiii).  The magnitude of the T(He) uncertainties limits one's ability to draw firm conclusions, but T(\oiii) does not appear to be generally biased high compared to T(He).  At the higher temperatures in our sample ($\gtrsim15,000\,\r{K}$), there is some evidence for a systematically higher value for T(\oiii) than T(He), but it is not conclusive.

More striking, perhaps, is the difference in T(He) and T(\oiii), with T(He) exceeding T(\oiii) in many cases, and occasionally dramatically lower, as shown in Figure \ref{figure:THe_TOIII}.  The large uncertainties for T(He) mean that the two temperatures agree within 2$\sigma$ for most of the targets, but there are also a few clear outliers with larger absolute and relative differences.  These large temperature discrepancies raise concern that T(He) may be systematically over or underestimated in those targets, biasing \y+.  In particular, T(He) exceeds T(\oiii) by 2680\,K or 17.0\% for Mrk\,71 and 2360\,K or 15.7\% for HS\s1222+3741, though it should be noted that HS\s1222+3741 is flagged for underlying helium absorption, and its best-fit parameter values may be less reliable and potentially systematically biased.  For two targets, T(He) is lower than T(\oiii) by a much greater degree than in the rest of the sample.  For AGC\s198691, T(He) is lower than T(\oiii) by 6030\,K or 30.1\%, and 4520\,K or 25.8\% for J0118+3512.  

\begin{figure}[t!]
\resizebox{\columnwidth}{!}{\includegraphics{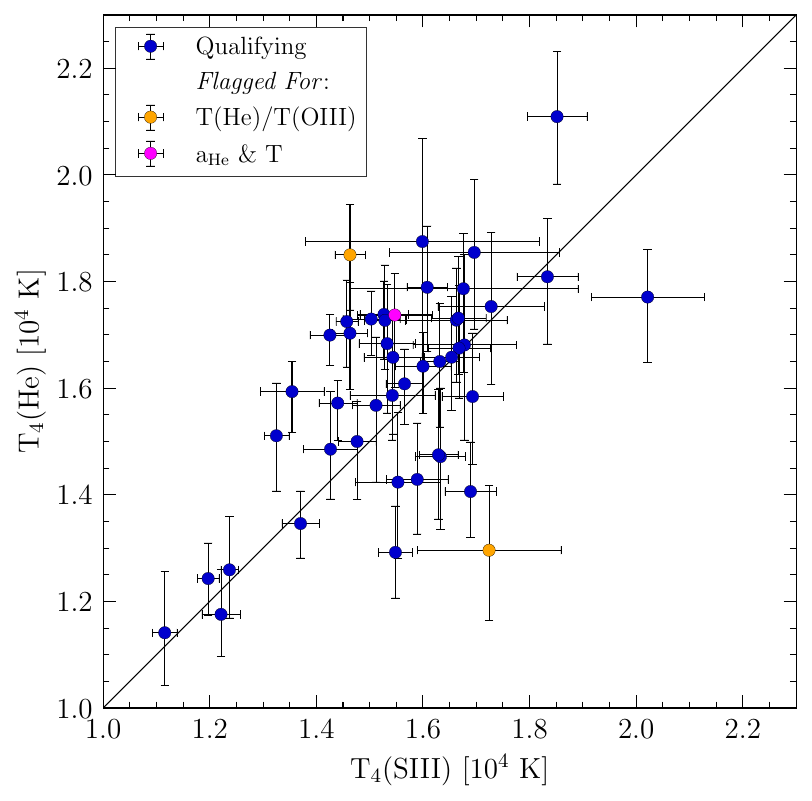}}
\caption{
The best-fit electron temperature from our model as determined (primarily) by the He emission lines, T(He), plotted versus the electron temperature determined by the [\siii] emission lines, T(\siii).  Our qualifying sample (47 targets) is plotted, though excluding one target without a [\siii] temperature.  The solid line plotted is for equality.  Four targets are flagged for the discrepancy in their T(He) and T(\oiii) temperatures (see Figure \ref{figure:THe_TOIII}), with two of the four plotted in orange, one that is additionally flagged for underlying helium absorption plotted in magenta, and the fourth not plotted because it lacks a T(\siii) value.  The two orange targets appear to be outliers when comparing their T(He) and T(\siii), with one higher and one lower, just as was seen for their T(He) and T(\oiii) in Figure \ref{figure:THe_TOIII}.  The magenta point's T(He) and T(\siii) are not discrepant, but it is already flagged for a$_{He}>0.5$.  
}
\label{figure:THe_TSIII}
\end{figure}

\begin{figure}[t!]
\resizebox{\columnwidth}{!}{\includegraphics{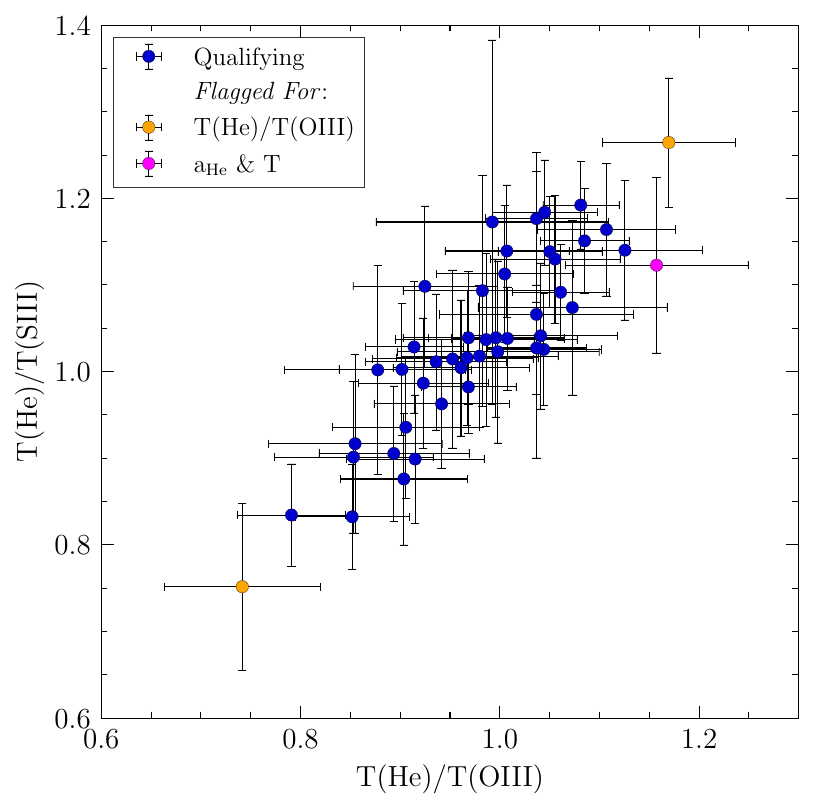}}
\caption{
T(He)/T(\siii) plotted versus T(He)/T(\oiii) for our qualifying sample (47 targets), though excluding 1 target lacking a [\siii] temperature.  A strong linear relationship is observed between the two temperature ratios, as expected based on the results presented in Paper II.  Four targets are flagged for the discrepancy in their T(He) and T(\oiii) temperatures (see Figure \ref{figure:THe_TOIII}), with two of the four plotted in orange, one that is additionally flagged for underlying helium absorption plotted in magenta, and the fourth not plotted because it lacks a T(\siii) value.  The two orange targets are outliers with one where T(He) exceeds both T(\oiii) and T(\siii) and one where it is lower than both.  The magenta point's T(He)/T(\siii) value is not discrepant, but it is already flagged for a$_{He}>0.5$.  This figure reinforces the flagging selections in Figure \ref{figure:THe_TOIII}.  
}
\label{figure:THe_TSIII_TOIII}
\end{figure}

The very tight linear relationship between T(\oiii) and T(\siii), as shown in Paper II, means that T(\siii) can also be used as a diagnostic for identifying T(He) discrepancies.  Figure \ref{figure:THe_TSIII} compares T(He) and T(\siii).  Though T(\siii) tends to be slightly lower than T(\oiii) (see Paper II), a similar pattern emerges.  As we saw comparing Mrk\s71's helium best-fit temperature to its [\oiii] temperature (Figure \ref{figure:THe_TOIII}), T(He) is also discrepantly elevated compared to T(\siii) for Mrk\s71.  Similarly, for J0118+3512, its T(He) is discrepantly low compared to T(\siii), just like it was compared to T(\oiii).  HS\s1222+3741 is not an outlier when comparing T(He) to T(\siii), unlike for T(\oiii), but HS\s1222+3741 is already flagged for underlying helium absorption.  The fourth outlier identified by T(\oiii), AGC\s198691, does not have a calculated [\siii] temperature because [\siii]\s\W6312 was not detected.  Figure \ref{figure:THe_TSIII_TOIII}, a plot of T(He)/T(\siii) vs.\ T(He)/T(\oiii), identifies those two unique T(He) outliers, Mrk\,71 \& J0118+3512, most clearly and demonstrates that they are shared outliers on each measure, T(\oiii) and T(\siii).

For these four identified outliers (Figure \ref{figure:THe_TOIII}), the temperature differences are larger than can be justified by any reasonable physical model.  The calculated helium abundance, y$^+$, depends strongly on the electron temperature.  Therefore, such large temperature differences in the best-fit T(He) value and the T(\oiii) and T(\siii) values may be introducing a significant systematic error in the best-fit helium abundance.  On this basis, we flag the four targets identified above: HS\s1222+3741 \& Mrk\,71 for $\r{T(He)/T(\oiii)} > 115\%$ and AGC\s198691 \& J0118+3512 for $\r{T(He)/T(\oiii)} < 75\%$.  HS\s1222+3741 is already flagged for $a_{He}>0.5$ and does not increase our number of flagged targets.

\subsection{Nitrogen-to-Oxygen Ratio (log(N/O))}

In a recent work, \citet{yana2024} report results from JWST observations of three high-$z$ ($z \sim 6$) galaxies, which show high values for their nitrogen-to-oxygen abundance ratio ($\r{log(N/O)} \gtrsim -1.0$).  They also show very strong \heins\ emission, which may signal significantly elevated helium abundances ($\r{y} = \r{He/H} \gtrsim 0.10$), compared to what would be expected for their metallicity, O/H.  Based on that correlation between high N/O and potentially high He/H, and the possibility for chemical evolution that deviates substantially from the rest of their sample, \citet{yana2025} exclude three targets from their sample with $\r{log(N/O)} > -1.0$.  

\begin{figure}[t!]
\resizebox{\columnwidth}{!}{\includegraphics{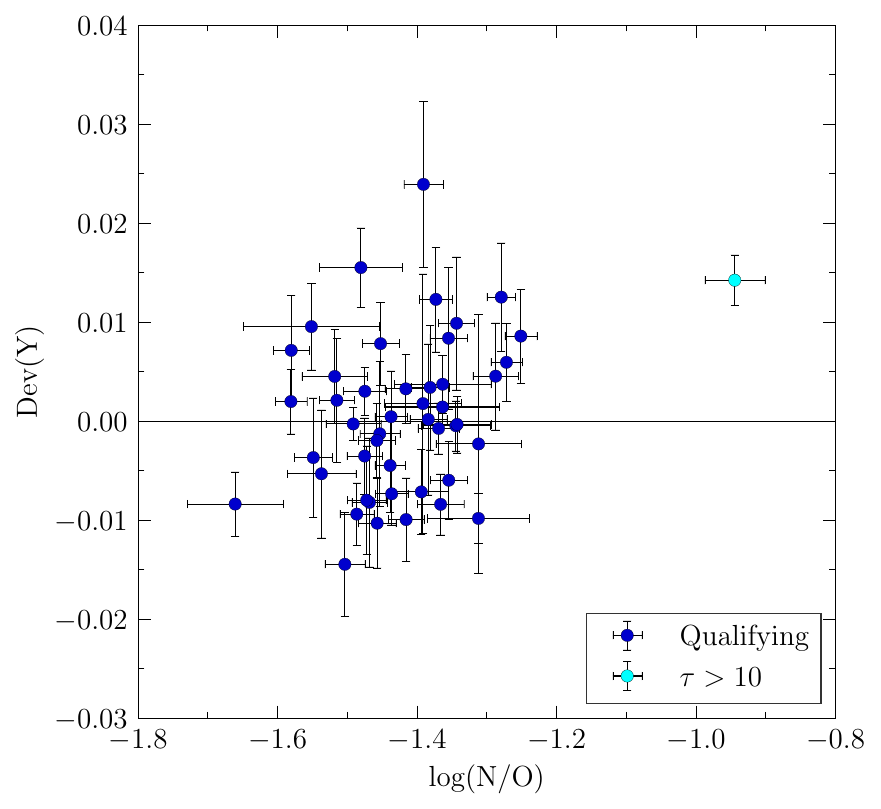}}
\caption{
The deviation in the helium abundance for our qualifying sample (47 targets) from their linear trend (i.e., their residuals) plotted versus log(N/O).  There is no clear evidence of systematic bias in the helium abundance related to the N/O abundance ratio for the targets in our sample.  The one target with an anomalously larger log(N/O) is flagged for optical depth ($\tau > 10$) and plotted in cyan.  
}
\label{figure:dY-logNO}
\end{figure}

We investigated flagging targets in our sample based on their N/O value, but, as shown in Figure \ref{figure:dY-logNO}, our sample did not exhibit any systematic effect on Y as a function of N/O.  Furthermore, there is only one target in our \X2 qualifying sample with $\r{log(N/O)} > -1.0$, and it is already flagged for optical depth.  Excluding this target, our sample has a nitrogen-to-oxygen abundance ratio in the range,  $-1.7 < \r{log(N/O)} < -1.2$.

The target in Figure \ref{figure:dY-logNO} with $\r{log(N/O)} > -1.0$, HS\s1028+3843, has, by far, the \textit{highest} best-fit optical depth in the sample, with $\tau = 13.25$.  This is very large, and unusually so, for a low-metallicity \hii region.  The galaxies in \citet{yana2024} with high N/O also show very high optical depths ($\tau>8$).  Similarly, two targets excluded due to their best-fit \X2 values exceeding their 95\% confidence level have $\r{log(N/O)} \gtrsim -1.0$, HS\s0837+4717 and WISEA\s J051902.72+000728.7, and their best-fit optical depth values are 9.46 and 8.57, respectively.  Also worthy of note, HS\s0837+4717 exhibits the strongest wings on its strong emission lines, as mentioned above (\S \ref{chi2cut}), and it returns $\y+ = 0.10907$.

As discussed above (\S \ref{tau}), high optical depth values carry the potential for significant systematic errors in the model and the determined helium abundance.  Both high log(N/O) and large optical depth values may indicate physical conditions in the \hii region that do not conform with the rest of the sample and the applied modeling.  Although there appears to be a correlation between high $\tau$ and high log(N/O), they may both be markers for \hii regions with atypical underlying dynamics or physical conditions.  To this end, we provide a more detailed discussion concerning the elevated N/O abundances measured in five of the LBT \Yp\ targets in Paper~VI.

\subsection{Additional Testing}

Similar to Figures \ref{figure:dY-tau}, \ref{figure:dY-aHe}, \& \ref{figure:dY-logNO}, we examined the deviation in the helium abundance from the sample trend versus each of our physical model parameters.  As mentioned in Section \ref{UAHe}, the underlying Balmer and Paschen stellar absorption parameters, $a_H$ \& $a_P$, did not exhibit any unusually large values, clear outliers, or systematic trend with the helium abundance, nor did the reddening, C(H$\beta$), or neutral hydrogen fraction, $\xi$.  Only in the cases of optical depth and temperature was there clear evidence of a potential systematic bias.  In the case of temperature, T(He), that relationship is expected since, all else being equal, higher temperatures increase the helium abundance, due to the emissivity ratio (E(He)/E(H$\beta$) for the three strongest helium lines---\hei \W\W4471, 5876, 6678)---decreasing with temperature.  As discussed in \ref{Tcomp}, we do flag based on the divergence of T(He) from T(\oiii) and T(\siii).  Similar to our model parameters, we also investigated for other parameters and measured values, such as W(H$\beta$), \n(\sii), He$^{++}$, O$^+$/O, $\eta = \frac{{\rm O}^+/{\rm O}^{+2}}{{\rm S}^+/{\rm S}^{+2}}$, \X2/dof, etc.  We did not find clear evidence for a systematic relationship between any of these other parameters and the difference in helium abundance (\D Y).


\section{Sample Results} \label{Results}


Following the \X2 qualifying cut and applied flags for potential systematic errors (\S \ref{chi2cut} \& \S \ref{Testing}), Table \ref{table:Flagged} presents a breakdown of our sample dataset.  For our starting sample of 54 targets, 7 were excluded due to best-fit (minimized) \X2 values exceeding their 95\% CL (see \S \ref{chi2cut}).  Of the remaining 47 targets, 6 are flagged, and there are 41 qualifying targets without any flags.  These 41 targets constitute our final qualifying dataset.  The best-fit parameter solutions with uncertainties for each of the 47 targets, qualifying and flagged, is provided in Table \ref{table:Params} above.


\begin{table}[ht!]
\footnotesize
\centering
\vskip .1in
\begin{tabular}{lr}
\hline\hline
The LBT \Yp\ Project				& 54   \\
\hline
\hspace{0.1in} $\chi^{2}>95\%$ CL		   		& 7/54  \\
\hline
Well-Defined Solutions	&    47 \\
\hline
\underline{\textit{Flagged}} \\
\hspace{0.1in} $\tau > 4$					&  1  \\
\hspace{0.1in} a$_{He} > 0.5$ \AA				&  2  \\
\hspace{0.1in} T(He)/T(\oiii) > 1.15			&  1  \\
\hspace{0.1in} T(He)/T(\oiii) < 0.75			&  2  \\
\hline
Subtotal:  Flagged					&  6 \\
\hline
Final Qualifying Dataset				& 41  \\
\hline
\end{tabular}
\caption{Sample Breakdown of Cuts and Flags}
\label{table:Flagged}
\end{table}


\subsection{Uncertainty Comparison} \label{HeComp}

Comparing our results to our previous work is complicated by the mixed sources of observations in our two most recent \Yp\ determinations \citep{aver2021,aver2022}.  To recap, \Proj\ began with the high-quality LBT MODS \& LUCI observations of Leo\,P \citep{skil2013,aver2021}.  The re-analysis of Leo\,P in \citet{aver2021} utilized a significantly expanded and improved physical model, enabled by the MODS \& LUCI observations.  Leo\,P served as a proof-of-concept for the advances in helium abundance determinations that could be made based on high-quality, high signal-to-noise, and broad wavelength coverage optical and NIR spectra.  Subsequently, AGC\s198691 was also re-analyzed based on a LBT MODS observation \citep{aver2022}.  Therefore, to properly compare with our results prior to \Proj, \citet{aver2015} is the appropriate comparison.  

Increasing the precision of galactic helium abundance determinations is one of the primary goals of this project.  Comparing to our previous results, the average relative uncertainty for \y+ in \citet{aver2015} was 4.0\%, while for \Proj, it is 2.3\%.  This 42\% reduction in the average He abundance uncertainty across qualifying targets without flags (41 targets) is due to multiple improvements in \Proj:  
\begin{enumerate}
    \item High signal-to-noise LBT MODS \& LUCI spectral observations, uniformly observed and reduced (See Papers II \& III); 
    \item High spectral resolution and broad wavelength coverage optical and NIR spectra, as enabled by MODS \& LUCI (See Paper I); 
    \item The full deployment of our expanded model, including additional, weaker He lines, Paschen lines, and higher Balmer lines, as enabled by the high resolution and broad wavelength coverage observations (See \citet{aver2021}); 
    \item Our improved flux uncertainty model based on the stable MODS spectral sensitivity function (See Paper I).  
\end{enumerate}

While all of our model parameters are more precisely determined, given the limited constraints on the neutral hydrogen fraction values, $\xi$, in our previous samples \citep{aver2012, aver2013, aver2015}, it is particularly encouraging to see the significant improvements in better constraining $\xi$ (for $\T4>1.4$).  Comparing to the \citet{aver2015} dataset, the average uncertainty for $\xi$ decreased by 90\%.  The inclusion of Paschen and higher Balmer emission lines significantly improves the determination of the primary hydrogen fit parameters, $C(H\beta)$, $a_H$, $a_P$, and $\xi$, which, in turn, contribute to the reduced uncertainty for \y+.

\subsection{Emissivity Comparison} \label{EmissComp}

\begin{figure}[t!]
\resizebox{\columnwidth}{!}{\includegraphics{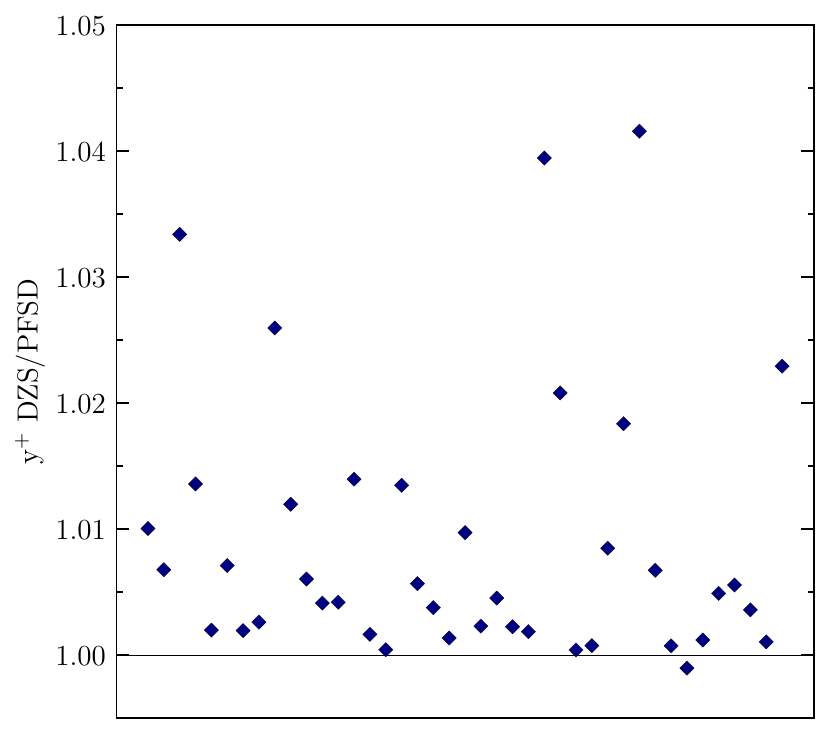}}
\caption{
The best-fit helium abundance, \y+, based on the updated He emissivities of \citet{delz2022}, relative to that of the previously employed He emissivities, \citet{port2012,port2013}.  Our final qualifying sample (41 targets) is plotted.  Due to the relatively minor changes in the emissivities at low density ($\sim$1\%), the effect on \y+ is relatively small, though it is systematic.  Overall, there is an increase in the resulting helium abundance from the new emissivities, with an average relative increase of 0.9\%.  
}
\label{figure:y+_DZS_PFSD}
\end{figure}

In this work, we updated the He emissivities employed in our model from the \citet{port2012,port2013} emissivities to the more recent emissivity calculations of \citet{delz2022} (\S \ref{Emiss}).  At the same time, the normalizing H$\beta$ emissivity, E(H$\beta$), was also updated using the new \citet{stor2015} H emissivities (\S \ref{Emiss}).  Figure \ref{figure:y+_DZS_PFSD} shows the change in \y+ from the \citet{delz2022} emissivities.  

As shown in Figure \ref{figure:y+_DZS_PFSD}, the effect on the helium abundance was relatively small, but systematic, with all but one target (40 of 41) showing a small increase in \y+. Across our final qualifying dataset, the average relative change in \y+ was an increase of 0.9\%.  As discussed in Section \ref{Emiss}, the most significant change for the \citet{delz2022} emissivities is a decrease in the \heil6678 emissivity (see Figures \ref{figure:DSZ-PFSD-T} \& \ref{figure:DSZ-PFSD-D}).  Though complicated by the full set of model parameter values changing, a decrease in the \heil6678 emissivity tends to increase the helium abundance, since the lower emissivity requires a larger abundance to produce the same flux.  

The updated absolute H$\beta$ emissivity increased by $\sim0.1\%$ to $\sim0.25\%$.  Since E(H$\beta$) normalizes the He emissivities, this change decreases the helium-to-hydrogen emissivity ratio uniformly for all helium lines, thus increasing \y+ by a commensurate amount.  The combination of the \citet{delz2022} emissivities change, primarily a decrease \heil6678, along with the E(H$\beta$) increase, yields the overall increase in the best-fit helium abundance, \y+, of 0.95\%.

\begin{figure}[t!]
\resizebox{\columnwidth}{!}{\includegraphics{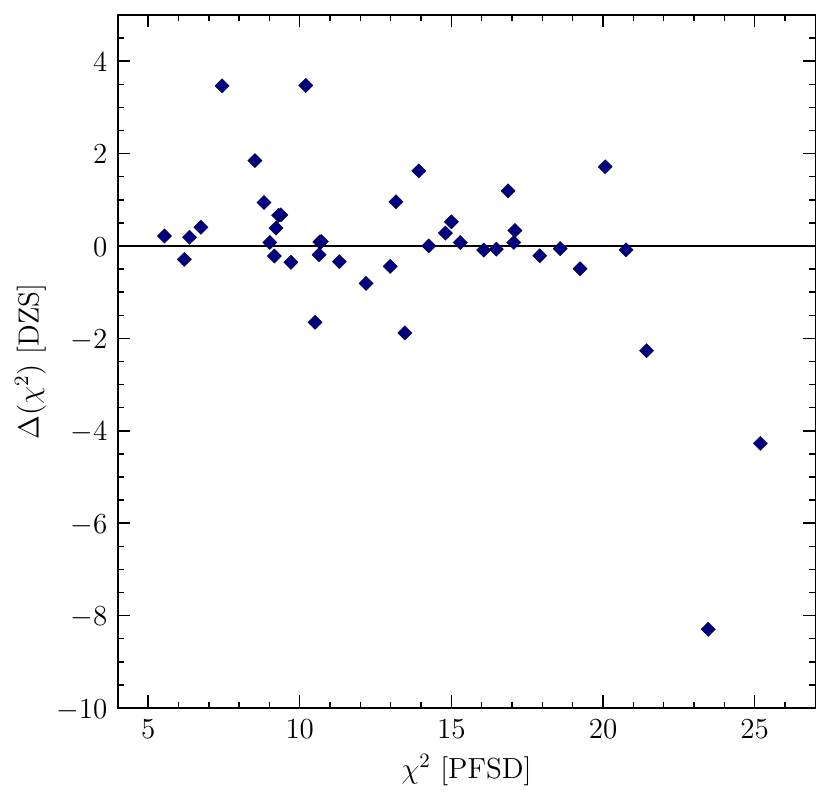}}
\caption{
The difference in the best-fit (minimized) \X2 between the updated He emissivities of \citet{delz2022} and the previously employed He emissivities of \citet{port2012,port2013} plotted versus the \citet{port2012,port2013} best-fit \X2.  Our final qualifying sample (41 targets) is plotted.  Due to the relatively minor changes in the emissivities at low density ($\sim$1\%), the effect on \X2 is relatively small, and for most targets the change is consistent with random variation (i.e., some small increases, some small decreases).  However, the two largest changes from the updated emissivities are decreases in \X2, indicating closer agreement between the model and observed fluxes.  
}
\label{figure:X2_DZS_PFSD}
\end{figure}

Figure \ref{figure:X2_DZS_PFSD} compares the best-fit \X2 for the \citet{port2012,port2013} and \citet{delz2022} emissivities.  As shown in Figure \ref{figure:X2_DZS_PFSD}, the best-fit \X2 value is largely unchanged for the majority of our sample.  That is not surprising given the relatively minor changes in the emissivities at low density (see Figure \ref{figure:DSZ-PFSD-D}).  However, it is encouraging that the two targets with the largest changes in their \X2 value show decreases.  The decreased best-fit \X2 for those targets from the \citet{delz2022} emissivities indicates improved agreement between the model and observed fluxes.

\subsection{Radiative Transfer Model Comparison} \label{RadTransComp}

As discussed in Section \ref{RadTrans}, we have replaced the radiative transfer calculations of \citet[][, BSS02]{benj2002} in our model with the recent, updated calculations \citet[][, KI25]{kuri2025}.  In this subsection, we track the effects of the new \citetalias{kuri2025} radiative transfer calculations from the previous \citetalias{benj2002} calculations, as incorporated using the trilinear interpolation from \citetalias{benj2002}.  

As shown in Figure \ref{figure:ftau_3889_7065} and discussed in Section \ref{RadTrans}, the magnitude of the radiative transfer effect for \heil7065---the line most strongly affected by radiative transfer---is increased in the new \citetalias{kuri2025} calculations compared to the previous \citetalias{benj2002} calculations.  As a result, to achieve a similar increase to \heil7065 from radiative transfer as the interpolation would predict, the corresponding optical depth will be lower.  Therefore, it is not surprising that the best-fit optical depth value for targets returning higher optical depths tended to decrease from the adoption of the new \citetalias{kuri2025} radiative transfer equations.  

\begin{figure}[t!]
\resizebox{\columnwidth}{!}{\includegraphics{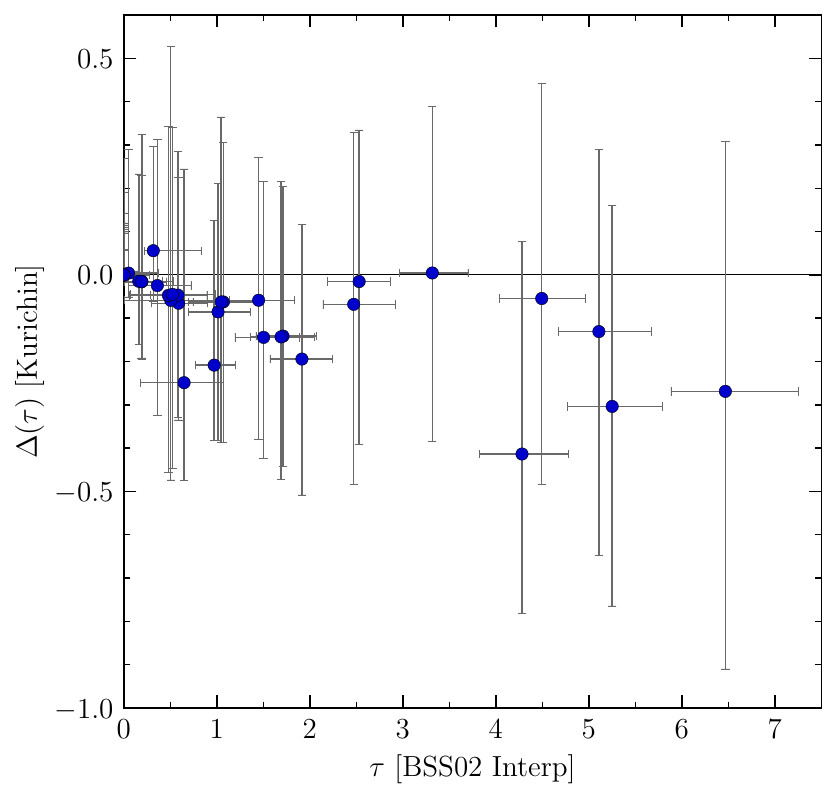}}
\caption{
The difference in the best-fit optical depth, $\tau$, between the \citet{kuri2025} radiative transfer fit equations and the \citet{benj2002} radiative transfer interpolation plotted versus $\tau$ based on the interpolation.  The value of $\tau$ corresponds to the optical depth for \3889, with the radiative transfer correction for each helium triplet emission line parameterized in terms of the optical depth for \3889.  Our final qualifying sample (41 targets) is plotted.  The best-fit $\tau$ tends to decrease, largely driven by an increase in the magnitude of the radiative transfer effect for \heil7065 (the line most strongly affected by radiative transfer, see Figure \ref{figure:ftau_3889_7065}). 
}
\label{figure:tau_RadTrans_Comp}
\end{figure}

This is shown in Figure \ref{figure:tau_RadTrans_Comp}, which plots the change in the best-fit $\tau$ from the \citetalias{kuri2025} equations compared to the best-fit $\tau$ from the \citetalias{benj2002} interpolation.  As $\tau$ increases, the divergence between the \citetalias{benj2002} interpolation and the \citetalias{kuri2025} fit equations tends to grow in most cases, hence, the divergence between the best-fit $\tau$ values also tends to grow.  Correspondingly, the magnitude of the decrease to the returned $\tau$ under the new \citetalias{kuri2025} radiative transfer model tends to grow proportionally to the $\tau$ value itself (i.e., a larger decrease for higher $\tau$).

\begin{figure}[t!]
\resizebox{\columnwidth}{!}{\includegraphics{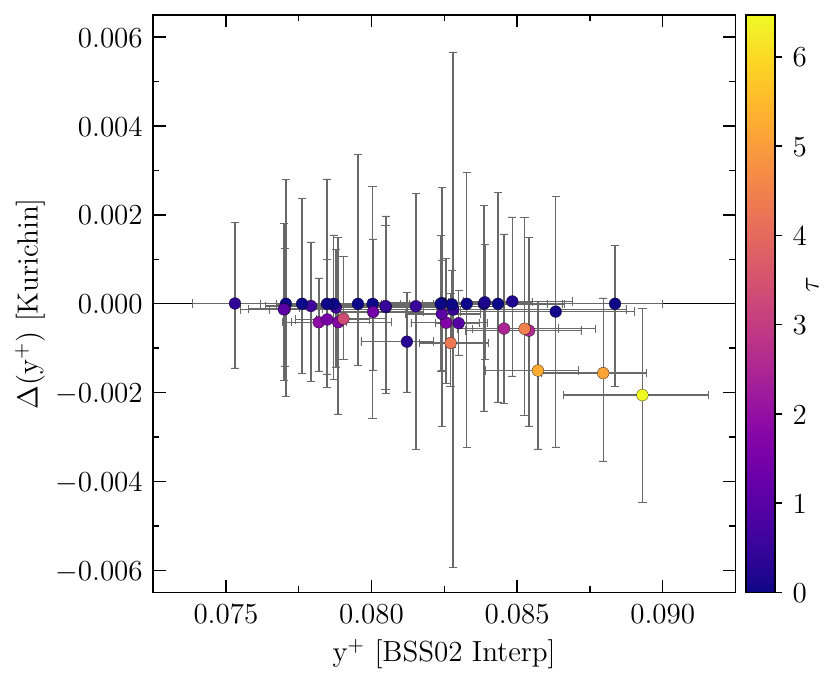}}
\caption{
The difference in the best-fit helium abundance, \y+, between the \citet{kuri2025} radiative transfer fit equations and the \citet{benj2002} radiative transfer interpolation plotted versus the interpolation \y+.  Our final qualifying sample (41 targets) is plotted, and the points are shaded by their best-fit optical depth ($\tau = \tau_{\ssn{3889}}$) from the \citetalias{benj2002} interpolation.  For targets with non-negligible optical depths, the best-fit \y+ decreases marginally, due to the decrease in the best-fit optical depth, $\tau$ (see Figure \ref{figure:tau_RadTrans_Comp}).  
}
\label{figure:y+_RadTrans_Comp}
\end{figure}

15 targets in our sample have best-fit optical depths with $\tau < 0.1$ (13 with $\tau \sim 0$) and are essentially unaffected by radiative transfer effects.  As a result, their physical parameter solutions were not significantly altered by the adoption of the new \citetalias{kuri2025} radiative transfer calculations.  For the remaining targets, the effects of the change vary depending on their previous best-fit solution.  As discussed above, the best-fit $\tau$ tended to decrease, and that decrease propagates through to the helium abundance, generally resulting in a decreased best-fit \y+.  This is shown in Figure \ref{figure:y+_RadTrans_Comp}, which compares the change in the best-fit \y+ from the \citetalias{kuri2025} equations to the best-fit $\tau$ from the \citetalias{benj2002} interpolation.  Because, for our relatively low optical depth systems, radiative transfer only has significant effects on \hei \W\W3889, 7065, 10830, the effect on \y+ is attenuated by the presence of the other helium emission lines.

\begin{figure}[t!]
\resizebox{\columnwidth}{!}{\includegraphics{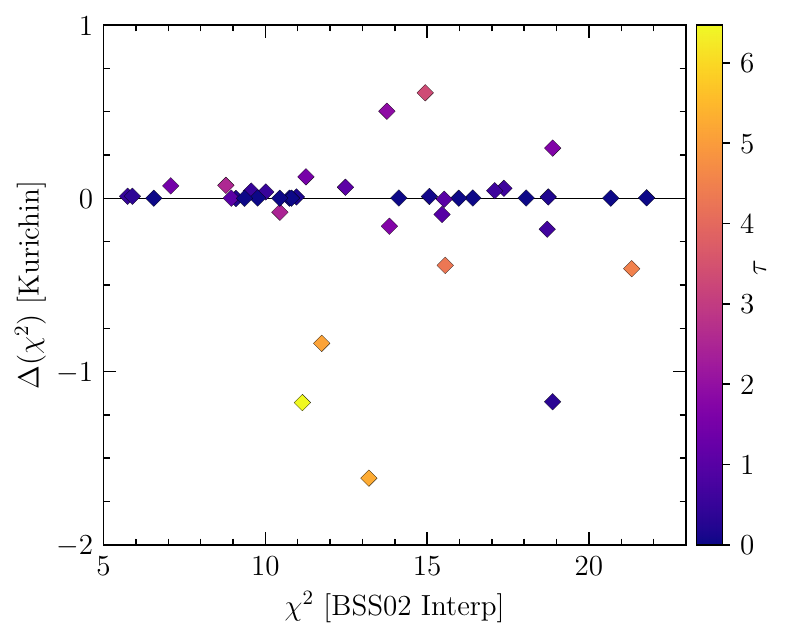}}
\caption{
The difference in the best-fit (minimized) \X2 between the \citet{kuri2025} radiative transfer fit equations and the \citet{benj2002} radiative transfer interpolation plotted versus the interpolation \X2.  Our final qualifying sample (41 targets) is plotted, and the points are shaded by their best-fit optical depth ($\tau = \tau_{\ssn{3889}}$) from the \citetalias{benj2002} interpolation.  Consistent with the similarities in the two radiative transfer calculations, most of the sample exhibits random scatter in their \D(\X2) about zero.  For five of the targets with higher values of optical depth, there are significant changes in their best-fit \X2, and they each show decreases. This provides evidence that the new \citetalias{kuri2025} calculations are an improvement upon the preceding \citetalias{benj2002} calculations, as expected.  
}
\label{figure:X2_RadTrans_Comp}
\end{figure}

To help evaluate the accuracy of the updated \citetalias{kuri2025} calculations compared to the previous \citetalias{benj2002} calculations, we compare the best-fit \X2 values returned for each.  This is shown in Figure \ref{figure:X2_RadTrans_Comp}, which plots the change in \X2 from the adopted \citetalias{kuri2025} fit equations versus the \X2 from the \citetalias{benj2002} interpolation.   If the new \citetalias{kuri2025} calculations are an improvement, one would expect better agreement between the fluxes predicted by the model and their measured values.

Overall, that is what we see.  For most of the lines and most relevant portions of parameter space, the new \citetalias{kuri2025} calculations are not substantially different from the preceding \citetalias{benj2002} calculations.  As such, it is not surprising to see that most of our sample exhibits random scatter in their \D(\X2) about zero with changes of less than $\pm 0.75$ in \X2.  However, for five of the targets with higher values of optical depth---four with $\tau>4$ and one $\tau>2$--- there are significant changes in their best-fit \X2, and they each show decreases, with three targets showing decreases greater than one unit in \X2.  Not coincidentally, the four targets with $\tau>4$ also exhibit the four largest decreases in their $\tau$ and \y+ best-fit values (see Figures \ref{figure:tau_RadTrans_Comp} \& \ref{figure:y+_RadTrans_Comp}).  

That all of the largest \X2 changes are significant decreases provides evidence that the \citetalias{kuri2025} calculations are an improvement and more accurate than the preceding \citetalias{benj2002} calculations.  This advance can be traced back to the improvements in modeling and updated atomic data in the \citetalias{kuri2025} calculations.  In particular, the significant decreases in \X2 for the higher optical depth targets provides evidence for improved reliability of the \citetalias{kuri2025} radiative transfer corrections at higher values of optical depth.  In turn, that lends support for not flagging the results that fall within the fit domain for the \citetalias{kuri2025} equations ($0 < \tau < 10$).

That all of the largest \X2 changes are significant decreases provides evidence that, as expected, the \citetalias{kuri2025} calculations, employing the latest atomic data, as well as improvements in modeling, are an improvement and more accurate than the preceding \citetalias{benj2002} calculations.


\section{Primordial Helium} \label{Yp}


In this section, we calculate the primordial helium abundance (mass fraction), \Yp.  To best determine \Yp, we have prioritized observations with the lowest metallicity, O/H.  The unprecedented number of low-metallicity, yet high quality, targets in our sample enables, for the first time, a robust weighted average determination (\S \ref{Mean}).   For comparison, we also perform a linear regression using the full qualifying sample without any flags (\S \ref{Regression}).  In Section \ref{YpComps}, we compare our results to previous determinations in the literature.  Before calculating \Yp, we first provide an overview of our sample and its characteristics in Section \ref{SampleOverview} below.

\subsection{Sample Overview} \label{SampleOverview}

\begin{figure*}[p]
\centering
\resizebox{0.9\textwidth}{!}{\includegraphics{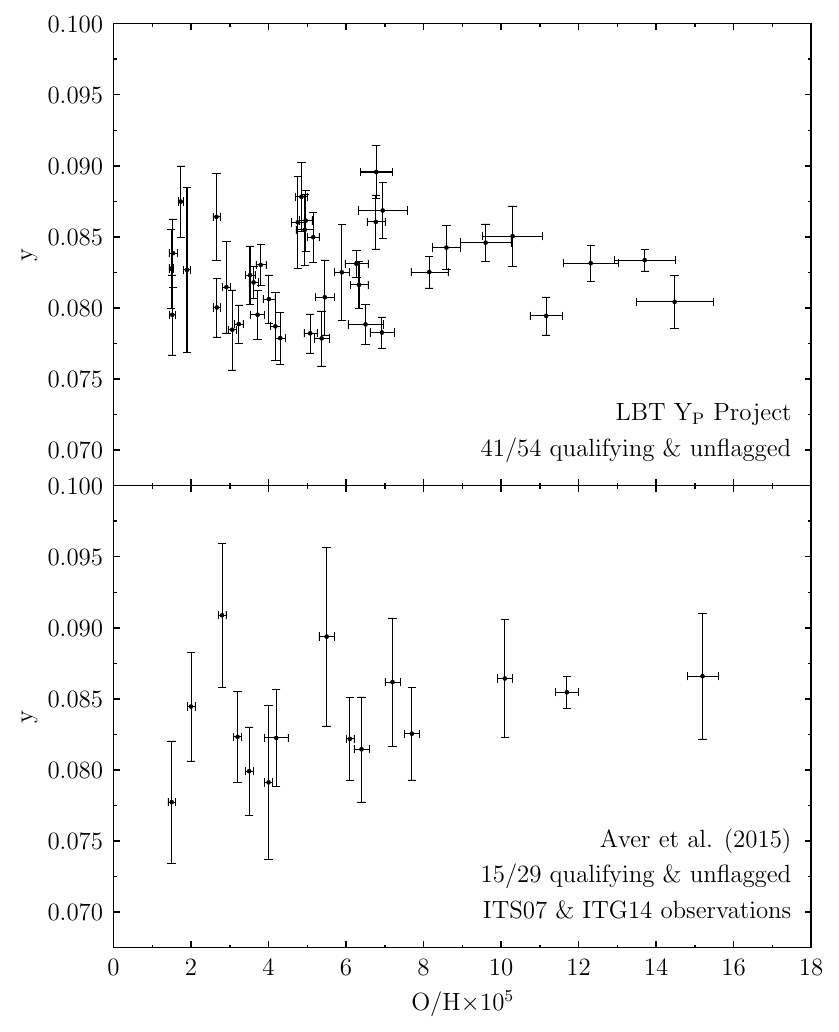}}
\caption{
Comparison of the \citet{aver2015} dataset with \Proj\ dataset presented in this work, with helium abundance (number density ratio, $\r{y=\frac{n(He)}{n(H)}}$) versus oxygen-to-hydrogen ratio (O/H) plotted.  The LBT \Yp\ Project's uniformly observed and reduced dataset exhibits reduced uncertainties, lower dispersion, and much denser sampling at low metallicities, as well as a greater sample size.  
}
\label{figure:LBT_Yp_AOS4}
\end{figure*}

From our initial dataset of 54 targets, there are 47 targets in the \X2 qualifying dataset, including 6 which are flagged (see \S \ref{Results}).  For the 41 qualifying points without any flags, Figure \ref{figure:LBT_Yp_AOS4} presents y vs.\ O/H for \Proj\ dataset, as well as the corresponding \citet{aver2015} dataset for comparison.  The values plotted in the upper panel are given in Table \ref{table:yVals}, with the O/H values taken directly from Paper II.  For targets with \heiil4686 detections, we calculate \ydp, as also reported in Table \ref{table:yVals}, to add to \y+\ to determine y.  The uncertainty for \ydp\ is determined by propagating the flux, \Te, C(H$\beta$), and $a_{H}$ uncertainties.  Neutral helium is not expected to be significant for our targets, given our low metallicity sample with hard ionization radiation fields.  This conclusion is supported by the radiation softness parameter, \citep[$\eta = \frac{{\rm O}^+/{\rm O}^{+2}}{{\rm S}^+/{\rm S}^{+2}}$,][]{VilchezPagel1988}, for our sample.  The average $\eta$ for our final qualifying dataset is 0.87, and \citet{berg2026} finds that contributions from He$^0$ become non-negligible for $\eta \gtrsim 2$.  Note that this is significantly smaller than the criterion of $\eta \gtrsim 10$ used by \citet{page1992} \citep[see discussion in][]{berg2026}.  

As discussed in Section \ref{HeComp} and as is immediately apparent in Figure \ref{figure:LBT_Yp_AOS4}, the increase in the number of targets in \Proj\ final dataset is dramatic, and, reassuringly, that increase is due in large part to a larger \X2 qualifying fraction.   As discussed in Sections \ref{chi2cut} \& \ref{Results}, 47 of 54 targets (87\%) pass the 95\% CL \X2 ``goodness-of-fit'' test, with 6 flagged, leaving 41 qualifying points without any flags (76\%).  Comparing to \citet{aver2015}, 19 of 31 targets (61\%) pass the 95\% \X2 test, with 3 of those 19 observations of the same target (SBS\s0335-052E) and 3 targets flagged (including 1 SBS\s0335-052E observation).  That left 16 qualifying observations without any flags, 15 of which were unique targets (15 of 29, 52\%).  Almost tripling from 15 to the current 41 points is a very significant advance in our dataset.

\begin{deluxetable*}{lrcccc}[pth!]
\label{table:yVals}
\centering
\tablecaption{Final Dataset Helium Abundance Values}
\tablehead{
Object  & O/H $\times$ 10$^5$    & 	$\y+=\r{He^+/H^+}$ 	      & $\ydp=\r{He^{++}/H^+}$     &   y~$=$~y$^{+}$~$+$~y$^{++}$   & Y 		   }
\startdata
\multicolumn{6}{c}{Final Dataset Not Flagged} \\
\hline
I~Zw~18~SE	&	1.49	$\pm$	0.06	&	0.08243	$\pm$	0.00278	&	0.00034	$\pm$	0.00005	&	0.08277	$\pm$	0.00278	&	0.2487	$\pm$	0.0063	\\
DDO~68	&	1.52	$\pm$	0.08	&	0.07706	$\pm$	0.00280	&	0.00246	$\pm$	0.00017	&	0.07952	$\pm$	0.00281	&	0.2412	$\pm$	0.0065	\\
Leo~P	&	1.53	$\pm$	0.11	&	0.08386	$\pm$	0.00241	&	0.00000	$\pm$	0.00000	&	0.08386	$\pm$	0.00241	&	0.2511	$\pm$	0.0054	\\
SBS~0335-052E	&	1.74	$\pm$	0.06	&	0.08470	$\pm$	0.00250	&	0.00279	$\pm$	0.00006	&	0.08748	$\pm$	0.00250	&	0.2591	$\pm$	0.0055	\\
J2104-0035	&	1.90	$\pm$	0.08	&	0.08267	$\pm$	0.00581	&	0.00000	$\pm$	0.00000	&	0.08267	$\pm$	0.00581	&	0.2484	$\pm$	0.0131	\\
LEDA~2790884	&	2.66	$\pm$	0.09	&	0.08615	$\pm$	0.00304	&	0.00026	$\pm$	0.00005	&	0.08641	$\pm$	0.00304	&	0.2567	$\pm$	0.0067	\\
J1044+0353	&	2.67	$\pm$	0.08	&	0.07844	$\pm$	0.00206	&	0.00159	$\pm$	0.00004	&	0.08003	$\pm$	0.00206	&	0.2424	$\pm$	0.0047	\\
SBS~0940+544	&	2.91	$\pm$	0.10	&	0.08146	$\pm$	0.00321	&	0.00000	$\pm$	0.00000	&	0.08146	$\pm$	0.00321	&	0.2456	$\pm$	0.0073	\\
UGC~4483	&	3.06	$\pm$	0.11	&	0.07847	$\pm$	0.00281	&	0.00000	$\pm$	0.00000	&	0.07847	$\pm$	0.00281	&	0.2387	$\pm$	0.0065	\\
SBS~1159+545	&	3.23	$\pm$	0.11	&	0.07813	$\pm$	0.00135	&	0.00074	$\pm$	0.00006	&	0.07886	$\pm$	0.00136	&	0.2397	$\pm$	0.0031	\\
J0133+1342	&	3.52	$\pm$	0.13	&	0.08042	$\pm$	0.00206	&	0.00189	$\pm$	0.00008	&	0.08231	$\pm$	0.00206	&	0.2475	$\pm$	0.0047	\\
HS~1353+4706	&	3.61	$\pm$	0.12	&	0.08036	$\pm$	0.00113	&	0.00144	$\pm$	0.00007	&	0.08180	$\pm$	0.00113	&	0.2464	$\pm$	0.0026	\\
J2213+1722	&	3.72	$\pm$	0.18	&	0.07787	$\pm$	0.00170	&	0.00166	$\pm$	0.00016	&	0.07952	$\pm$	0.00171	&	0.2411	$\pm$	0.0039	\\
HS~0122+0743	&	3.80	$\pm$	0.13	&	0.08214	$\pm$	0.00145	&	0.00089	$\pm$	0.00006	&	0.08303	$\pm$	0.00145	&	0.2491	$\pm$	0.0033	\\
SBS~1415+437	&	4.01	$\pm$	0.16	&	0.07870	$\pm$	0.00170	&	0.00193	$\pm$	0.00005	&	0.08063	$\pm$	0.00170	&	0.2437	$\pm$	0.0039	\\
SBS~1211+540	&	4.18	$\pm$	0.12	&	0.07761	$\pm$	0.00237	&	0.00109	$\pm$	0.00007	&	0.07871	$\pm$	0.00237	&	0.2392	$\pm$	0.0055	\\
WJ1327+4022	&	4.30	$\pm$	0.13	&	0.07532	$\pm$	0.00182	&	0.00256	$\pm$	0.00013	&	0.07788	$\pm$	0.00182	&	0.2373	$\pm$	0.0042	\\
HS~1442+4250	&	4.75	$\pm$	0.16	&	0.08327	$\pm$	0.00323	&	0.00275	$\pm$	0.00011	&	0.08602	$\pm$	0.00323	&	0.2558	$\pm$	0.0071	\\
J0807+3414	&	4.85	$\pm$	0.16	&	0.08725	$\pm$	0.00241	&	0.00057	$\pm$	0.00008	&	0.08782	$\pm$	0.00241	&	0.2597	$\pm$	0.0053	\\
UM~133	&	4.93	$\pm$	0.23	&	0.08434	$\pm$	0.00250	&	0.00115	$\pm$	0.00007	&	0.08549	$\pm$	0.00250	&	0.2546	$\pm$	0.0056	\\
J1323-01325	&	4.96	$\pm$	0.16	&	0.08480	$\pm$	0.00215	&	0.00134	$\pm$	0.00004	&	0.08615	$\pm$	0.00215	&	0.2560	$\pm$	0.0047	\\
J1331+4151	&	5.08	$\pm$	0.17	&	0.07690	$\pm$	0.00136	&	0.00132	$\pm$	0.00007	&	0.07822	$\pm$	0.00137	&	0.2381	$\pm$	0.0032	\\
SBS~1420+544	&	5.15	$\pm$	0.15	&	0.08421	$\pm$	0.00176	&	0.00077	$\pm$	0.00003	&	0.08498	$\pm$	0.00176	&	0.2534	$\pm$	0.0039	\\
SBS~1249+493	&	5.37	$\pm$	0.19	&	0.07686	$\pm$	0.00194	&	0.00100	$\pm$	0.00009	&	0.07786	$\pm$	0.00194	&	0.2372	$\pm$	0.0045	\\
UGC~5541	&	5.45	$\pm$	0.25	&	0.08004	$\pm$	0.00264	&	0.00071	$\pm$	0.00008	&	0.08075	$\pm$	0.00264	&	0.2439	$\pm$	0.0060	\\
SBS~1030+583	&	5.89	$\pm$	0.19	&	0.07953	$\pm$	0.00337	&	0.00298	$\pm$	0.00010	&	0.08251	$\pm$	0.00337	&	0.2478	$\pm$	0.0076	\\
SBS~1331+493	&	6.27	$\pm$	0.30	&	0.08275	$\pm$	0.00096	&	0.00037	$\pm$	0.00003	&	0.08311	$\pm$	0.00096	&	0.2492	$\pm$	0.0022	\\
UM~161	&	6.33	$\pm$	0.23	&	0.07987	$\pm$	0.00165	&	0.00175	$\pm$	0.00009	&	0.08162	$\pm$	0.00165	&	0.2458	$\pm$	0.0037	\\
UM~461	&	6.50	$\pm$	0.45	&	0.07870	$\pm$	0.00140	&	0.00015	$\pm$	0.00002	&	0.07884	$\pm$	0.00140	&	0.2395	$\pm$	0.0032	\\
WJ1136+4709	&	6.77	$\pm$	0.23	&	0.08489	$\pm$	0.00190	&	0.00116	$\pm$	0.00007	&	0.08605	$\pm$	0.00190	&	0.2557	$\pm$	0.0042	\\
KUG~1138+327	&	6.78	$\pm$	0.41	&	0.08836	$\pm$	0.00186	&	0.00120	$\pm$	0.00005	&	0.08956	$\pm$	0.00186	&	0.2634	$\pm$	0.0040	\\
HS~0134+3415	&	6.93	$\pm$	0.31	&	0.07777	$\pm$	0.00109	&	0.00050	$\pm$	0.00002	&	0.07827	$\pm$	0.00109	&	0.2381	$\pm$	0.0025	\\
WJ2310-0211	&	6.95	$\pm$	0.63	&	0.08640	$\pm$	0.00198	&	0.00046	$\pm$	0.00002	&	0.08686	$\pm$	0.00198	&	0.2575	$\pm$	0.0044	\\
SBS~1152+579	&	8.15	$\pm$	0.48	&	0.08183	$\pm$	0.00111	&	0.00069	$\pm$	0.00002	&	0.08253	$\pm$	0.00111	&	0.2478	$\pm$	0.0025	\\
SBS~1437+370	&	8.59	$\pm$	0.37	&	0.08237	$\pm$	0.00155	&	0.00187	$\pm$	0.00007	&	0.08425	$\pm$	0.00155	&	0.2516	$\pm$	0.0035	\\
SBS~0946+558	&	9.60	$\pm$	0.65	&	0.08392	$\pm$	0.00131	&	0.00067	$\pm$	0.00003	&	0.08459	$\pm$	0.00131	&	0.2523	$\pm$	0.0029	\\
J1148+2546	&	10.30	$\pm$	0.77	&	0.08399	$\pm$	0.00214	&	0.00105	$\pm$	0.00006	&	0.08504	$\pm$	0.00214	&	0.2533	$\pm$	0.0047	\\
HS~0029+1748	&	11.17	$\pm$	0.42	&	0.07869	$\pm$	0.00133	&	0.00075	$\pm$	0.00004	&	0.07944	$\pm$	0.00134	&	0.2406	$\pm$	0.0031	\\
SBS~0948+532	&	12.31	$\pm$	0.71	&	0.08219	$\pm$	0.00127	&	0.00096	$\pm$	0.00004	&	0.08315	$\pm$	0.00127	&	0.2490	$\pm$	0.0029	\\
SBS~1135+581	&	13.71	$\pm$	0.79	&	0.08256	$\pm$	0.00074	&	0.00080	$\pm$	0.00004	&	0.08336	$\pm$	0.00075	&	0.2494	$\pm$	0.0017	\\
SBS~0926+606	&	14.48	$\pm$	0.99	&	0.08043	$\pm$	0.00188	&	0.00000	$\pm$	0.00000	&	0.08043	$\pm$	0.00188	&	0.2427	$\pm$	0.0043	\\
\hline
\multicolumn{6}{c}{Final Dataset with Flags} \\
\hline
AGC~198691	&	1.00	$\pm$	0.12	&	0.08076	$\pm$	0.00572	&	0.00000	$\pm$	0.00000	&	0.08076	$\pm$	0.00572	&	0.2441	$\pm$	0.0131	\\
J0118+3512	&	3.08	$\pm$	0.20	&	0.07520	$\pm$	0.00223	&	0.00059	$\pm$	0.00020	&	0.07579	$\pm$	0.00224	&	0.2325	$\pm$	0.0053	\\
UGC~6456	&	4.45	$\pm$	0.21	&	0.08375	$\pm$	0.00281	&	0.00000	$\pm$	0.00000	&	0.08375	$\pm$	0.00281	&	0.2507	$\pm$	0.0063	\\
HS~1028+3843	&	6.38	$\pm$	0.21	&	0.08786	$\pm$	0.00118	&	0.00106	$\pm$	0.00004	&	0.08892	$\pm$	0.00118	&	0.2620	$\pm$	0.0026	\\
HS~1222+3741	&	6.78	$\pm$	0.28	&	0.09171	$\pm$	0.00396	&	0.00179	$\pm$	0.00012	&	0.09349	$\pm$	0.00396	&	0.2718	$\pm$	0.0084	\\
Mrk~71	&	7.33	$\pm$	0.24	&	0.08369	$\pm$	0.00108	&	0.00027	$\pm$	0.00001	&	0.08397	$\pm$	0.00108	&	0.2511	$\pm$	0.0024	\\
\enddata
\end{deluxetable*}


The dramatic improvement in \Proj's dataset is also demonstrated by the significant reduction in the target uncertainties, which is also clearly evident in Figure \ref{figure:LBT_Yp_AOS4}.  Compared to \citet{aver2015}, the average uncertainty for y decreased by 47\%.  

As Figure \ref{figure:LBT_Yp_AOS4} shows, our target coverage is greater for lower metallicities, reflecting our target selection priorities.  That low-metallicity target coverage vastly exceeds \citet{aver2015}, with 33 targets compared to 12 for \OHe.  Restricting to even lower metallicities, \OHc, \Proj\ dataset still retains an impressive 15 targets, while \citet{aver2015} contained only 6.  

As seen in Figure \ref{figure:LBT_Yp_AOS4}, for \OHe, there appears to be a degree of intrinsic scatter in our dataset, which is not uncommon with astronomical observations.  The sampling for \OHge\ is limited.  With relatively sparse sampling, the degree of intrinsic scatter in that higher metallicity region may not be fully captured by our dataset.  Additionally, the still relatively large dispersion compared to the relatively sparse sampling limits our ability to constrain the rate of chemical evolution, \Dy.  This provides further support for preferring a weighted mean determination for \Yp\ (See \S \ref{Mean} below).  

As noted above, at low metallicities (\OHe), where the bulk of our targets are, there is evidence for significant intrinsic dispersion.  This can arise from at least two different causes.  First, if we have underestimated our uncertainties, this could result in the appearance of an intrinsic scatter.  Alternatively, the intrinsic scatter could be produced by the galaxies following slightly different paths in chemical evolution.  We cannot rule out a small underestimate of our uncertainties, but it is worth discussing what might lead to a true intrinsic dispersion in y at a given value of O/H.  

Due to the correlation of metallicity with stellar mass \citep[e.g.,][]{berg2012}, at the lowest values of O/H, the targets are all relatively low mass galaxies.  At these low masses, chemical enrichment is extremely inefficient \citep[e.g., $\sim$5\% of oxygen is retained in Leo\,P,][]{McQuinn2015b}, and thus may be subject to stochastic effects. The nucleosynthetic production of helium and oxygen are expected to have different timescales.  Helium production is thought to be significantly driven by AGB stars in the mass range of 4 -- 8 M$_{\odot}$, while oxygen production through supernovae is dominated by stars more massive than 8 M$_{\odot}$ \citep[e.g.,][, and their Figure 4 in particular]{Weller2025}.  Thus, the evolution of a single galaxy in the y-O/H diagnostic diagram could take a zigzagged path, as was originally envisioned for nitrogen based on a similar timescale argument \citep{Garnett1990}.

Additionally, many of the lowest metallicity star forming galaxies are thought to have exceptionally low metallicities due to the recent infall of relatively pristine gas, as proposed by \citet{Ekta2010} and discussed by \citet{mcQu2020}.  Given the above, it is perhaps not surprising to detect an intrinsic dispersion in the values of y at the lowest values of O/H.  For the purpose of determining \Yp, the dispersion has the effect that the presumed slope in y versus O/H is more difficult to detect or constrain at the lowest values of O/H.

\subsection{Weighted Average for Y$_p$} \label{Mean}

The goal of our project is to determine \Yp, and determining \Yp\ does not require determining the slope, \Dy\ (or \DY), as long as one can sufficiently limit the metallicity baseline, as is discussed further below.  Correspondingly, as discussed in Paper I and Section \ref{Sample}, we prioritized the lowest metallicity targets over a more even coverage across our metallicity range.  Given its dispersion, our sample does not contain sufficient higher metallicity observations to determine the slope, \Dy, with confidence, as is evident in Figure \ref{figure:LBT_Yp_AOS4} and will be further explored in Section \ref{Regression}.  Indeed, from Figure \ref{figure:LBT_Yp_AOS4}, it is evident that our dataset does not exhibit a significant degree of enrichment for low metallicities. However, enabled by our observing strategy, the unprecedented number of low metallicity, yet high quality optical and NIR observations in our sample allows for a robust weighted average determination of \Yp\ with unprecedented precision.  

\begin{figure}[t!]
\resizebox{\columnwidth}{!}{\includegraphics{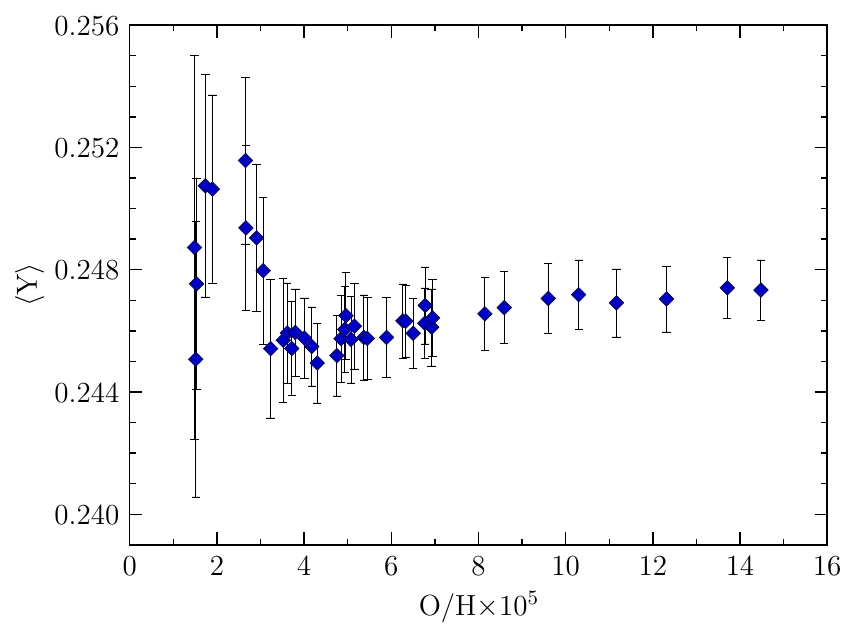}}
\caption{
The weighted average value for the helium abundance mass fraction, \Yavg, versus increasing baseline cutoff values for the oxygen-to-hydrogen ratio, O/H, used for calculating the weighted average.  The series of weighted average values is calculated from our final qualifying sample (41 targets), and that weighted average is calculated for y and then converted to Y for easier comparison.  Though \Yavg\ varies and there is some increase at higher metallicities, the weighted average does not show any significant systematic increase as the metallicity baseline increases, especially within the O/H = $2.3 - 7 \times 10^{-5}$ band and within uncertainties.  Correspondingly, our dataset does not show any clear evidence for the presence of a significant chemical enrichment slope in our dataset.  
}
\label{figure:Yavg_OH}
\end{figure}

Over an extended range in metallicity, we expect the helium abundance to increase with metallicity.  As a result, the mean of the helium abundance, \Yavg, with O/H $< x$, would be expected to increase with $x$.  However, in our data, no clear systematic increase in \Yavg\ with O/H is observed.  This is demonstrated in Figure \ref{figure:Yavg_OH}, which plots the weighted mean, \Yavg, calculated for progressively increased O/H cutoffs.  This limits the bias in \Yp\ that one would introduce due to a positive slope from chemical evolution.  For example, if one assumes a slope of $\DY = 46$, as determined by the relationship for higher metallicity \citep{kuri2021}, the increase in \Yp\ introduced by the weighted average would be $\sim$0.0009 (i.e., the midpoint enrichment value of Y over a domain of \OHc), which is significantly less than the uncertainty we find in our result below.

As Figure \ref{figure:Yavg_OH} shows, the running weighted average, \Yavg, for increasing O/H baselines, is relatively constant once a suitable number of points is included.  Indeed, within the uncertainties, there is little change in the mean for a cutoff of O/H $< 3.3 \times 10^{-5}$ (10 points) and that for \OHe\ (33 points).  This makes the result relatively insensitive to the choice of cutoff, which is reassuring.  In addition, there is little evidence for the introduction of any significant bias from the weighted average. 

To determine \Yp, we first calculate \yavg, the weighted average of the helium abundance (number density ratio) for our \OHc\ subsample, to determine \yp, the primordial helium abundance.  Then, we convert \yp\ to \Yp, the primordial helium abundance mass fraction, for easier comparison with the CMB prediction and astrophysical determinations from the literature.

To ensure that our result for \Yp\ is not subject to the effects of significant chemical evolution, we restrict our metallicity range to \OHc.  This limits the bias in \Yp\ that one would introduce due to a positive slope from chemical evolution (if present).  The weighted average of $y$ for those 15 qualifying targets without any flags and \OHc\ yields, 
\beq
\yp = 0.08146 \pm 0.00058,
\label{eq:yp_avg}
\eeq
The primordial helium abundance, \yp, may be converted to the corresponding mass fraction, \Yp, via,  
\beq
\Yp = \frac{4 \yp}{1 + 4 \yp}, 
\eeq
which yields,
\beq
\Yp = 0.2458 \pm 0.0013.
\label{eq:Yp_avg}
\eeq

The uncertainty in Eqs.\:(\ref{eq:yp_avg}) \& (\ref{eq:Yp_avg}) includes a S-factor correction for dispersion.  This correction and our specific procedure for calculating the mean and uncertainty is described in detail in Appendix \ref{Sfact}.  These results carry an S-factor, $S=1.15$, and, while this correction is non-negligible, it also indicates that there is not a substantial amount of dispersion in this data subset.  

Eq.\:(\ref{eq:Yp_avg}) is our primary result for the primordial helium abundance (mass fraction).  Our \Yp\ determination achieves a relative uncertainty of 0.5\%.  This is an unprecedented precision and represents the realization of \Proj's primary goal. 

\begin{figure}[t!]
\resizebox{\columnwidth}{!}{\includegraphics{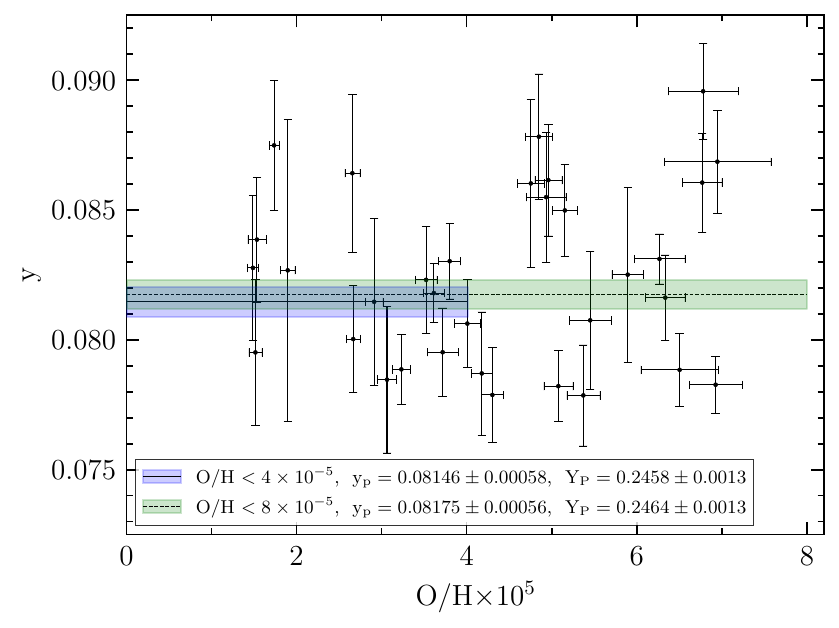}}
\caption{
The helium abundance (number density ratio, $\r{y=\frac{n(He)}{n(H)}}$) for the 33 targets from our final qualifying dataset with \OHe, plotted versus the oxygen-to-hydrogen ratio (O/H). The horizontal shaded bands show the weighted average, \yavg,  with $\pm 1 \sigma$ uncertainties.  The weighted average for the set of 15 targets with \OHc\ is shown by the darker (blue) shaded band and the weighted average for the set of 33 targets with \OHe\ is shown by the light (green) shaded band.  Despite the doubling of the metallicity baseline (and more than doubling of the number of points, from 15 to 33), the weighted average results are very similar, demonstrating the robustness of the weighted average value with respect to sample cutoff.  For each set, the weighted average results are included (\yp), as well as the corresponding \Yp\ values. 
}
\label{figure:y_avg_OH_8}
\end{figure}

The insensitivity of our result to O/H cutoff is underscored by Figure \ref{figure:y_avg_OH_8}, which plots y vs.\ O/H for our dataset with \OHe, overlaid with the weighted average results for \OHc\ and \OHe.  Despite doubling the metallicity baseline from \OHc\ to \OHe, resulting in more than double the number of points (33 vs.\ 15), the weighted average result does not change significantly:  $\yp = 0.08175 \pm 0.00056$ and $\Yp = 0.2464 \pm 0.0013$.  The uncertainty here is the same as in Eq.\:(\ref{eq:Yp_avg}), despite more than twice as many points being averaged.  This is due to the increased dispersion (easily seen in Figure \ref{figure:y_avg_OH_8}), and the uncertainty carries a more significant correction of $S = 1.76$.  To be conservative, we restrict to \OHc\ for our weighted average calculation.  One could argue for an even more restrictive baseline, but that would cut against the need to have a significant number of points for a robust weighted average determination.

\subsection{Linear Regression to Y$_p$} \label{Regression}

Although our primary result for \Yp\ is given in Eq.\:(\ref{eq:Yp_avg}), for completeness and comparison, we also perform a linear regression to determine \Yp.  We calculate a regression of $y$, versus O/H, the oxygen abundance, from the full sample of qualifying targets without any flags, to extrapolate to the primordial value, \yp. From \yp, we then calculate \Yp, the primordial helium abundance mass fraction.  The regression using $y$, rather than $Y$, the helium abundance mass fraction, removes the dependence on $Z$ (through $\r{Y}=\frac{4y(1-Z)}{1+4y}$) in determining \Yp, as advocated by \citet{hsyu2020}.  

\begin{figure}[t!]
\resizebox{\columnwidth}{!}{\includegraphics{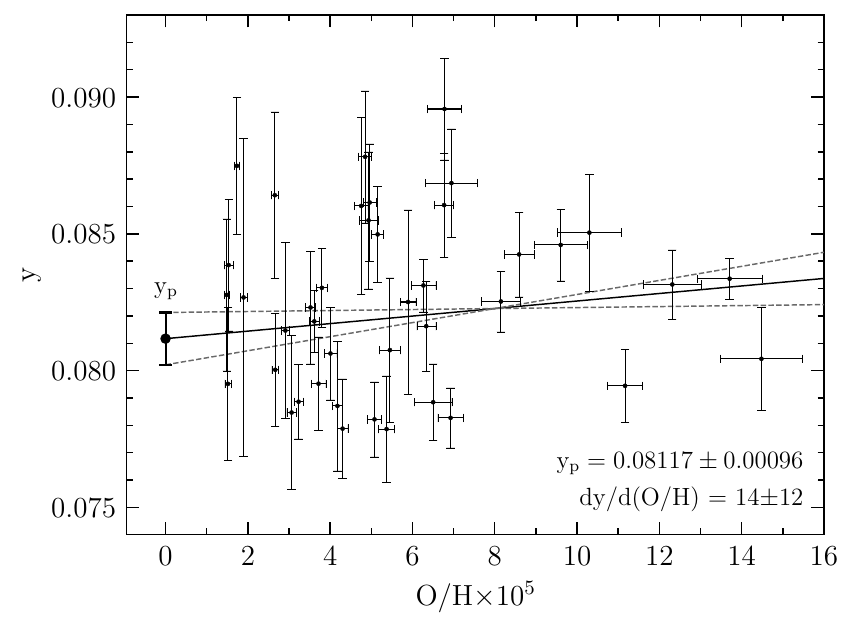}}
\caption{
A linear regression of the Helium abundance (number density ratio, $\r{y=\frac{n(He)}{n(H)}}$) versus oxygen-to-hydrogen ratio (O/H) for the complete final qualifying sample (41 points).  The fit line slope, \Dy, along with its $\pm 1 \sigma$ deviations, and intercept, \yp, are given.  
}
\label{figure:y_OH_Reg}
\end{figure}

Our method for regression to \yp\ is briefly described in Appendix \ref{Sfact}.  We use a weighted least squares linear regression with uncertainties on both variables (y \& O/H).  Regression based on the 41 targets in our final qualifying dataset yields,
\beq
\yp = 0.08117 \pm 0.00096,
\label{eq:yp_reg}
\eeq
with a slope $\Dy = 14 \pm 12$. 
This corresponds to a primordial helium mass fraction of
\beq
\Yp = 0.2451 \pm 0.0022.
\label{eq:Yp_reg}
\eeq
The quoted uncertainties include an S-factor, as described in Appendix \ref{Sfact}, which in this case is $S=1.70$. The result of the linear regression is shown in Figure \ref{figure:y_OH_Reg}.

While Eq.\:(\ref{eq:Yp_reg}) agrees well with our preferred \Yp\ result, the weighted average, as given in Eq.\:\ref{eq:Yp_avg}, the data do not allow for an unambiguous determination of a slope, \Dy. The determination of the slope is almost entirely reliant on the sparse points with \OHge, as can been seen in Figure \ref{figure:y_OH_Reg}.  With only 8 of 41 points in the upper half of regression's metallicity baseline, our ability to reliably extract the slope, \Dy, is severely limited.  This point is reinforced by the dispersion evident in those 8 points.  Indeed, without the points with \OHge, a determination of the slope of the 33 lowest metallicity targets yields $\Dy = 15 \pm 36$ ($S= 1.79$), clearly indicating that taking the weighted mean is preferable for the lower metallicity subset of the data. 

Given the indications of intrinsic scatter in our dataset, one can further investigate using the \textsc{python linmix} package\footnote{\url{https://github.com/jmeyers314/linmix}}, which is briefly described in Appendix \ref{Sfact}.  Using \textsc{linmix} for the finaly qualifying dataset (41 points) returns $\yp = 0.08184 \pm 0.00110$, with a slope $\Dy = 8 \pm 16$, and intrinsic scatter $\sigma_{int} = 0.00262 \pm 0.00048$. Once again we see that a non-zero slope can not be determined with any confidence. The returned intrinsic scatter value indicates significant intrinsic scatter in the dataset.  However, it is not clear if the intrinsic random scatter in our dataset is well-characterized by a single parameter across the dataset, or whether it may vary with metallicity.  

The \textsc{linmix} \yp\ translates to $\Yp = 0.2466 \pm 0.0025$.  This result agrees well with Eq.\:(\ref{eq:Yp_reg}), though the difference illustrates the effect of fitting for intrinsic scatter, or not, on the resulting slope and intercept.  The weighted least squares regression returns a noticeably different slope and intercept than the Bayesian MCMC \textsc{linmix} result.  This difference in result, depending on regression algorithm, further cautions against the reliability of using linear regression to determine \Yp\ for our dataset.

\subsection{Y$_p$ Comparisons} \label{YpComps}

The weighted average (Eq.\:(\ref{eq:Yp_avg})) for \Yp\ agrees well with the  standard BBN (SBBN) value, $\Yp = 0.2467 \pm 0.0002$ \citep{yeh2022,yeh2023}, based on the Planck determined baryon density (\citetalias{Planck2018} \citeyear{Planck2018}).  Eq.\:(\ref{eq:Yp_avg}) also agrees well with the SBBN-independent, direct Planck estimation of $\Yp = 0.239 \pm 0.013$ (\citetalias{Planck2018} \citeyear{Planck2018}), using temperature, polarization, and lensing data, which is not surprising given the relatively large uncertainty on this value from the CMB alone. 

\citet{aver2015}, our last result prior to \Proj\ (See \S \ref{Results}), determined $\Yp = 0.2449 \pm 0.0040$ with a slope of $\DY = 79 \pm 43$.  Our \Yp\ results here, both from the weighted average and linear regression, are in very good agreement with the \citet{aver2015} result.  The more striking comparison is for the uncertainty.  For the linear regression, both results are based on the qualifying sample excluding flagged targets, and the model and analysis methodology employed are quite similar.  However, the fruits of this project are evident in the resulting uncertainty.  Compared to \citet{aver2015}, the uncertainty on \Yp\ is reduced by 45\% based on linear regression (Eq.\:(\ref{eq:Yp_reg})), to achieve a relative uncertainty of $< 1\%$.  

The regression dataset in \citet{aver2015} only included 15 targets, compared to 41 targets in this work.  Furthermore, the average uncertainty for its galactic helium abundance determinations was larger (see \S \ref{HeComp} \& \S \ref{SampleOverview}).  The combination of decreased uncertainties for the sample points and the increased number of sample points is the primary source of the decreased uncertainty for \Yp.  

However, that rather pedestrian statement masks the requirements for achieving those decreased target uncertainties.  As discussed in Section \ref{Results}, it rests on high SNR, high resolution, and broad wavelength coverage observations, uniformly observed and reduced, and then combined with an expanded physical model.  Figure \ref{figure:LBT_Yp_AOS4} (introduced above in \S \ref{SampleOverview}) compares the \citet{aver2015} dataset with \Proj\ dataset presented in this work.  As discussed in \S \ref{SampleOverview}, Figure \ref{figure:LBT_Yp_AOS4} demonstrates the greater number of targets, reduced uncertainties, lower dispersion, and much denser sampling at low metallicities.  These advances are the foundation of \Proj's significantly higher precision \Yp\ determination.  

Though \citet{aver2022} includes the first two targets from \Proj, Leo\,P and AGC\s198691, comparing to those results is still of interest, since they are our most recent.  Based on 17 targets, \citet{aver2022} determined $\Yp = 0.2448 \pm 0.0033$ with a slope of $\DY = 80 \pm 38$.  Since 15 of the 17 points were from \citet{aver2015}, it is not surprising that the intercept value was largely unchanged from \citet{aver2015}.  However, the inclusion of Leo\,P, with its extremely low metallicity and reduced uncertainty for Y, helped to better constrain the intercept.  Our \Yp\ results here are again in good agreement with the \citet{aver2022} determination, but with the linear regression \Yp\ uncertainty decreased by 33\%.  

The datasets in \citet{aver2015} and \citet{aver2022} did not allow for a reliable weighted average determination based on a low-metallicity subset, due to the paucity of very low metallicity targets in its dataset (even with Leo\,P and AGC\s198691 added).  Comparing our weighted average result here (Eq.\:(\ref{eq:Yp_avg})) to the \citet{aver2015} result for \Yp, the uncertainty shrinks by more than half, decreasing by 67\%.  Comparing to \Yp\ from \citet{aver2022}, the uncertainty for our weighted average \Yp\ determination is reduced by 60\%.  

As illustrated by the comparisons above, \Proj's weighted average determination is a significant advance in primordial helium abundance determinations, achieving a relative uncertainty of $0.5\%$.  Furthermore, it highlights the value and diagnostic power provided by assembling a dataset with many very low metallicity targets.  

\begin{figure}[t!]
\resizebox{\columnwidth}{!}{\includegraphics{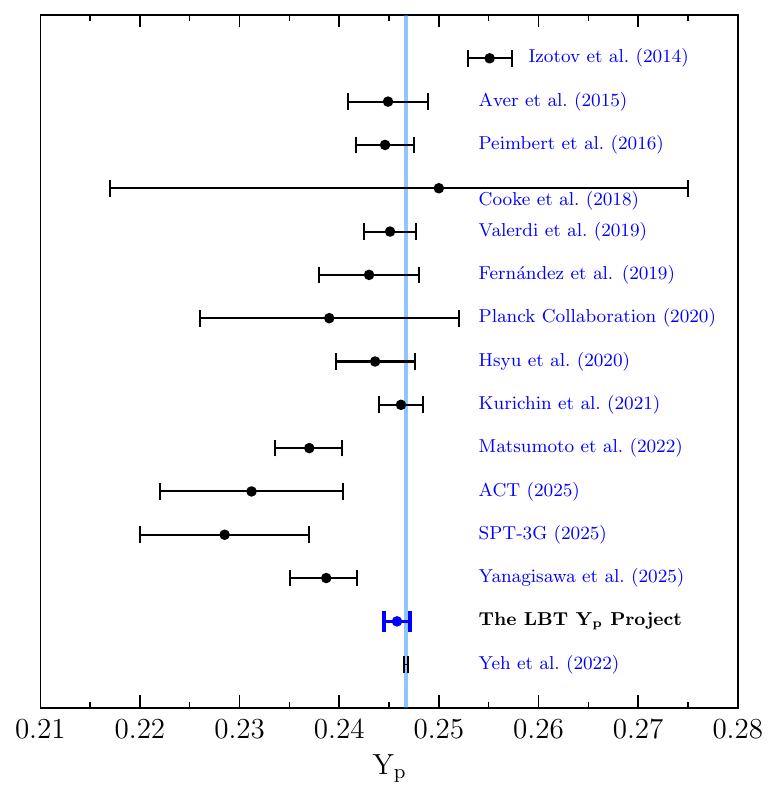}}
\caption{
A comparison of recent \Yp\ results.  The SBBN result based on the Planck determined baryon density (\citetalias{Planck2018} \citeyear{Planck2018}) is plotted as a vertical band in light blue \citep{yeh2022,yeh2023}.  The uncertainty for the \Proj\ result is reduced, yet it remains in good agreement with most of the recent results.  
}
\label{figure:Yp_Comp}
\end{figure}

A comparison of our result with other recent determinations is found in Table \ref{table:YpComp} and plotted in Fig.~\ref{figure:Yp_Comp}.  It is encouraging that almost all of the most recent results agree.  The only one not in agreement with all of the others within the 2-$\sigma$ level, including with our results presented here, is \citet{izot2014}\footnote{\citet{kuri2021} re-analyze \citet{izot2014} and find a value in closer agreement with the other determinations.}. \citet{aver2015} includes a detailed description of the differences in the methodology and dataset cuts between \citet{izot2014} and our model and approach. The recent CMB-based determinations from the \citetalias{Atacama2025} (\citeyear{Atacama2025}) and the \citetalias{SPT3G2025} (\citeyear{SPT3G2025}) show notably lower values for \Yp, but, due to their larger uncertainties, they still agree with the other more recent results, including our results here, within the 2-$\sigma$ level.  For \hii region based determinations, apart from the aforementioned \citet{izot2014}, they all agree with each other and with our results presented in this work within the 1-$\sigma$ level, except for \citet{mats2022} \& \citet{yana2025}, but those results do agree with the others and with our results here at 2-$\sigma$.


\begin{deluxetable*}{llrr}[ht!]
\label{table:YpComp}
\centering
\tablecaption{Recent Primordial Helium Abundance Results} 
\tablehead{Citation & Y$_\textrm{P}$ & N & Method }
\startdata
\citet{izot2014} &   0.2551 $\pm$ 0.0022     &   28      &   H~II Region     \\
\citet{aver2015} &   0.2449 $\pm$ 0.0040     &   15      &   H~II Region     \\
\citet{peim2016} &   0.2446 $\pm$ 0.0029     &   5       &   H~II Region     \\
\citet{cook2018b} &   0.250$^{+0.033}_{-0.025}$   &   1   &   Absorption Line     \\
\citet{vale2019} &   0.2451 $\pm$ 0.0026     &   1       &   H~II Region     \\
\citet{fern2019} &   0.243 $\pm$ 0.005       &   16      &   H~II Region     \\
\citetalias{Planck2018} (\citeyear{Planck2018}) &   0.239 $\pm$ 0.013   & -  &   CMB     \\
\citet{hsyu2020} &   0.2436$^{+0.0039}_{-0.0040}$    &   54  &   H~II Region     \\
\citet{kuri2021} &   0.2462 $\pm$ 0.0022     &   120      &   H~II Region     \\
\citet{mats2022} &   0.2370$^{+0.0034}_{-0.0033}$       &   59      &   H~II Region     \\
\citetalias{Atacama2025} (\citeyear{Atacama2025}) &   0.2312 $\pm$ 0.0092   & -  &   CMB     \\
\citetalias{SPT3G2025} (\citeyear{SPT3G2025}) &   0.2285 $\pm$ 0.0085   & -  &   CMB     \\
\citet{yana2025} &   0.2387$^{+0.0036}_{-0.0031}$    &   68  &   H~II Region     \\
The LBT \Yp\ Project    &   0.2458 $\pm$ 0.0013     &   15      &   H~II Region     \\
\hline
\citet{yeh2022} &   0.2467 $\pm$ 0.0002 &           &   SBBN + CMB            \\
\enddata
\end{deluxetable*}


\section{Future Investigations} \label{Future}


Two avenues for future investigation and potentially improving our \hii region modeling are directly motivated by our results above:  our temperature model and the "Case B" assumption for the He singlet emission lines.  Each could potentially introduce a systematic error in our analysis, and further investigation is warranted.  However, those explorations and any potential wholesale updates to our current temperature or radiative transfer models is beyond the scope of this work.  The subsections below discuss the evidence for potential deficiencies in our current temperature model (\S \ref{TempFut}), as well as expanding upon the previous discussion (\S \ref{Eduardo}) of the discrepant He singlet emission lines, \hei \W5016 \& \W7281 (\S \ref{EduardoFut}).  

\subsection{Temperature Model} \label{TempFut}

As discussed in Section \ref{Tcomp}, T(He), a fit parameter in our model, as determined by the He emission lines, would be expected to closely track T(\oiii), as determined by the [\oiii] emission lines.  However, due to potential temperature inhomogeneities, T(\oiii) could be biased upward compared to T(He) \citep{peim2000, peim2002, oliv2004, aver2011, mend2025}.  This is the primary motivation for including T(He) as a model parameter to fit from the He emission lines themselves, rather than adopting T(\oiii) as the appropriate \Te\ for calculating the He \& H emissivities and He radiative transfer \citep{peim2000,oliv2004}.  
However, as shown in Figure \ref{figure:THe_TOIII}, our results do not demonstrate a systematic discrepancy between T(\oiii) and T(He).  There is some evidence for a systematically higher value for T(\oiii) than T(He) for the higher temperature qualifying points in our sample, but it is not definitive.  

Therefore, the results from our dataset suggest further investigation of the temperature structure and relationships for the \hii regions in our sample.  That work will build upon the detailed temperature analysis in Paper II and focus on improving our temperature model for the He emission.  It is possible that T(\oiii) is not a biased measurement for T(He), and if so, the much higher precision of T(\oiii) offers an avenue for decreasing the uncertainty on the helium abundance.  In addition, previous studies have advocated for the importance of temperature structure and potential inhomogeneities in \hii regions \citep{peim1967,peim2000, peim2002}, and further investigation of these effects could lead to potential improvements in our model.

\subsection{He\,{\small \textit{I}} \W5016 \& \W7281 Testing} \label{EduardoFut}

Section \ref{Eduardo} discussed the systematic discrepancies in \hei\W5016 \& \W7281.  This was first reported by \citet{izot2014}, and \citet{mend2025} suggest deviations from ``Case B'' due to photon loss and/or temperature inhomogeneities as plausible causes.  Further examination of the effects of potential temperature inhomogeneities (or lack thereof) is already motivated by the comparison of the T(He) and T(\oiii) for our sample, as discussed above (\S \ref{TempFut}).  If the observed \hei\W5016 \& \W7281 fluxes are systematically decreased due to photon loss caused by \hi absorption or escape from the \hii region, that would have consequences for all of the He singlet lines, including \hei\W\W4922, 6678, \& 4388, though to a significantly reduced degree.  We did not observe the same systematically low fluxes for those three He lines (compared to the predicted fluxes), as we did for \hei\W5016 \& \W7281.  However, given the potential for the systematic error, further study is warranted.  \citet{mend2025} create a toy model which combines ``Case A'' and ``Case B'' emission to test the hypothesis of deviation from ``Case B'' due to photon loss.  For the \citet{mend2025} sample of \hii regions, though not for star-forming galaxies or planetary nebulae, this toy model's predictions are consistent with the observed discrepancy.  Expanded investigation of the assumption of ``Case B'' for \hii region modeling building off the ``Case A + B'' toy model may be a natural place to start.


\section{Conclusions} \label{Conclusion}


The primary goal of \Proj\ is to determine the primordial helium abundance with the highest precision yet achieved.  Through a carefully designed observing program leveraging the high resolution, high SNR, and broad wavelength coverage offered by the LBT, we have achieved that goal, determining \Yp\ with a precision of 0.5\%.  Our dataset is the largest high quality optical and NIR sample for helium abundance analysis yet assembled.  Our dataset shows significantly improved statistical reliability compared to previous datasets, and we have taken care to flag targets for potential systematic errors.  Along the way, we have updated our model to employ the latest atomic data, as well as incorporating refinements to our methodology.  Here is a brief summary of the major advances and results presented in this paper:  

\begin{enumerate}
    \item The latest helium emissivity calculations from \citet{delz2022} have been incorporated, using a higher-resolution temperature and density grid.  Similarly, the latest hydrogen emissivity calculations from \citet{stor2015} have been incorporated using a higher-resolution temperature and density grid.  The H emissivities are almost identical in the low-density regime relevant for this work, and the new He emissivities are broadly similar to those that preceded them.  This consistency provides greater confidence in the results from our model.   
    \item Our radiative transfer model has been updated with the (very) recent radiative transfer calculations by \citet{kuri2025}.  These new calculations are based on the latest atomic data and significant modeling improvements.  Demonstrating the improvements in these new radiative transfer calculations, our results show better agreement between our model and the observations, especially for higher optical depth systems.   
    \item Our model was expanded to incorporate three additional Paschen series lines (P13, P14, and P15) for greater diagnostic power and flexibility.  We improved our treatment of the blended emission H8 + \3889 line, and, due to the exponential temperature dependence for collisional excitation of neutral hydrogen, we excluded the neutral hydrogen fraction as a fit parameter for temperatures below 14,000\,K, where collisional enhancement is negligible.  Based on the recent findings of \citet{mend2025}, we excluded \hei \W5016 \& \W7281 from our analysis, due to systematic discrepancies in their observed fluxes compared to standard ``Case B'' modeling and assumed temperature homogeneity.  
    \item Many previous helium abundance analyses exhibited the disturbing result that a large fraction, typically the majority, of the observations failed a 95\% CL \X2 test.  From our starting dataset of 54 observations, 47 passed a 95\% CL \X2 test, equating to 87\% of the dataset.  This is a very significant improvement upon previous datasets.  Furthermore, of the 7 targets excluded by the \X2 test, the majority exhibited behavior in their spectra which would make them unsuitable, such that their failure was not surprising.  This dramatic improvement in the statistical reliability of our dataset can be directly attributed to our uniformly observed and reduced spectra combined with our high fidelity flux measurements and error model.  Our sample's much higher qualifying fraction bolsters our confidence in our applied model and thus the results.  
    \item We examined our dataset for potential systematic biases and conservatively flagged targets where there were concerns they may introduce systematic error into our results.  6 targets in total were flagged, 1 for optical depth, 2 for underlying helium absorption, and 3 for the divergence between their helium and [\oiii] temperatures.  
    \item Following directly from the quality of our observations and measurements, we were able to deploy our full model employing up to 24 lines, with the effect that the average uncertainty on the helium abundance (y) for the \Proj\ dataset decreased by 47\% compared to our previous dataset and analysis.  
    \item We specifically targeted the lowest metallicity \hii regions with sufficient signal strength to produce high SNR spectra.  As a result, our final dataset for determining \Yp\ is a subset of the total sample, consisting of the 15 targets with the lowest metallicities, \OHc.  This singular advance allows us to determine \Yp\ using a weighted average, thus removing the assumption of linear enrichment.  We note that, due to our reduced uncertainties on the individual data points, we detect a significant dispersion in y within our sample.    
    \item Figure \ref{figure:y_avg_OH4} presents our final, low metallicity dataset, with y plotted versus O/H, and our resulting \Yp\ determination.  Our resulting value, $\Yp = 0.2458 \pm 0.0013$, attains the unprecedented precision of 0.5\%.  This value is in good agreement with the CMB prediction, and the implications of this result will be explored in the next paper in this series (Paper~V).

    \begin{figure}[t!]
    \resizebox{\columnwidth}{!}{\includegraphics{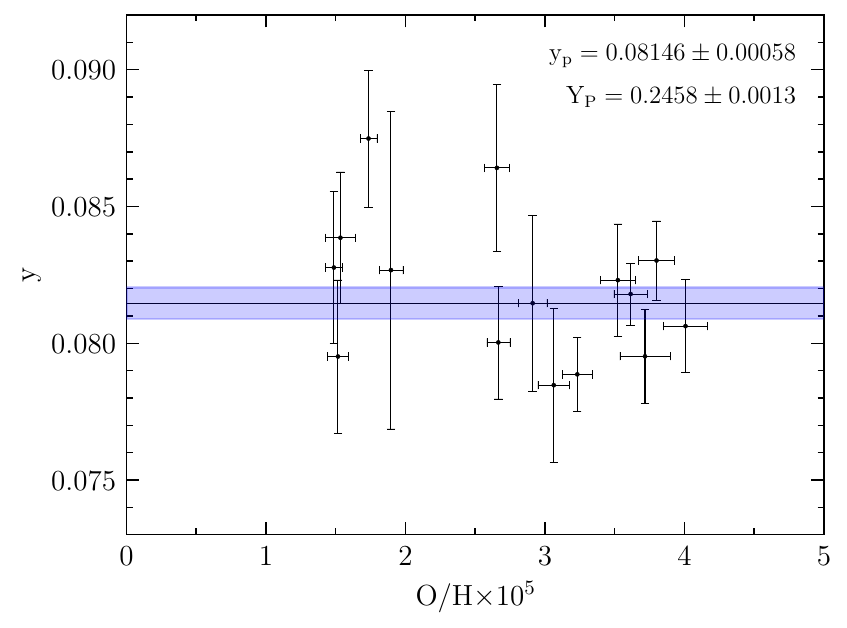}}
    \caption{
    The helium abundance (number density ratio, $\r{y=\frac{n(He)}{n(H)}}$) for the 15 targets in our final qualifying dataset with \OHc\ plotted versus the oxygen-to-hydrogen ratio (O/H).  The weighted average, \yavg, with $\pm 1 \sigma$ uncertainties is shown by the darker (blue) shaded band.  The weighted average result for \OHc\ is provided (\yp), as well as the corresponding \Yp\ value. 
    }
    \label{figure:y_avg_OH4}
    \end{figure}

    \item Our results demonstrate the power of a large dataset of high quality spectra of very low metallicity targets. As such, adding more high SNR, high resolution optical and NIR observations of the lowest metallicity targets remains our highest priority for further improvements in our \Yp\ determination.  We think our work demonstrates that is the future for primordial helium abundance determinations.  

\end{enumerate}



\begin{acknowledgements}

This work was supported by funds provided by NSF Collaborative Research Grants AST-2205817 to RWP, AST-2205864 to EDS, and AST-2205958 to EA. 
Our team workshop at OSU in July 2024 was sponsored in part by OSU's Center for Cosmology and AstroParticle Physics (CCAPP). The work of KAO was supported in part by DOE Grant DE-SC0011842 at the University of Minnesota.  EA benefited greatly from multiple visits to the University of Minnesota and is grateful to the University of Minnesota and the William I.\ Fine Theoretical Physics Institute for the support.  

The authors owe special thanks to Giulio Del~Zanna and Peter Storey for providing their He and H emissivity calculations, respectively, on a finer grid.  Our team greatly appreciated the invitation, discussion, and feedback at the May 2025 CERN and EuCAPT workshop, \textit{News, Opportunities and Challenges in Big Bang Nucleosynthesis}.  

This paper uses data taken with the MODS spectrographs built with funding from NSF grant AST-9987045 and the NSF Telescope System Instrumentation Program (TSIP), with additional funds from the Ohio Board of Regents and the Ohio State University Office of Research.
This paper made use of the modsIDL spectral data reduction pipeline developed in part with funds provided by NSF Grant AST-1108693.

This work is based on observations made with the Large Binocular Telescope. The LBT is an international collaboration among institutions in the United States, Italy and Germany. LBT Corporation Members are: The University of Arizona on behalf of the Arizona Board of Regents; Istituto Nazionale di Astrofisica, Italy; LBT Beteiligungsgesellschaft, Germany, representing the Max-Planck Society, The Leibniz Institute for Astrophysics Potsdam, and Heidelberg University; The Ohio State University, and The Research Corporation, on behalf of The University of Notre Dame, University of Minnesota, and University of Virginia. Observations have benefited from the use of ALTA Center (alta.arcetri.inaf.it) forecasts performed with the Astro-Meso-Nh model. Initialization data of the ALTA automatic forecast system come from the General Circulation Model (HRES) of the European Centre for Medium Range Weather Forecasts.

This research used the facilities of the Italian Center for Astronomical Archive (IA2) operated by INAF at the Astronomical Observatory of Trieste.

All LBT observations for this project were executed remotely beginning during the global COVID-19 pandemic, when we all learned together how to operate the LBT from basements, spare bedrooms, and home offices. We are most grateful for the tireless efforts of the Mount Graham and Tucson technical and observing support staff of the Large Binocular Telescope Observatory who were critical to making it all work smoothly, keeping telescopes and instruments operating in top form under trying circumstances. The 130 hours of high-quality, consistent spectrophotometric data acquired for this project would not have been possible without them.

EDS, KAO, EA, DAB, NSJR, and JHM would like to acknowledge and thank Stanley Hubbard for his generous gift to the University of Minnesota that allowed the University to become a member of the LBT collaboration.

This research has made use of the NASA/IPAC Extragalactic Database (NED) which is operated by the Jet Propulsion Laboratory, California Institute of Technology, under contract with the National Aeronautics and Space Administration and the NASA Astrophysics Data System (ADS). 

\end{acknowledgements}

\facilities{LBT (MODS), LBT (LUCI)}
\software{
\texttt{astropy} \citep{astr2013, astr2018, astr2022},
\texttt{jupyter} \citep{kluy2016},
\texttt{modsIDL} \citep{crox2019},
\texttt{modsCCDRed} \citep{pogg2019},
\texttt{PypeIt} \citep{proc2020},
\texttt{PyNeb} \citep{luri2012,luri2015},
\texttt{numpy} \citep{harr2020},
\texttt{PypeIt} \citep{pypeit:joss_pub, pypeit:zenodo}
}


\clearpage


\appendix \label{Appendix}
\restartappendixnumbering

The extensive Appendices in \citet{aver2021} provide our entire model and the data employed, including the model equations for the helium and hydrogen flux ratios.  This Appendix provides the corresponding updates to the equations and data provided in the \citet{aver2021} Appendices.  This Appendix should be read as a supplement to the \citet{aver2021} Appendices to bring them up to date with our model and data as employed in this work.  

\vspace{0.5cm}

\section{The Blended Emission Line H8 + \3889} \label{Appendix:Blended}

As discussed in Section \ref{Model}, the treatment The blended emission line H8 + \3889 has been revised.  Correspondingly, Eq.\:(A.8) from \citet{aver2021} is replaced by the following:

\beq
\begin{split}
\frac{F \sr{(He3889\!+\!H8)}}{F \sr{(H\beta)}}  = & \; y^{+} \frac{E \sr{(He3889)}}{E \sr{(H\beta)}}    
\frac{\frac{W ({\rm H}\beta)+ \mbox{\ns $a$}_{\ssn{H}} ({\rm H}\beta)}{W ({\rm H}\beta)}} {\frac{W {\rm (He3889)}+\mbox{\ns $a$}_{\ssn{He}} {\rm (He3889)}}{W {\rm (He3889)}}} \,\, \times \\
& f_{\tau} {\sr ({\rm He}3889)} \frac{1}{1+\frac{C}{R}{\sr ({\rm H}\beta)}}10^{-f  {\sxr ({\rm He}3889)} \,C{\sxr ({\rm H}\beta)}} ~~ + \\
& \frac{E {\sr (H8)}}{E {\sr (H\beta)}}
\frac{\frac{W ({\rm H}\beta)+ \mbox{\ns $a$}_{\ssn{H}} ({\rm H}\beta)}{W ({\rm H}\beta)}} {\frac{W {\rm (H8)}+\mbox{\ns $a$}_{\ssn{H}} {\rm (H8)}}{W {\rm (H8)}}} 
\frac{1+\frac{C}{R} {\sr({\rm H}8)}}{1+\frac{C}{R} {\sr ({\rm H}\beta)}} 10^{-f  {\sxr ({\rm H}8)} \,C{\sxr ({\rm H}\beta)}}
\end{split}
\label{eq:F_3889H8_EW}
\eeq

Following the steps in the \citet{aver2021} Appendix, Eq.\:(\ref{eq:F_3889H8_EW}) can be re-expressed to remove $W(\lambda)$ entirely, based on the constrained flux of the continuum at each wavelength, $h(\lambda)$.  This allows one to solve for a consistent emission line ratio relative to H$\beta$ and yields a simplified equation for the flux ratio, replacing Eq.\:(A.17) from \citet{aver2021}:

\beq
\begin{split}
\frac{F \sr{(He3889\!+\!H8)}}{F \sr{(H\beta)}} = & \; \left(y^{+}\frac{E \sr{(He3889)}}{E \sr{(H\beta)}} {f_{\tau}\sr{(He3889)}} + \frac{E \sr{(H8)}}{E \sr{(H\beta)}}(1+\tfrac{C}{R} \sr{(H8)} \dr{)}\right)\\ 
 & \times \frac{W \sr{(H\beta})+\mbox{\ns $a$}_{\ssn{H}} \sr{(H\beta)}}{W \sr{(H\beta)}}\frac{1}{1+\frac{C}{R} \sr{(H\beta)}}
 10^{-f  {\sxr ({\rm He3889+H8})} \,C{\sxr ({\rm H}\beta)}} \\ 
 & \; - \frac{(\mbox{\ns $a$}_{\ssn{He}} \sr{(He3889)}+\mbox{\ns $a$}_{\ssn{H}} \sr{(H8)} \dr{)}}{W \sr{(H\beta)}}\frac{h \sr{(He3889\!+\!H8)}} {h \sr{(H\beta)}}
\label{eq:F_3889H8}
\end{split}
\eeq

For both Eqs.\:(\ref{eq:F_3889H8_EW}) \& (\ref{eq:F_3889H8}), the only change is the removal of the $f_{\tau}(\sr{He}3889)$ term from the H8 flux contribution.  The modeling and treatment of the blended emission line \3889 + H8 had been improved in \citet{aver2021} by treating it as a blended line in the model, rather than subtracting the theoretical flux expected for H8 from the measured, blended flux.  As a result, the theoretical flux was then found by summing the contributions from \3889 and H8.  In calculating the H8 flux, the \3889 radiative transfer term was included due to the two lines only being separated by 0.4~\AA.  However, even with estimated Doppler broadening of $\sim$0.4~\AA, only a small fraction of the H8 emission would have a wavelength that would match the \3889 absorption wavelength.  Furthermore, in terms of the relative effect of that absorption for H8 compared to \3889, it would be diluted by the much smaller number of He atoms per unit volume compared to H atoms (more than ten times less).  Correspondingly, the absorption of H8-emitted photons by He is expected to be relatively rare with negligible impacts on the H8 or He fluxes (\hei \W\W3889, 7065, 10830).

\section{He Emissivities} \label{Appendix:HeEmiss}

The He emissivities used in this work are based on the calculations of \citet{delz2022}.  They are parameterized and incorporated identically to the He emissivities in \citet{aver2021}, which are from \citet{aver2013}.  The only change for this work is the updated emissivity grid point values, based on the improved results of \citet{delz2022}, as discussed in Section \ref{Emiss}.  Those emissivities, as calculated on a higher-resolution temperature and density grid by G.\:Del~Zanna (Private Communication), are attached with this work.  

The finer parametric grid is the same as was implemented in \citet{aver2013} and is summarized in the \citet{aver2021} Appendix.  The helium emissivities are functions of the electron temperature and density, \Te\ \& \n, respectively.  The temperatures for the finer grid are in 250\,K increments from 10,000\,K to 25,000\,K (inclusive), and 31 electron densities are irregularly spaced from 1 to 10,000~\cm\ (also inclusive), with tighter sampling at the lower densities most relevant for helium abundance analysis.  

The helium emissivity ratio, $\frac{E(\lambda)}{E(H\!\beta)}$, is then found by interpolation on the finer parametric grid.  Measured and theoretical fluxes for \10830 are calculated relative to P$\gamma$, rather than H$\beta$.  Correspondingly, the emissivity ratio for \10830, $\frac{E(\lambda)}{E(P\!\gamma)}$, is calculated using the P$\gamma$ emissivity, instead of the H$\beta$ emissivity, and then interpolated on the finer parametric grid.

\section{H Emissivities} \label{Appendix:HEmiss}

The H emissivities for this work are based on the calculations of \citet{stor2015}.  They are parameterized and incorporated identically to \citet{aver2021}, but the fits are now based on the updated results of \citet{stor2015}, as discussed in Section \ref{Emiss}.  Those emissivities, as calculated on a higher-resolution temperature and density grid by P.\:Storey (Private Communication), are attached with this work.  

For all the Balmer lines and for the Optical/NIR Paschen lines in the LBT/MODS spectrum (P8 - P15), the emissivity ratio is calculated relative to H$\beta$, $\frac{E(\lambda)}{E(H\!\beta)}$.  For the IR Paschen lines from the LBT/LUCI spectrum (P$\beta$ \& P$\delta$), the emissivity ratio is calculated relative to P$\gamma$, $\frac{E(\lambda)}{E(P\gamma)}$, though P$\beta$ \& P$\delta$ are not currently used in our analysis.  Using the definition $\T4 = \r{T}/10^{4}$, the functional form of the fits is as follows,

\beq
\frac{E(\lambda)}{E(H\!\beta|P\gamma)} = \sum_{ij} c_{ij} (\log(\T4))^{i} (\log(\n))^{2j}.  
\label{eq:HEmiss}
\eeq

Table \ref{table:HEmiss} lists the coefficients, $c_{ij}$ and replaces Table 6 in the \citet{aver2021} Appendix.



\begin{table}[ht!]
\footnotesize
\centering
\vskip .1in
\begin{tabular}{lcccc}
\hline\hline
\!\!Line                           & 
{i$\downarrow$}			&
{j$\rightarrow$}  & & \\
\hline
& & 0 & 1 & 2 \\
\hline
H$\alpha$	&	0	&	2.87	&	-1.36E-3	&	-5.84E-6	\\
	&	1	&	-0.506	&	4.08E-3	&	1.69E-5	\\
	&	2	&	0.335	&	-4.19E-3	&	-1.14E-5	\\
H$\gamma$	&	0	&	0.468	&	3.83E-5	&	9.02E-7	\\
	&	1	&	2.90E-2	&	-1.23E-4	&	-2.64E-6	\\
	&	2	&	-1.71E-2	&	1.13E-4	&	1.15E-6	\\
H$\delta$	&	0	&	0.259	&	2.67E-5	&	7.54E-7	\\
	&	1	&	2.19E-2	&	-8.91E-5	&	-2.61E-6	\\
	&	2	&	-1.34E-2	&	7.12E-5	&	1.51E-6	\\
H8	&	0	&	0.105	&	9.69E-6	&	6.33E-7	\\
	&	1	&	9.48E-3	&	-4.47E-5	&	-1.83E-6	\\
	&	2	&	-6.43E-3	&	3.23E-5	&	9.78E-7	\\
H9	&	0	&	7.31E-2	&	4.42E-6	&	7.59E-7	\\
	&	1	&	6.29E-3	&	-3.06E-5	&	-1.83E-6	\\
	&	2	&	-4.50E-3	&	1.95E-5	&	1.11E-6	\\
H10	&	0	&	5.30E-2	&	-3.07E-7	&	1.01E-6	\\
	&	1	&	4.26E-3	&	-2.05E-5	&	-1.95E-6	\\
	&	2	&	-3.22E-3	&	1.38E-5	&	1.12E-6	\\
H11	&	0	&	3.97E-2	&	-4.80E-6	&	1.37E-6	\\
	&	1	&	2.93E-3	&	-1.12E-5	&	-2.30E-6	\\
	&	2	&	-2.34E-3	&	5.26E-6	&	1.53E-6	\\
H12	&	0	&	3.05E-2	&	-9.48E-6	&	1.84E-6	\\
	&	1	&	2.06E-3	&	-2.73E-6	&	-2.80E-6	\\
	&	2	&	-1.73E-3	&	-5.29E-8	&	1.84E-6	\\
P$\beta$	&	0	&	1.81	&	-4.25E-4	&	-6.62E-6	\\
	&	1	&	-0.173	&	1.08E-3	&	1.73E-5	\\
	&	2	&	8.55E-2	&	-9.08E-4	&	-1.26E-5	\\
P$\delta$	&	0	&	0.613	&	5.55E-5	&	1.50E-6	\\
	&	1	&	2.85E-2	&	-1.31E-4	&	-4.29E-6	\\
	&	2	&	-1.30E-2	&	7.84E-5	&	3.56E-6	\\
P8	&	0	&	3.66E-2	&	-1.45E-6	&	-9.59E-8	\\
	&	1	&	-8.54E-3	&	-9.02E-7	&	1.16E-7	\\
	&	2	&	-1.05E-3	&	6.19E-6	&	-2.23E-8	\\
P9	&	0	&	2.54E-2	&	-5.02E-7	&	-8.02E-9	\\
	&	1	&	-5.61E-3	&	-2.19E-6	&	-1.08E-7	\\
	&	2	&	-9.68E-4	&	5.76E-6	&	1.17E-7	\\
P10	&	0	&	1.84E-2	&	-2.99E-7	&	5.47E-8	\\
	&	1	&	-3.95E-3	&	-1.61E-6	&	-2.89E-7	\\
	&	2	&	-7.90E-4	&	3.47E-6	&	3.22E-7	\\
P11	&	0	&	1.38E-2	&	-5.26E-7	&	1.21E-7	\\
	&	1	&	-2.91E-3	&	-4.06E-7	&	-4.62E-7	\\
	&	2	&	-6.21E-4	&	1.53E-6	&	4.83E-7	\\
P12	&	0	&	1.06E-2	&	-9.50E-7	&	1.94E-7	\\
	&	1	&	-2.23E-3	&	1.24E-6	&	-6.65E-7	\\
	&	2	&	-4.83E-4	&	-5.68E-7	&	6.89E-7	\\
P13	&	0	&	8.32E-3	&	-1.56E-6	&	2.82E-7	\\
	&	1	&	-1.76E-3	&	3.26E-6	&	-9.03E-7	\\
	&	2	&	-3.73E-4	&	-3.09E-6	&	9.37E-7	\\
P14	&	0	&	6.66E-3	&	-2.32E-6	&	3.86E-7	\\
	&	1	&	-1.42E-3	&	5.45E-6	&	-1.18E-6	\\
	&	2	&	-2.89E-4	&	-5.42E-6	&	1.21E-6	\\
P15	&	0	&	5.41E-3	&	-3.20E-6	&	5.07E-7	\\
	&	1	&	-1.17E-3	&	7.71E-6	&	-1.47E-6	\\
	&	2	&	-2.25E-4	&	-7.59E-6	&	1.49E-6	\\
\hline
\end{tabular}
\caption{Fit coefficients for the hydrogen emissivities, $c_{ij}$, in Eq.\:(\ref{eq:HEmiss})}
\label{table:HEmiss}
\end{table}


As discussed in Sections \ref{Emiss} \& \ref{EmissComp}, the fit equations for the emissivities of H$\beta$ and P$\gamma$, E(H$\beta$) \& E(P$\gamma$), have been updated based on the newest \citet{stor2015} calculations and the higher-resolution density and temperature grid.  Eq.\:(\ref{eq:HB}) replaces Eq.\:(A.21) from \citet{aver2021}, and Eq.\:(\ref{eq:PG}) replaces Eq.\:(A.22).

\beq
\begin{split}
E(H\!\beta) = & \; \Biggl{[} -2.79258\times10^5 + 3.69812\times10^4~\ln T \\ 
              & ~~ - 1.47201\times10^3~(\ln T)^2 + \frac{6.98598\times10^5}{\ln T} \Biggr{]} ~\frac{1}{T}\\
              & \times \Bigl{[} 1.0 + 87.55976~(\log_{10} n_e)^2 \times T^{\;-1.34756} \Bigr{]} \\
              & \times 10^{-25} \textrm{ ergs cm}^{3} \textrm{ s}^{-1}.
\label{eq:HB}
\end{split}
\eeq

\beq
\begin{split}
E(P\!\gamma) = & \; \Biggl{[} 6.04710\times10^4 + 8.96867\times10^2~\ln T \\ 
               & ~~ - 2.59600\times10^2~(\ln T)^2 - \frac{3.27429\times10^5}{\ln T} \Biggr{]} ~\frac{1}{T}\\
               & \times \Bigl{[} 1.0 + 3.37185\times10^3~(\log_{10} n_e)^2 \times T^{\;2.11469} \Bigr{]} \\
               & \times \rm{SIGN}(3.0\times10^4 - 5.0\times10^4~\log_{10} n_e - T)  \\
               & \times (\lvert3.0\times10^4 - 5.0\times10^4~\log_{10} n_e - T\rvert)^{1/4}  \\
               & \times 10^{-26} \textrm{ ergs cm}^{3} \textrm{ s}^{-1}.
\label{eq:PG}
\end{split}
\eeq

\section{Radiative Transfer} \label{Appendix:RadTrans}

The radiative transfer calculations employed in this work come from \citet{kuri2025}, as discussed in Section \ref{RadTrans}. Radiative transfer affects the He triplet lines, and the correction to the He fluxes for radiative transfer is given by $f_\tau(\lambda)$, and depends on the \hii region's optical depth, $\tau$, as well as electron density and temperature, \n\ \& \Te.  The calculations are parameterized in terms of the optical depth for \3889 at line center, such that the model parameter, $\tau$, represents the optical depth for \3889 (i.e., $\tau = \tau_{\ssn{3889}}$).    \citet{kuri2025} helpfully provide fit equations for $f_\tau(\lambda)$ over the parameter ranges $8,000\,\r{K} < \Te < 22,000\,\r{K}$, $1\,\cm < \n < 10,000~\cm$, and $0 < \tau < 10$ with an accuracy of $\lesssim0.1\%$.  Those fit equations replace Eq.\:(A.24) from \citet{aver2021}, and are reproduced below from \citet{kuri2025}, with the fit coefficients reproduced in Table \ref{table:RadTrans}.  

\beq
    f_\tau(n_e, T) = \frac{1 + a(n_e, T) \times\tau}{1 + b \times\tau}, 
    \label{eq:RadTrans}
\eeq
with
\beq
a(n_e, T) = \sum_{i=0}^5 A_i(T) x^i~,~~x=\log_{10}\left(\frac{n_e}{10^2\,\cm}\right).  \nonumber
\eeq


\begin{table}[ht!]
\centering
\begin{tabular}{lrrrr}
\hline
& \multicolumn{1}{c}{$B_0$} & \multicolumn{1}{c}{$B_1$} & \multicolumn{1}{c}{$B_2$} & \multicolumn{1}{c}{$B_3$}  \\
\hline

& \multicolumn{4}{c}{$\lambda 3889$, $b= $ 5.854E-02}  \\
\hline
$A_0$ & 7.399E-04 & 2.506E-04 & -2.762E-04 & 1.422E-04  \\
$A_1$ & 1.806E-04 & -3.057E-04 & -8.839E-05 & 1.853E-04  \\
$A_2$ & -9.546E-05 & 1.822E-05 & -3.319E-04 & 2.257E-04  \\
$A_3$ & -9.595E-05 & -3.434E-05 & -2.756E-04 & 4.107E-04  \\
$A_4$ & 4.975E-05 & -1.233E-04 & 2.953E-05 & 1.700E-04  \\
$A_5$ & -2.042E-06 & 4.591E-05 & 4.476E-05 & -1.138E-04  \\
\hline

& \multicolumn{4}{c}{$\lambda 4026$, $b= $ -9.988E-03}  \\
\hline
$A_0$ & -9.095E-03 & 7.274E-04 & 3.047E-04 & -7.155E-05  \\
$A_1$ & 1.007E-04 & -7.632E-05 & 9.027E-05 & -4.993E-04  \\
$A_2$ & 3.797E-06 & -5.135E-06 & 1.064E-04 & -6.808E-04  \\
$A_3$ & -3.394E-05 & 1.106E-05 & -1.811E-04 & -3.631E-04  \\
$A_4$ & 1.486E-05 & -2.175E-05 & -1.611E-04 & 1.065E-04  \\
$A_5$ & -1.713E-06 & 4.692E-06 & 6.341E-05 & 7.705E-05  \\
\hline

& \multicolumn{4}{c}{$\lambda 4471$, $b= $ 2.118E-03}  \\
\hline
$A_0$ & 4.204E-03 & 9.874E-04 & 8.248E-04 & -3.394E-04  \\
$A_1$ & -5.567E-04 & 4.938E-04 & -4.527E-04 & -8.866E-04  \\
$A_2$ & 4.332E-05 & -4.700E-05 & 8.129E-05 & -1.352E-03  \\
$A_3$ & 1.922E-04 & -2.217E-04 & -6.444E-05 & -4.032E-04  \\
$A_4$ & -8.242E-05 & 5.455E-05 & -2.957E-04 & 4.713E-04  \\
$A_5$ & 7.268E-06 & 5.672E-06 & 9.692E-05 & -2.058E-06  \\
\hline

& \multicolumn{4}{c}{$\lambda 5876$, $b= $ 4.068E-02}  \\
\hline
$A_0$ & 4.507E-02 & 3.753E-03 & 1.882E-03 & -1.899E-03  \\
$A_1$ & -2.264E-04 & -1.975E-04 & -1.024E-03 & -5.707E-03  \\
$A_2$ & 8.758E-05 & -3.356E-04 & -1.676E-03 & -5.033E-03  \\
$A_3$ & 9.324E-05 & -3.737E-04 & -1.951E-03 & 4.196E-03  \\
$A_4$ & -7.755E-05 & -8.592E-05 & -3.971E-04 & 4.138E-03  \\
$A_5$ & 1.304E-05 & 8.815E-05 & 5.352E-04 & -1.834E-03  \\
\hline

& \multicolumn{4}{c}{$\lambda 7065$, $b= $ 5.461E-02}  \\
\hline
$A_0$ & 2.361E-01 & -8.759E-02 & -5.182E-02 & 8.783E-02  \\
$A_1$ & -1.132E-02 & -6.569E-02 & -2.734E-02 & 1.896E-01  \\
$A_2$ & -1.440E-02 & -7.036E-02 & 5.927E-02 & 1.441E-01  \\
$A_3$ & -6.555E-03 & 6.632E-03 & 1.216E-01 & -1.965E-01  \\
$A_4$ & 1.941E-03 & 3.070E-02 & 1.702E-02 & -1.396E-01  \\
$A_5$ & 9.679E-04 & -8.075E-03 & -3.082E-02 & 7.805E-02 \\
\hline

& \multicolumn{4}{c}{$\lambda 10830$, $b= $ 6.368E-02}  \\
\hline
$A_0$ & 7.667E-02 & -5.558E-03 & -6.043E-03 & 1.144E-02  \\
$A_1$ & -5.435E-03 & -1.455E-02 & 1.262E-02 & 3.525E-03  \\
$A_2$ & -3.630E-03 & 1.509E-03 & 3.039E-02 & -3.820E-02  \\
$A_3$ & 1.018E-03 & 7.582E-03 & -8.689E-03 & -3.769E-03  \\
$A_4$ & 1.281E-03 & -4.800E-04 & -1.722E-02 & 2.311E-02  \\
$A_5$ & -4.396E-04 & -9.710E-04 & 6.709E-03 & -6.684E-03  \\
\hline

\end{tabular}
\caption{
Fit coefficiencts $b$ and $B_j^{(i)}$ for the optical depth correction function Eq.\:(\ref{eq:RadTrans}), as given in and reproduced from \citet{kuri2025}.
}
\label{table:RadTrans}
\end{table}


\section{Wavelength-Dependent Underlying Stellar Absorption} \label{Appendix:UA}

\citet{aver2021} includes coefficients for scaling relative underlying stellar absorption contributions to the set of He emission lines employed in that work.  Those coefficients were calculated using spectral energy distributions from BPASS \citep{eldr2009, eldr2017, stan2018}.  Because Leo\,P did not include detections for \hei \W4388 \& \W7281, scaling coefficients for those emissions lines were not included.  The scaling coefficients for those two lines are 0.707 \& 0.165, respectively.  As discussed above (\S \ref{Eduardo}), \heil 7281, along with \heil 5016, are not utilized in our analysis here, due to evidence of systematic bias in these lines \citep{mend2025}.

\section{Hydrogen Emission Lines Paschen Series P13 - P15} \label{Appendix:P13-15}

As enabled by the high resolution and broad wavelength coverage of MODS, Paschen series emission lines P8 through P12 were observed and utilized for the analysis of Leo P in \citet{aver2021}.  Many of the \hii regions in our galactic survey for this work have sufficient SNR to allow clear detections of P13, P14, and P15.  As a result, they were added to our model.  The H emissivity fit equation coefficients for P13 - P15 have already been included along with the other employed H emission lines in Table \ref{table:HEmiss} above.  

As with all of our other emission lines, the reddening coefficients, $f(\lambda)$, found in the terms $10^{-f(\lambda)C(H\beta)}$, are calculated from the extinction fits of \citet{card1989} using $R=3.1$ ($R=\frac{A(V)}{E(B-V)}$).  

For modeling underlying absorption in terms of equivalent width, the Paschen lines are scaled relative to P$\gamma$.  For the Paschen lines, the ratios were all very close to unity.  The deviations from unity in the averages between the Paschen lines were not established to be significant, but instead the deviations may be dominantly due to random variation.  As a result, the underlying absorption correction (in terms of equivalent of width) for all of the Paschen lines, including P13, P14, \& P15, is taken to be the same, with no variation by line/with wavelength.  

For the contribution to the hydrogen fluxes from neutral hydrogen collisional excitation, the relative amount of collisional-to-recombination emission can be calculated as follows,

\beq
\frac{C}{R}(\lambda) = {\xi_{4}} \times 10^{-4} \times \frac{K_{eff}}{\alpha_{eff}} = {\xi_{4}} \times \sum_{i}a_{i}\exp(-\frac{b_{i}}{\T4}){\T4}^{c_{i}}, 
\label{eq:CR2}
\eeq

where $\xi = \frac{\r{n(\hi)}}{\r{n(\hii)}}$.  That parametrization is unchanged from \citet{aver2021}.  Also as in \citet{aver2021}, collisional excitations to P13, P14, \& P15 are extrapolated according to a scaling law  due to the available collisional excitation calculations not including $\r{n} > 8$ \citep{aver2021,fern2016}.  

For P13, P14, \& P15, Table \ref{table:HIcoll} lists the coefficients for constructing $\frac{K_{eff}}{\alpha_{eff}} \times 10^{-4}$, which is the quantity that multiplies $\xi_{4}$ and provides the increasing collisional enhancement factor for increasing temperature ($\T4 = \r{T}/10^{4}$).  Table \ref{table:HIcoll} augments Table 8 from \citet{aver2021}.


\begin{table}[ht!]
\footnotesize
\centering
\vskip .1in
\begin{tabular}{lcccc}
\hline\hline
Line	&	Coefficient	&	Terms \\
\hline
P13	&	a	&	2.5140	\\
	&	b	&	-15.689	 \\
	&	c	&	1.0223	\\
\hline	
P14	&	a	&	2.4148	\\
	&	b	&	-15.702	\\
	&	c	&	1.0305	\\
\hline	
P15	&	a	&	2.3287	\\
	&	b	&	-15.712	\\
	&	c	&	1.0374	\\
\hline
\end{tabular}
\caption{Coefficients for the Neutral Hydrogen Collisional Correction for P13-P15 in Eq.\:(\ref{eq:CR2}).  This table augments Table 8 from \citet{aver2021}.}
\label{table:HIcoll}
\end{table}

\section{Accounting for A Large Dispersion in Data}
\label{Sfact}

A large quantity of lower quality data (e.g., SDSS observations) not only increases the vulnerability to systematic uncertainties, but artificially reduces the statistical uncertainty in the determination of \Yp.

When calculating the weighted mean, we apply a correction to the calculated uncertainty in the mean due to dispersive data.  Given a set of helium abundances, $Y_i \pm \sigma_i$,
the weighted mean is given by 
\beq
{\bar Y} \pm {\bar \sigma} = \frac{\Sigma_i w_i Y_i}{\Sigma_i w_i} \pm (\Sigma_i w_i)^{-\frac12} \, , 
\eeq
where 
\beq
w_i = \frac{1}{\sigma_i^2}.
\label{wi}
\eeq
Given enough sample points, even with relatively large $\sigma_i$, an artificially small value of ${\bar \sigma}$ will result.
The amount of dispersion can be evaluated relative to the mean value by calculating the $\chi^2$ per degree of freedom, which is simply,
\beq
S^2 \equiv \frac{\chi^2}{N-1} = \frac{\Sigma_i w_i ({\bar Y} - Y_i)^2 } {N-1} \, ,
\eeq
where $N$ is the number of independent helium measurements. In the event that $S^2 > 1$ (but not very large), we correct the calculated uncertainty
so that the mean and uncertainty is given by
\beq
{\bar Y} \pm {\bar \sigma}\times S \, .
\eeq
This is the standard procedure used in the Review of Particle Physics \citep{rpp}.
If $S \le 1$, no correction is made. 
If $S \gg 1$, one should question the validity of averaging the data. 

For linear regressions, we compute the intercept (corresponding to primordial helium) and its uncertainty as well as the slope (and its uncertainty) using the uncertainties in both $Y_i$ and the oxygen abundance, $X_i \pm \sigma_{X_i}$.  
More specifically we use York's method \citep{1966CaJPh..44.1079Y} as described in detail in \citep{1989AmJPh..57..642R,1992AmJPh..60...59R}.
These also include the factor of $S$ when computing uncertainties. The $S$ factor is now given by
\beq
S^2 = \frac{\Sigma_i w_i (a + b X_i -Y_i)^2}{N-2} \, ,
\eeq
where now 
\beq
w_i = \frac{w_{X_i} w_{Y_i}}{w_{X_i}+b^2 w_{Y_I}} \, ,
\eeq
and $a$ and $b$ are the derived intercept and slope respectively. The individual weights are derived as in Eq.~(\ref{wi}) for $X$ and $Y$ separately.  We are assuming that the uncertainties are uncorrelated. The expressions for the slope (which is determined by the minimization of a cubic equation in the slope), intercept and their respective uncertainties are somewhat cumbersome and we refer the reader to the references given above. 

For linear regressions, we also use the program \textsc{linmix}, a Bayesian MCMC approach to linear regression allowing for uncertainties in both x and y \citep{kell2007}.  \textsc{linmix} determines its returned slope and intercept from the median values from 10,000 Markov Chains.  \textsc{linmix} has the additional attribute that it fits for an intrinsic scatter in the data.  For typical datasets, the results from York's method and \textsc{linmix} should be comparable.



\newpage
\clearpage

\bibliographystyle{aasjournalv7}
\bibliography{yp_total_bib}

\clearpage

\end{document}